\documentclass[fleqn,usenatbib]{mnras}

\usepackage{newtxtext,newtxmath}

\usepackage[T1]{fontenc}

\DeclareRobustCommand{\VAN}[3]{#2}
\let\VANthebibliography\thebibliography
\def\thebibliography{\DeclareRobustCommand{\VAN}[3]{##3}\VANthebibliography}

\usepackage{graphicx}	
\usepackage{amsmath}	
\usepackage{xcolor}
\usepackage{float}
\usepackage{siunitx}
\usepackage{setspace}
\usepackage{geometry}
\usepackage{graphicx}
\usepackage{natbib}
\usepackage{pdfpages}
\usepackage{subcaption}
\usepackage[toc,page]{appendix}
\usepackage{threeparttable}
\usepackage{bm}

\newcommand{\Halpha}{H$\rm \alpha$ }
\newcommand{\Porb}{$\rm P_{\rm orb}$ }

\title{Investigating Kozai-Lidov Oscillations and Disc Tearing in Be Star Discs}

\author[M.W. Suffak et al.]{
M.W. Suffak,$^{1}$\thanks{E-mail: msuffak@uwo.ca}
C.E. Jones,$^{1}$
A.C. Carciofi,$^{2}$
\\
$^{1}$Department of Physics and Astronomy, Western University, London, ON N6A 3K7, Canada\\
$^{2}$Instituto de Astronomia, Geof\'isica e Ci\'encias Atmosf\'ericas, Universidade de S\~ao Paulo, S\~ao Paulo, SP 05508-090, Brazil\\
}

\date{}

\pubyear{}

\begin{document}
\label{firstpage}
\pagerange{\pageref{firstpage}--\pageref{lastpage}}
\maketitle

\begin{abstract}
Recent simulations of Be stars in misaligned binary systems have revealed that misalignment between the disc and binary orbit can cause the disc to undergo Kozai-Lidov (KL) oscillations or disc-tearing. We build on our previous suite of three-dimensional smoothed particle hydrodynamics simulations of equal-mass systems by simulating eight new misaligned Be star binary systems, with mass-ratios of 0.1 and 0.5, or equal-mass systems with varying viscosity. We find the same phenomena occur as previously for mass ratios of 0.5, while the mass ratio of 0.1 does not cause KL oscillations or disc-tearing for the parameters examined. With increased viscosity in our equal-mass simulations, we show that these phenomena and other oscillations are damped out and do not occur. We also briefly compare two viscosity prescriptions and find they can produce the same qualitative disc evolution. Next, we use the radiative transfer code \textsc{hdust} to predict observable trends of a KL oscillation, and show how the observables oscillate in sync with disc inclination and cause large changes in the polarization position angle. Our models generate highly complex line profiles, including triple-peak profiles that are known to occur in Be stars. The mapping between the SPH simulations and these triple-peak features gives us hints as to where they originate. Finally, we construct interferometric predictions of how a gap in the disc, produced by KL oscillations or disc-tearing, perturbs the visibility versus baseline curve at multiple wavelengths, and can cause large changes to the differential phase profile across an emission line.
\end{abstract}




\section{Introduction}
\label{sec:intro}

Classical Be stars (Be stars) are rapidly rotating, main-sequence or slightly evolved B type stars whose spectrum has, or has had, Balmer lines in emission \citep{Collins1987}. These emission lines emanate from a gaseous circumstellar decretion disc, which has formed via material ejected from the equator of the star. The mass-ejection mechanism is likely a combination of rapid rotation and non-radial pulsations \citep{Baade2016}. In addition to Balmer emission lines, Be star discs produce noticeable excess continuum emission, and a linear polarization signature \citep{Halonen2013}. Interferometry has also been used to confirm the flat geometry of Be star discs \citep{Quirrenbach1997}, detect spiral density waves in these discs \citep{Carciofi2009}, refine binary orbits and detect companions \citep{tycner2011revised, Klement2021}, and approximate the emitting area of different wavelengths within a disc \citep{millan2010spectro}. See \cite{rivinius2013classical} for a thorough review.

The viscous decretion disc (VDD) model of \cite{Lee1991} is the widely accepted best model of Be star discs, and has been successfully implemented in many studies of single stars \citep[e.g.][]{Klement2015,Ghoreyshi2018,Marr2021} as well as larger statistical studies \citep[e.g.][]{Vieira2015, Rimulo2018, Rubio2023}. In the VDD model, mass that is injected into the disc moves outwards via viscous forces. It is now well established that the disc can alternate between being an outward flowing decretion disc, and an inward flowing accretion disc, depending on whether the mass injection from the star to the disc is occurring or not. The effect that the disc has on the observables as it grows and dissipates depends on the inclination that the disc is being viewed \citep{Haubois2012, Haubois2014}. For a face-on disc, the continuum will brighten as the disc grows and dim as the disc dissipates, while for an edge-on view, the opposite will occur for a sufficiently dense disc. Linear polarization will increase and decrease as the disc grows and reaccretes, however it is negligible for a face-on disc due to symmetry and peaks at an inclination of \ang{70} for a flat disc \citep{Wood1996}. The shape of emission lines will be single-peaked for a face-on disc, and double-peaked for any other inclination \citep{Struve1931}, while the strength of the line is dependent on the disc density and continuum emission level. Interferometric measurements of a Be star disc will usually yield an approximately Gaussian curve when plotting visibility versus baseline \citep[for example,][]{Quirrenbach1994}, while the differential phase versus wavelength is expected to present an S-shaped curve across an emission line due to the rotation of the disc and separate velocities emitting from different parts of the disc \citep{Stee1996, Faes2013}.

Be stars, like all massive stars, are frequently found in binary systems \citep{Oudmaijer2010}, and it has been suggested that all Be stars may be in binary systems \citep{Klement2019}. As such, there have been many studies examining how a binary companion can affect the evolution of a Be star disc, from an analytical perspective \citep[][for example]{Okazaki1991, Ogilvie2008, Martin2011}, using hydrodynamic simulations \citep[such as][]{Okazaki2002, Martin2014, Cyr2017, Suffak2022}, and directly modelling observations \citep[][etc.]{Bjorkman2002, Escolano2015, Silaj2016, Suffak2020, Marr2022}. From these studies, we now understand that the effect a binary companion has on a disc not only depends on its mass, orbital period and eccentricity, but also on whether the companion's orbit is coplanar or misaligned with the disc, if the disc is orbiting prograde or retrograde with respect to the orbit of the disc itself, and of course the value of the disc viscosity parameter, $\alpha_{\rm ss}$, plays a substantial role. In the ideal coplanar, prograde orbit, the companion truncates (i.e., limits the radial extent of) the disc, and can also induce asymmetric $m$ = 2 density modes within the disc \citep{Okazaki2002}, causing emission lines to have cyclic V/R ratios \citep{panoglou2018discs}. If the binary orbit is misaligned to the disc, these spiral density enhancements are shown to have a tighter winding \citep{Cyr2020}. The disc is also less truncated with increasing misalignment \citep{Brown2019}, and can undergo precession, as well as the phenomena of Kozai-Lidov (KL) oscillations and disc-tearing given certain conditions \citep{Martin2019,Suffak2022,Suffak2024}. Alternatively, coplanar retrograde orbits have been shown to lack a truncation effect \citep{Lubow2015}, and also cause the disc to be unstable to tilting \citep{Overton2024}. The highly eccentric Be star, $\delta$ Sco, likely has a retrograde binary companion \citep{che2012imaging}, which has been shown to have a minimal impact on the disc at each periastron passage due to the short interaction time associated with an eccentric retrograde orbit \citep{Rast2024}.

Despite many Be stars being confirmed to be in binary systems, there has been a startling lack of main-sequence companions detected among them \citep{Bodensteiner2020}. The binary companions of Be stars have frequently been found to be faint stripped companions; either subdwarf OB stars, white dwarfs, or neutron stars \citep{Klement2024}. This has provided evidence that many Be stars are formed through binary mass transfer \citep{Dodd2024}, with mass from the (once more massive) component adding excess angular momentum to the Be star, allowing it to achieve rapid rotation which, in turn, is believed to be the key factor allowing Be stars to shed excess angular momentum and mass at its equator to create the viscous decretion disc.

Among Be binary systems, a subset of Be/X-ray binary systems has emerged, involving a Be star and a compact object companion \citep{Reig2011}. The interaction of the compact object with the disc can produce large X-ray outbursts either periodically (Type I) or sporadically (Type II) \citep{Martin2014b}. A number of Be/X-ray binaries are thought to have misaligned orbits, likely due to a supernova kick when the neutron star is formed \citep{Martin2011}. This misalignment not only affects the X-ray emission, but also drives superorbital period variations in the continuum \citep{Martin2024}.

In our previous work \citep[][hereafter Paper I]{Suffak2022}, we computed and analyzed six smoothed particle hydrodynamics (\textsc{sph}) simulations of misaligned equal-mass Be star binary systems, where we varied the misalignment angle between \ang{20}, \ang{40} and \ang{60}, as well as the orbital period between 30 and 300 days. These simulations showed how the disc can tilt, warp, and tear while mass-injection into the disc is occurring, and then precess and undergo KL oscillations while the mass-injection mechanism is turned off. This study has garnered attention from both the Be star and the Be/X-ray communities. Consequently, we are now extending our previous investigation to compare the behaviour of systems with unequal masses and varying viscosity. 

We also build on Paper I by examining how the observables of these systems change over time. This was partially accomplished in \cite{Suffak2024}, where we used an interface between the \textsc{sph} code and the 3-dimensional (3D) non-local thermodynamic equilibrium (nLTE) Monte Carlo radiative transfer code \textsc{hdust} \citep{carciofi2006non} to compute observables of the disc-tearing \textsc{sph} model from Paper I. We found the trends of the disc-tearing model remarkably close to the trends of the Be star Pleione (28 Tau), which had previously been hypothesized to have a tearing disc \citep{Marr2022, Martin2022}. Our results also showed that the outer disc in the tearing-disc configuration contributes substantially to the \Halpha line shape, and to the polarization position angle.

In this paper, we will first discuss the evolution of the systems in comparison to their equal-mass counterparts. We then utilize \textsc{hdust} in the same manner as \cite{Suffak2024} to predict observables of an asymmetric disc during KL oscillations. We also present interferometric predictions of a torn disc and a KL oscillating disc. Section \ref{sec:methods} details our \textsc{sph} simulation parameters, Section \ref{sec:vary_q} shows the evolution of discs in systems of different mass ratio, while Section \ref{sec:vary_alpha} presents results of varying viscosity and comparing viscosity prescriptions. Section \ref{sec:obsverables} examines the observables produced from \textsc{hdust}, Section \ref{sec:interferometry} presents our interferometric predictions, and Sections \ref{sec:discussion} and \ref{sec:conclusions} contain our discussion and conclusions.

\section{Methodology}
\label{sec:methods}

\subsection{\textsc{sph}}
\label{sec:sph}

The \textsc{sph} code used in this study has also been highlighted in \cite{Okazaki2002}, \cite{panoglou2016discs}, and \cite{Cyr2017}, among other works. It was developed by \cite{Benz1990Book} and \cite{Benz1990}, and refined by \cite{Bate1995} to include a second order Runge-Kutta-Fehlburg integrator. It was adapted for use with Be star decretion discs by \cite{Okazaki2002}.

Our simulations start with no disc, and then at every individual timestep, 5000 equal-mass particles are injected into the equatorial plane of the primary star at a radial distance of 1.04 $\rm R_*$. This is equivalent to 500,000 particles injected over an orbital period. We continue particle injection for 100 orbital periods, then turn particle injection off to allow the disc to dissipate. Note that most of the injected mass almost immediately falls back onto the primary star, and only a small fraction remains in the disc.

We define the disc viscosity, $\nu$, to mimic the Shakura-Sunyaev viscosity \citep{Shakura1973}
\begin{equation}
    \nu = \alpha_{\rm ss} c_{\rm s} H,
\end{equation}
where $H$ is the theoretical disc scale height ($c_s$/$\Omega$), $c_{\rm s}$ is the isothermal sound speed, and $\alpha_{\rm ss}$ is a dimensionless scaling parameter defined as the ratio of the turbulent velocity to the sound speed, which we use as a free input parameter. Internally, the code uses the standard \textsc{sph} artificial viscosity introduced by \cite{Monaghan1983}, with two artificial parameters $\alpha_{\rm sph}$ and $\beta_{\rm sph}$, but when implementing the Shakura-Sunyaev viscosity we set $\beta_{\rm sph}$ = 0, so that for a chosen constant value of $\alpha_{\rm ss}$
\begin{equation}
    \alpha_{\rm sph} = 10\, \alpha_{\rm ss} \frac{H}{h},
    \label{eq:alpha_conv}
\end{equation}
where $h$ is the average smoothing length (the distance over which a particle's quantities are smoothed) of two particles when calculating viscous forces. This change was first implemented and described in more detail by \cite{Okazaki2002}.

\begin{figure}
    \centering
    \includegraphics[scale=0.35]{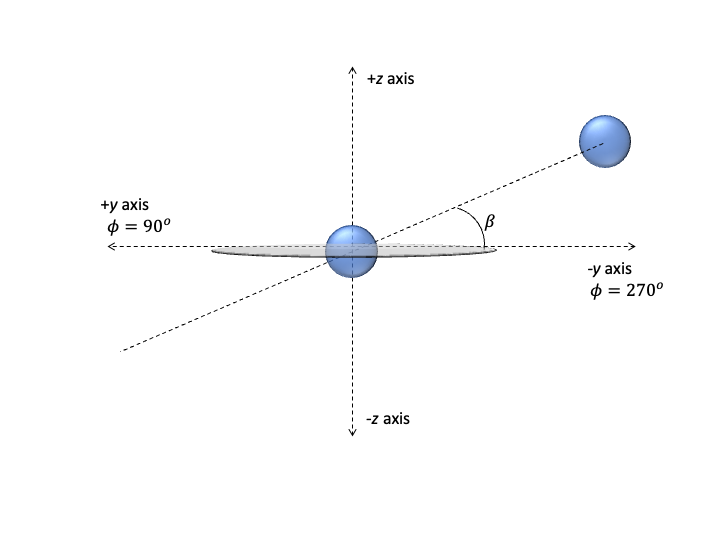}
    \caption{Schematic of our simulation setup with a misaligned binary companion. The primary and secondary stars are shown as blue spheres, while the disc is shown in light grey in the $xy$ plane. Labelled are the $y$ and $z$ axes, the misalignment angle $\beta$, and azimuthal angles $\phi$ of \ang{90} and \ang{270}, which are aligned with the positive and negative $y$-axis, respectively.}
    \label{fig:tilted_sec_schem}
\end{figure}

The geometry of our simulation is such that the spin axis of the primary star is aligned with the $z$-axis, so the equatorial plane of the primary star, and initial plane of the disc where the particles are injected, is the $xy$-plane. When the secondary star's orbit is misaligned, the orbit is rotated about the $x$-axis, and so the misalignment angle $\beta$ is measured from the $y$-axis. A schematic showing this setup is presented in Figure \ref{fig:tilted_sec_schem}. We define the azimuthal angle such that $\phi\,=\,\ang{0}$ along the +$x$-axis and the polar angle such that $\theta\,=\,\ang{0}$ in the +$z$ direction.

\begin{table}
\centering
\caption{Simulation parameters of our base \textsc{sph} models.}
\label{tab:sph_sim_params}
\begin{threeparttable}[c]
\renewcommand{\TPTminimum}{\linewidth}
\makebox[\linewidth]{
\vspace{3mm}
\begin{tabular}[c]{ccc}
\hline\hline
    Parameter & Symbol & Value  \\
    \hline
    Be star mass & $M_{\rm p}$ & 8 $\rm M_\odot$ \\
    Be star radius & $R_{\rm p}$ & 5 $\rm R_\odot$ \\
    Be star effective temperature & $T_{\rm eff}$ & 20000 K \\
    Mass ratio & $q$ & 1.0 \\
    Disc temperature & $T_{\rm d}$ & 12000 K \tnote{a} \\
    Viscosity parameter & $\alpha_{\rm ss}$ & 0.1 \\ 
    Mass injection rate & $\dot{M}_{\rm inj}$ & $10^{-8} \, \rm M_\odot yr^{-1}$ \\
    Particle injection radius & $r_{\rm inj}$ & $1.04 \, \rm R_p$ \\
    Misalignment Angle & $\beta$ & \ang{40}/\ang{60} \\ 
    Orbital Period & $P_{\rm orb}$ & 30 days\\
\hline
\end{tabular}}
\begin{tablenotes}
\item[a] The disc temperature is set to 60\% of the primary star's effective temperature. This value was found by \cite{carciofi2006non} to be the average temperature in the isothermal regions of the disc.
\end{tablenotes}
\end{threeparttable}
\end{table}

\begin{table}
\centering
\caption{Parameters of our new simulations that vary from the base model parameters in Table \ref{tab:sph_sim_params}.}
\label{tab:sph_new_sim_params}
\renewcommand{\TPTminimum}{\linewidth}
\makebox[\linewidth]{
\vspace{3mm}
\begin{tabular}[c]{cccc}
\hline\hline
    Model Number & $\beta$ & $\alpha_{\rm ss}$ & $q$  \\
    \hline
    1 & \ang{40} & 0.1 & 0.1 \\
    2 & \ang{40} & 0.1 & 0.5 \\
    3 & \ang{40} & 0.5 & 1.0 \\
    4 & \ang{40} & 1.0 & 1.0 \\
    5 & \ang{60} & 0.1 & 0.1 \\
    6 & \ang{60} & 0.1 & 0.5 \\
    7 & \ang{60} & 0.5 & 1.0 \\
    8 & \ang{60} & 1.0 & 1.0 \\
\hline
\end{tabular}}
\end{table}

The parameters from two of our simulations in Paper I, which we call our base models, are presented in Table \ref{tab:sph_sim_params}. The model with a $\ang{40}$ misalignment exhibited disc-tearing, while the $\ang{60}$ misaligned model underwent KL oscillations during disc dissipation. In eight new models, we individually vary the mass ratio $q$ (0.1, 0.5, or 1.0), and the Shakura-Sunyaev viscosity parameter $\alpha_{\rm ss}$ (0.1, 0.5, or 1.0) from our base model values. The specific parameter combination of the misalignment angle, mass ratio, and $\alpha_{\rm ss}$ for our eight new models is presented in Table \ref{tab:sph_new_sim_params}. 

To analyze the disc evolution, we calculate orbital parameters for individual particles using their position and velocity vectors, and then average those parameters over all particles in the disc. We start with calculating the particle's angular momentum, $\bm{j}$, relative to the primary star as
\begin{equation}
    \bm{j} = (\bm{v} - \bm{v_p}) \times (\bm{r} - \bm{r_p}),
\end{equation}
where $\bm{v}$ and $\bm{r}$ refer to the velocity and position vectors, and the subscript $\bm{p}$ refers to the primary star. The inclination, $i$, and eccentricity, $e$, of a particle is then calculated by
\begin{eqnarray}
    i &=& \arccos{\frac{j_z}{|\bm{j}|}}, \\
    e &=& \frac{\bm{v} \times \bm{j}}{G M_p} - \frac{\bm{r}}{|\bm{r}|},
\end{eqnarray}
with $G$ the gravitational constant and $M_p$ the mass of the primary star. We also track the total disc mass simply by summing the mass of every active particle in the simulation at a given time.

\subsection{\textsc{hdust}}
\label{sec:hdust}

To predict observables of our \textsc{sph} simulations at any timestep, we use the 3-dimensional non-local thermodynamic equilibrium Monte Carlo radiative transfer code \textsc{hdust} \citep{carciofi2006non}. We have developed an interface that is capable of converting the particle distribution of an \textsc{sph} simulation into a computational grid that \textsc{hdust} can accept \citep[see][]{Suffak2024}, which accounts for particle velocities in addition to their density. The grid is defined in spherical coordinates with 50 radial cells, 50 azimuthal cells, and 50 polar cells.

For the stellar parameters that \textsc{hdust} requires, we input the primary star as having a rotation critical fraction, $W$, of 0.7, which is average for Be stars \citep{rivinius2013classical}, a mass of 8 $\rm M_\odot$, polar radius of 4 $\rm R_\odot$ (consistent with $W\,=\,0.7$ to give an equatorial radius of 5 $\rm R_\odot$), and a luminosity of 2300 $\rm L_\odot$.

\section{Varying Mass Ratio}
\label{sec:vary_q}

\begin{figure}
    \centering
    \includegraphics[scale=0.42]{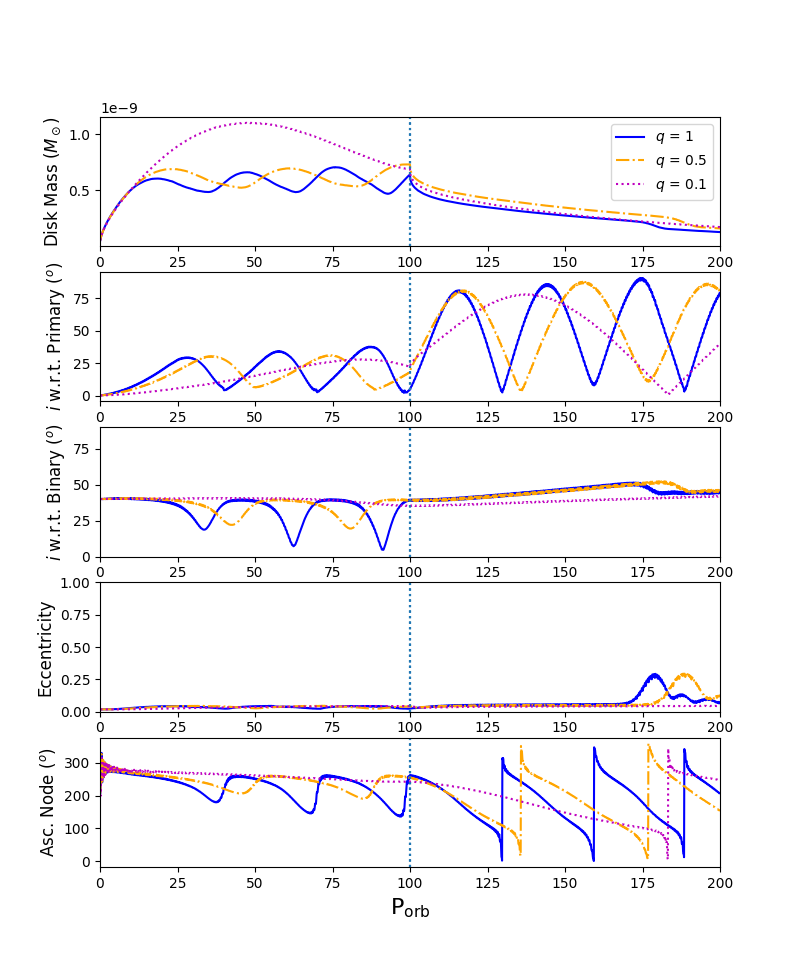}
    \caption{Top to bottom, total disc mass, disc inclination with respect to the primary equatorial plane, disc inclination with respect to the binary orbital plane, disc eccentricity, and the longitude of the ascending node of the disc, for models with a $\ang{40}$ misalignment angle and varying mass ratio as indicated by the legend. The $x$-axis is in units of binary orbital periods.}
    \label{fig:40deg_diffMR}
\end{figure}

\begin{figure}
    \centering
    \includegraphics[scale=0.42]{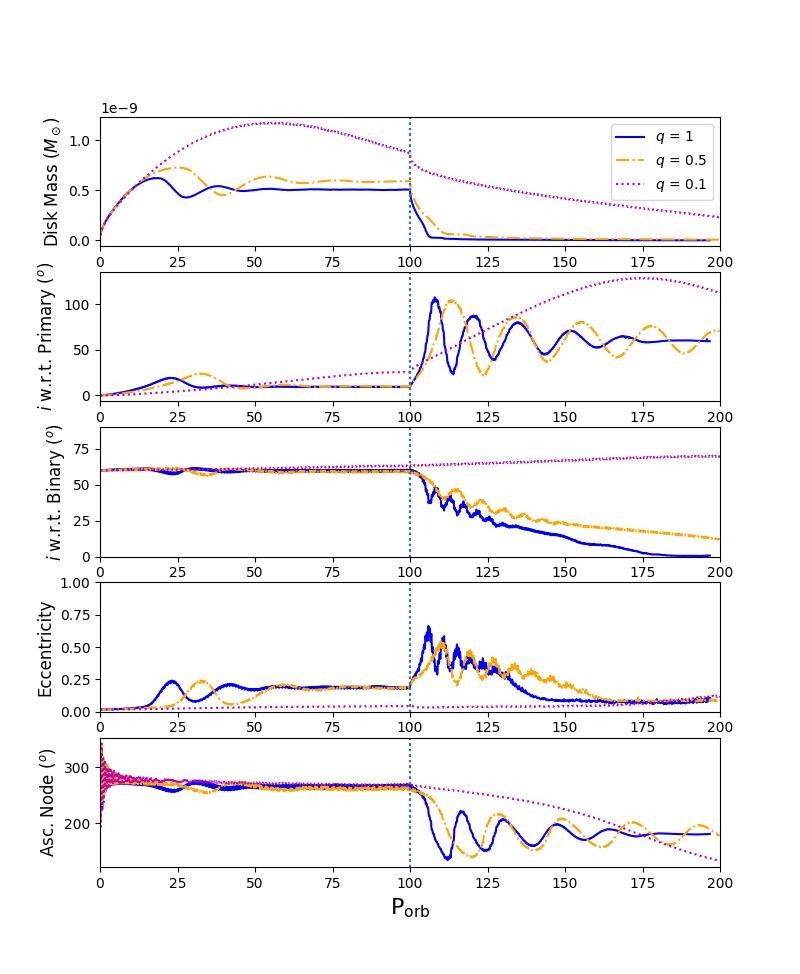}
    \caption{Same format as Figure \ref{fig:40deg_diffMR}, but for a misalignment of $\ang{60}$.}
    \label{fig:60deg_diffMR}
\end{figure}

The effect on the disc of varying the binary mass ratio with a misalignment of $\ang{40}$ and $\ang{60}$ is shown in Figures \ref{fig:40deg_diffMR} and \ref{fig:60deg_diffMR} respectively. We show the time evolution of the disc mass, inclination with respect to the primary equatorial plane and secondary's orbital plane, eccentricity, and longitude of the ascending node, with mass ratios of 1, 0.5, and 0.1 overlaid on the same plot. We see similar trends across both misalignment angles, with the $q$ = 0.5 models having the same, but slightly delayed, evolution to the equal-mass case, while the $q$ = 0.1 models display much slower, less dramatic changes.

For the $\ang{40}$ misalignment, episodes of disc-tearing are seen in Figure \ref{fig:40deg_diffMR} in the form of repeated oscillations of the disc mass (top panel) and disc inclinations with respect to both the primary equatorial plane and binary orbital plane (second and third panels, respectively). We see that disc-tearing occurs for $q$ = 0.5 with a slightly longer tearing-recombining period than in the equal mass case. While for $q$ = 0.1, no evidence of tearing occurs, and the change in disc inclination takes two to three times longer to peak than the higher mass ratio cases. The lack of disc tearing when $q$ = 0.1 is in agreement with the approximation of the tearing radius, $R_{\rm break}$, by \citep{Dogan2015}
\begin{equation}
    R_{\rm break} \geq \bigg(\frac{4\alpha}{\sin2\beta}\frac{H}{R_{\rm out}}\frac{M_{\rm p}}{M_{\rm s}}\bigg)^{1/3} a_{\rm b},
    \label{eq:disc_tearing}
\end{equation}
where $M_{\rm s}$ and $a_{\rm b}$ are the mass and semi-major axis of the binary, and $H$ is the scale height of the disc at the disc outer radius $R_{\rm out}$. For $q$ = 1 and 0.5, $R_{\rm break}$ = 4.7 $R_{\rm p}$ and 5.9 $R_{\rm p}$, respectively, while for $q$ = 0.1, $R_{\rm break}$ = 10.2 $R_{\rm p}$, which is very near the outer edge of the disc.

The total disc mass for $q$ = 0.1 peaks at more than double the other models, as the disc is able to radially extend much more due to the lower torque of the companion. The decrease of the disc mass is directly tied to the increase in disc inclination with respect to the primary star, as discussed by \cite{Martin2011}. The measured accretion rates of disc material onto the primary star oscillate in the same manner as the disc inclination with respect to the primary equatorial plane.

After mass-injection into the disc is turned off at 100 orbital periods, all of the discs precess about the binary's angular momentum vector in approximate accordance with the equation \citep{Larwood1996}
\begin{equation}
    \omega_{\rm p} = - \frac{3GM_{\rm s}}{4a_{\rm b}^3} \cos\beta \frac{\int^{R_{\rm out}}_{R_{\rm in}} \Sigma R^3 dR}{\int^{R_{\rm out}}_{R_{\rm in}} \Sigma R^3 \sqrt{\frac{GM_{\rm p}}{R^3}} dR},
    \label{eq:disc_precess}
\end{equation}
where the integral is computed with respect to the radial position, $R$, over the whole disc. Due to substantial warps and asymmetries in the disc at 100 $\rm P_{\rm orb}$, the measured precession rates do not immediately agree with Equation \ref{eq:disc_precess}, however after mass-injection has been off for 10 to 20 orbital periods, the disc flattens and Equation \ref{eq:disc_precess} predicts the observed precession rates of 138 \Porb for $q$ = 0.1, 38 \Porb for $q$ = 0.5, and 27 \Porb for $q$ = 1, which agree well with the measured precession from the slopes of the ascending node versus time. The discs show no substantial eccentricity regardless of mass ratio until they have largely dissipated (after 175 orbital periods), when the $q$ = 1 and $q$ = 0.5 models show signs of KL oscillations.

The $\ang{60}$ misalignment also shows much slower changes with the $q$ = 0.1 case, which again has a much larger disc mass than the higher mass ratios. Here there is no disc-tearing in any model, but the $q$ = 1 and $q$ = 0.5 cases create an oscillation in disc eccentricity early in the simulation, after about 25 orbital periods during mass-injection. The increase in eccentricity to 0.25 and the very slight decrease in inclination with respect to the binary orbital plane is consistent with the conservation of angular momentum perpendicular to the binary plane seen in KL oscillations, given by \citep{Lidov1962}
\begin{equation}
    \sqrt{1-e^2}\cos i_b \approx \rm constant,
\end{equation}
where $e$ is the eccentricity of a test particle, and $i_b$ is the inclination relative to the binary plane. Therefore, these oscillations while mass injection is occurring are KL oscillations, although the oscillation is damped and eventually the eccentricity settles to a value of roughly 0.2. The $q$ = 0.1 model shows no sign of eccentricity growth throughout the simulation. After mass injection is turned off, all three models undergo precession, and the $q$ = 1 and $q$ = 0.5 models undergo KL oscillations as evidenced by the opposing changes in eccentricity and inclination with respect to the binary orbit. The $q$ = 0.5 model has a longer KL period than the equal mass case. For a test particle, the analytical KL timescale is \citep{Kiseleva1998}
\begin{equation}
    \frac{\tau_{\rm KL}}{P_{\rm orb}} \approx \frac{M_{\rm p} + M_{\rm s}}{M_{\rm s}}\frac{P_{\rm orb}}{P_{\rm p}}(1-e_{\rm s}^2)^{3/2},
    \label{eq:particle_KL}
\end{equation}
where $e_{\rm s}$ denotes the eccentricity of the companion star, and $P_{\rm p}$ is the orbital period of the test particle. This can be generalized into an estimate for a global disc KL oscillation \citep{Martin2014}
\begin{equation}
    \langle \tau_{\rm KL} \rangle \approx \frac{\int^{R_{\rm out}}_{R_{\rm in}} \Sigma R^3 \sqrt{\frac{GM_{\rm p}}{R^3}}dR}{\int^{R_{\rm out}}_{R_{\rm in}} \tau_{\rm KL}^{-1} \Sigma R^3 \sqrt{\frac{GM_{\rm p}}{R^3}} dR},
    \label{eq:disc_KL}
\end{equation}
with $\Sigma$ the disc surface density, $G$ the gravitational constant, and again the integrals are computed with respect to the radial position, $R$, over the whole disc. Equation \ref{eq:disc_KL} predicts KL periods of 32 and 40 orbital periods for the $q$ = 1 and $q$ = 0.5 models respectively, however our measured initial KL oscillation period, using the location of the first peak in the eccentricity, is 12 and 15 $\rm P_{orb}$ for $q$ = 1 and $q$ = 0.5. This difference can be explained by the findings of \cite{Martin2019}, who found a particle with an initial eccentricity of 0.2 would have a KL period 2.7 times lower than the prediction of Equation \ref{eq:particle_KL}. Applying this factor of 2.7 to the predictions of Equation \ref{eq:disc_KL} results in the exact initial KL periods found in our simulations. We do note that this oscillation period shortens as the disc dissipates due to the changing surface density profile of the disc.

\section{Varying Viscosity}
\label{sec:vary_alpha}

\subsection{Shakura-Sunyaev Viscosity}
\label{sec:shakura_visc}

\begin{figure}
    \centering
    \includegraphics[scale=0.42]{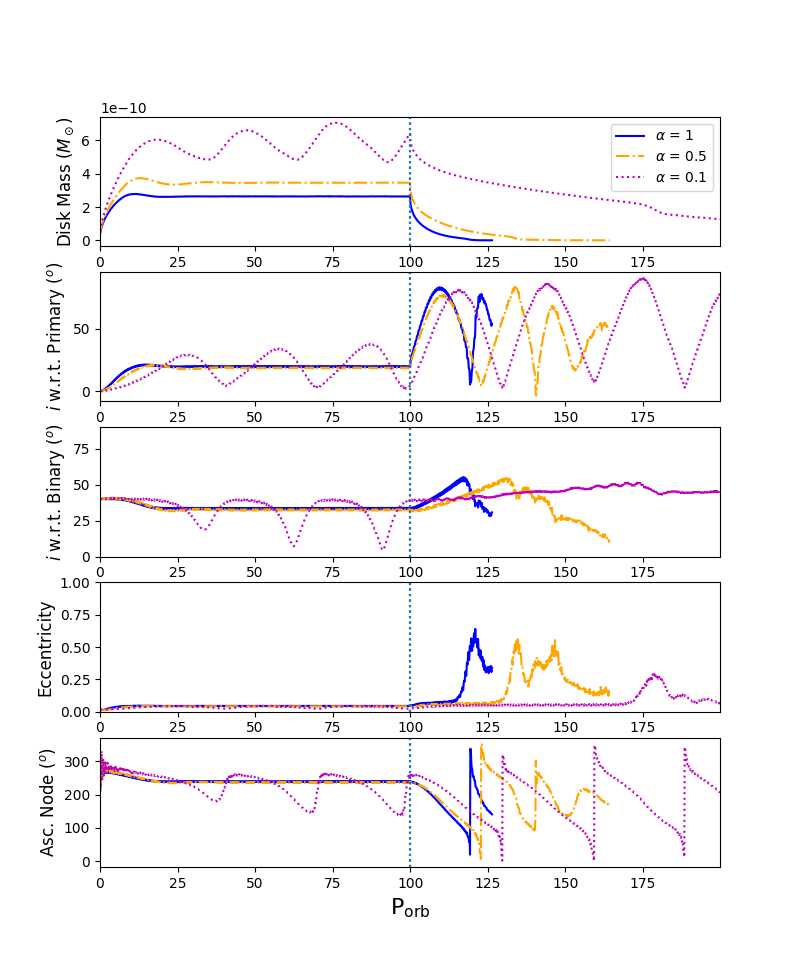}
    \caption{Same format as Figure \ref{fig:40deg_diffMR}, but varying viscosity as indicated by the legend.}
    \label{fig:40deg_diffalpha}
\end{figure}

\begin{figure}
    \centering
    \includegraphics[scale=0.42]{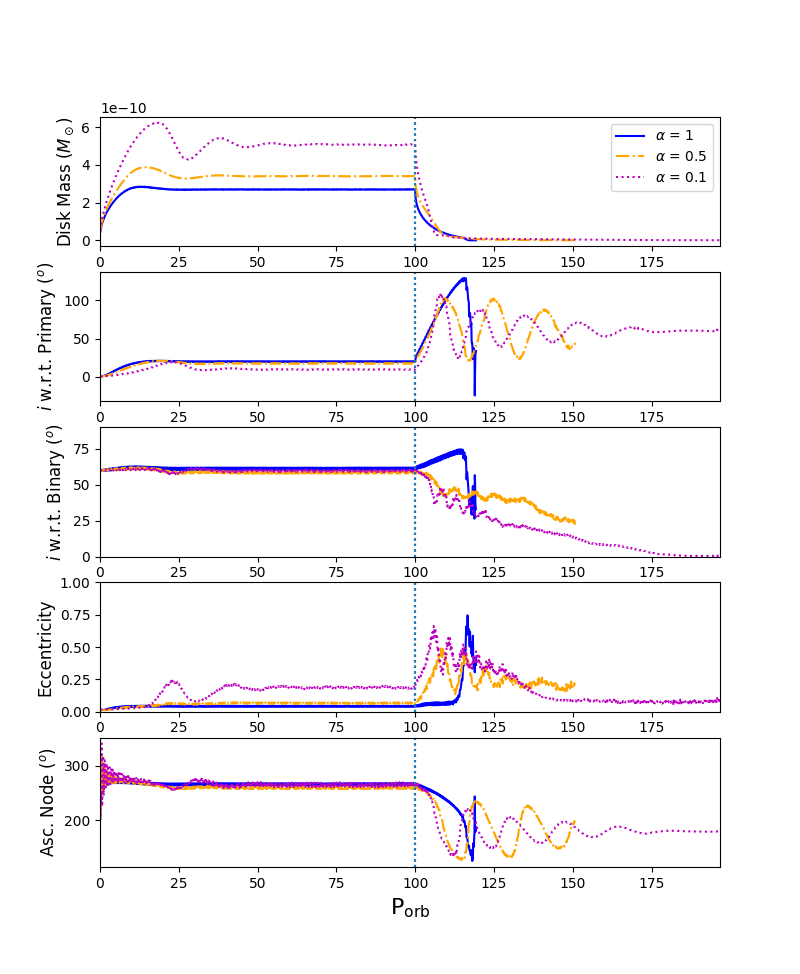}
    \caption{Same format as Figure \ref{fig:40deg_diffalpha}, but for a misalignment of $\ang{60}$.}
    \label{fig:60deg_diffalpha}
\end{figure}

Effects of varying viscosity are shown in Figures \ref{fig:40deg_diffalpha} and \ref{fig:60deg_diffalpha}, where we show the same format of plot as Figures \ref{fig:40deg_diffMR} and \ref{fig:60deg_diffMR}, but with varying viscosity instead of mass ratio. The models in these Figures are all equal mass systems, and the viscosity parameter $\alpha_{\rm ss}$ is varied between 0.1, 0.5, and 1. 

The patterns seen with increasing $\alpha_{\rm ss}$ are due to the viscous torque being directly proportional to the value of $\alpha_{\rm ss}$ \citep{Papaloizou1983}, hence a higher $\alpha_{\rm ss}$ corresponds to a more efficient angular momentum transport within the disc (i.e., the disc can better communicate with itself). With both the $\ang{40}$ and $\ang{60}$ misalignment, a higher viscosity can be seen to dampen or eliminate oscillations that occur in the $\alpha_{\rm ss}$ = 0.1 discs. In the $\ang{40}$ models (Figure \ref{fig:40deg_diffalpha}), both higher $\alpha_{\rm ss}$ values show no signs of disc-tearing, and also grow less massive with increasing $\alpha_{\rm ss}$. This lower mass is consistent with the expected surface density for a Be star disc \citep[equation 27 of][for example]{Bjorkman2005}, where the surface density is inversely proportional to $\alpha_{\rm ss}$, and thus a larger $\alpha_{\rm ss}$ results in a less massive disc. The inclination of the disc with respect to the primary equatorial plane still increases but the higher $\alpha_{\rm ss}$ models quickly achieve a steady inclination value. After mass-injection is turned off, all models precess about the binary angular momentum vector, however the disc dissipates faster with increasing $\alpha_{\rm ss}$, and thus the disc behaviour becomes erratic quite quickly for $\alpha_{\rm ss}$ values of 0.5 and 1.

The same can be seen in the $\ang{60}$ models in Figure \ref{fig:60deg_diffalpha}, where the higher $\alpha_{\rm ss}$ discs are lower in mass, and quickly reach a steady inclination. The high $\alpha_{\rm ss}$ models also do not show any disc eccentricity evolution while mass-injection is on, unlike in the $\alpha_{\rm ss}$ = 0.1 case. When dissipation begins at 100 orbital periods, we find the $\alpha_{\rm ss}$ = 0.5 model also undergoes KL oscillations with an initial period of 16 $\rm P_{orb}$, 33\% smaller than the prediction of Equation \ref{eq:disc_KL} likely due to it's small but non-zero eccentricity at the start of dissipation, while the highest $\alpha_{\rm ss}$ model dissipates very quickly and only shows slight precessing behaviour before the disc is mostly dissipated. Note that in Figures \ref{fig:40deg_diffalpha} and \ref{fig:60deg_diffalpha}, models for $\alpha$ = 1 and 0.5 stop short of 200 orbital periods due to faster dissipation than when $\alpha$ = 0.1, since the number of particles gets to be too small for the simulation to continue.

\subsection{SPH Artificial Viscosity}
\label{sec:artif_visc}

As described in \autoref{sec:sph}, our choice to set the Shakura-Sunyaev viscosity parameter, $\alpha_{\rm ss}$, in our models means that the internal artificial viscosity, $\alpha_{\rm sph}$, is calculated via \autoref{eq:alpha_conv}, and thus varies with each particle interaction. As this conversion is dependent on the theoretical scale height, which scales with radius as $r^{3/2}$, areas in the simulation where the disc does not follow this scale height prescription may have inflated artificial viscosity values (for example, when the disc becomes warped and tilted, or if there is a disc that forms around the companion star). This could particularly be an issue far from the central star where the theoretical scale height grows quite large.

To test this possibility, we computed another set of \textsc{sph} models with parameters identical to those in \autoref{tab:sph_sim_params} and \autoref{tab:sph_new_sim_params} with the exception that instead of defining $\alpha_{\rm ss}$, we set $\alpha_{\rm sph}$ to be a constant value 10 times what $\alpha_{\rm ss}$ was for each corresponding model. So if $\alpha_{\rm ss}$ was 0.1, then we set $\alpha_{\rm sph}$ = 1, and so forth.

\begin{figure}
    \centering
    \includegraphics[width=0.5\textwidth]{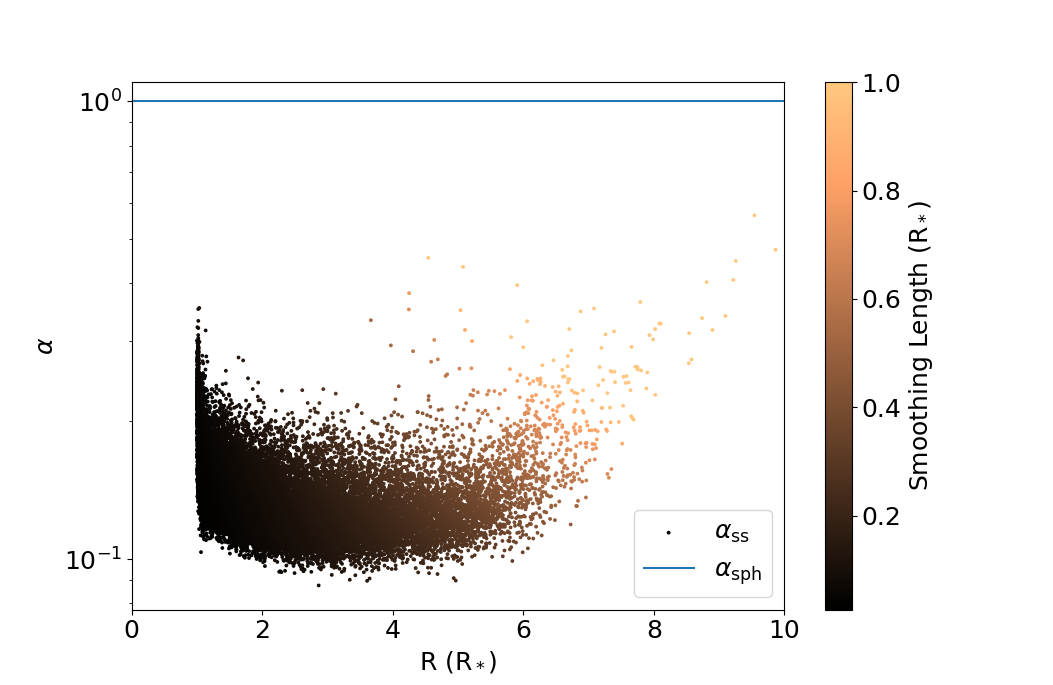}
    \caption{Shakura-Sunyaev viscosity parameter, calculated using \autoref{eq:alpha_conv}, versus radius for each particle in a simulation with a constant $\alpha_{\rm sph}$ = 1 as indicated by the solid line. The points representing each particle are coloured by their smoothing length. This is for an equal-mass binary orbit inclined by $\ang{40}$, after 85 orbital periods.}
    \label{fig:alpha_v_r}
\end{figure}

We find negligible differences in the disc evolution in models with high viscosity values ($q$ = 1; $\alpha_{\rm sph}$ = 5, 10), or with low mass ratios ($q$ = 0.1; $\alpha_{\rm sph}$ = 1) compared to their Shakura-Sunyaev counterparts. For the models with equal and half mass ratios and low viscosity ($\alpha_{\rm sph}$ = 1, $\alpha_{\rm ss}$ = 0.1), we see significant damping of oscillations while mass-injection into the disc is occurring. Most notably, we find disc tearing does not occur in models where it did previously. This can be explained by the conversion from $\alpha_{\rm sph}$ to $\alpha_{\rm ss}$ plotted in \autoref{fig:alpha_v_r}. We can see that the use of a constant artificial viscosity parameter leads to a non-constant $\alpha_{\rm ss}$, and furthermore that the resulting values of $\alpha_{\rm ss}$ are higher than the constant value of 0.1 that we had set in our original models, implying that $H/h\,<\,1$ in \autoref{eq:alpha_conv}. A higher $\alpha_{\rm ss}$ means the viscous torque in the disc is larger and so the disc is able to stay intact. If we lower $\alpha_{\rm sph}$ such that the effective $\alpha_{\rm ss}$ is lower than 0.1, then we do find that the disc is able to undergo tearing. Increasing the number of particles to increase spatial resolution and make $H/h\,\approx\,1$ would also lower the effective $\alpha_{\rm ss}$.

\begin{figure}
    \centering
    \includegraphics[width=0.5\textwidth]{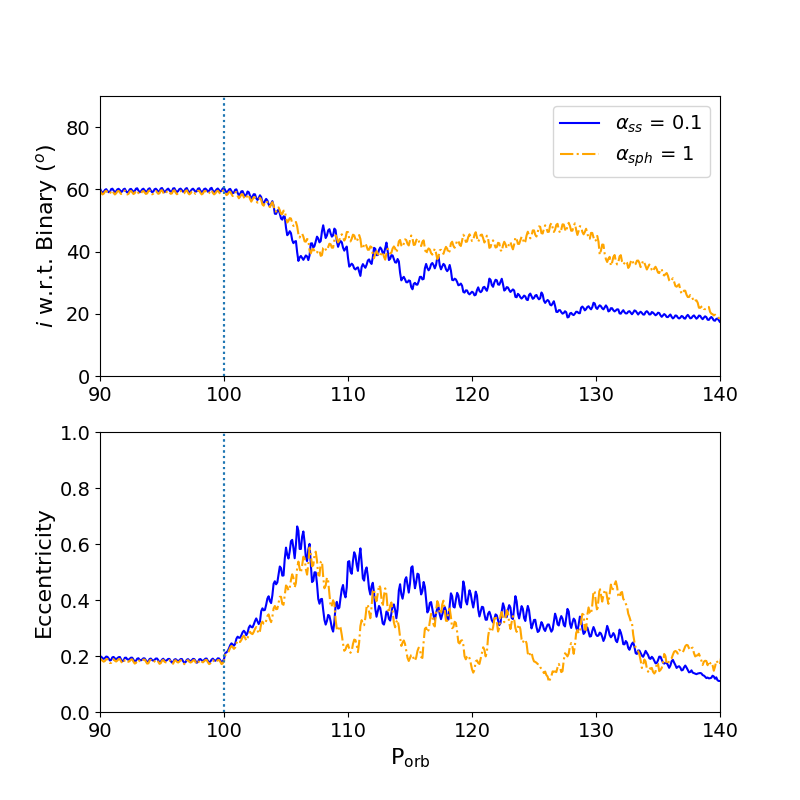}
    \caption{Disc inclination with respect to the binary orbit, and disc eccentricity versus time, for models with an equal-mass binary companion, and a constant $\alpha_{\rm ss}$= 0.1 (blue solid line) or a constant $\alpha_{\rm sph}$ = 1 (yellow dashed line). The binary orbit is inclined by $\ang{60}$. The dotted vertical line indicates the start of disc dissipation}
    \label{fig:KL_alphacomp1}
\end{figure}

\begin{figure}
    \centering
    \includegraphics[width=0.5\textwidth]{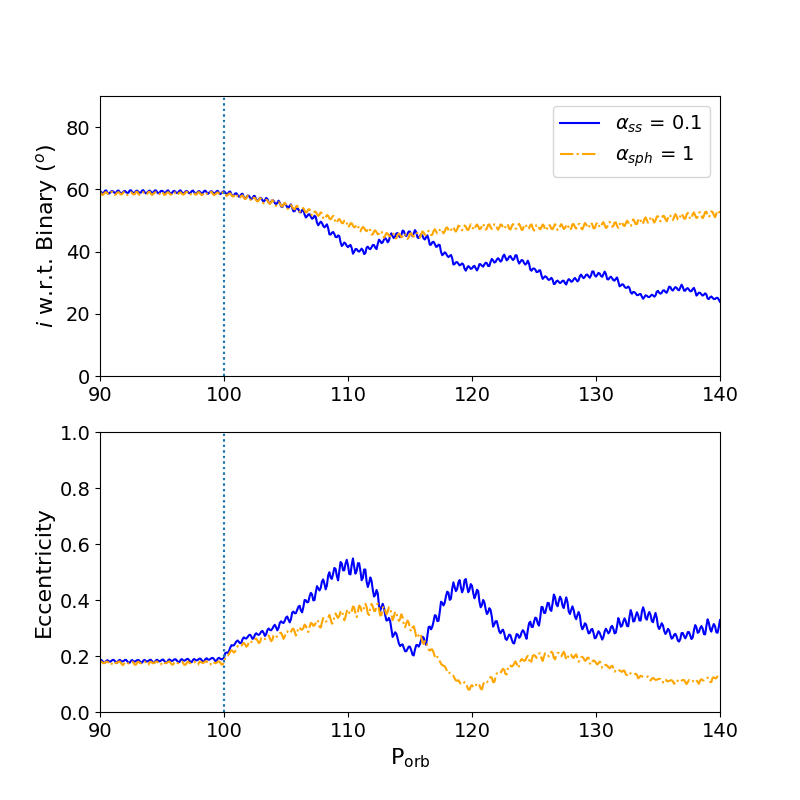}
    \caption{Same format as \autoref{fig:KL_alphacomp1}, but for a half mass ratio binary system ($q$ = 0.5).}
    \label{fig:KL_alphacomp2}
\end{figure}

The relationship between $\alpha_{\rm sph}$ and $\alpha_{\rm ss}$ shown in \autoref{fig:alpha_v_r} is similar across all models with a constant $\alpha_{\rm sph}$. This also has an effect on the occurrence of KL oscillations in the models with a $\ang{60}$ binary orbit inclination. \autoref{fig:KL_alphacomp1} shows that KL oscillations still occur in the equal-mass binary simulation with a constant $\alpha_{\rm sph}$, but with a slightly longer oscillation period than its constant $\alpha_{\rm ss}$ counterpart. Alternatively, in \autoref{fig:KL_alphacomp2} the half-mass ratio simulation does not display KL oscillations for a constant $\alpha_{\rm sph}$. Both of these differences can be explained by the different viscosity prescription resulting in slightly different disc configuration at the onset of disc dissipation due to the effectively higher disc viscosity. 

\section{Predicted Observations of a KL Oscillation}
\label{sec:obsverables}

To predict observations of a KL oscillation, we took our base \textsc{sph} model with a $\ang{60}$ misalignment and computed observables with \textsc{hdust} at the start of every orbital period from 100 to 120 $\rm P_{orb}$. The observables are computed at 152 unique observing positions, defined by two angles: the polar angle $\theta$, with $\theta = \ang{0}$ along the +$z$-axis, and the azimuthal angle $\phi$ where $\phi = \ang{0}$ is along the +$x$ direction. $\theta$ ranges from $\ang{0}$ to $\ang{180}$ in steps of 10 degrees, while $\phi$ ranges from $\ang{0}$ to $\ang{315}$, with steps every 45 degrees.

The actual observable trends are dependent on each specific observing angle, and the projected area of the disc facing the observer at a given timestep, so it is not practical to go through every computed observing orientation. Instead, here we present an examination of two general cases - a pole-on and an equator-on observing position. For the interest of the reader we show results from polar observing angles $\ang{30}$, $\ang{60}$, and $\ang{90}$, at all computed azimuthal angles, in Appendix \ref{sec:appendix}.

\begin{figure*}
    \centering   
    \includegraphics[scale = 0.5]{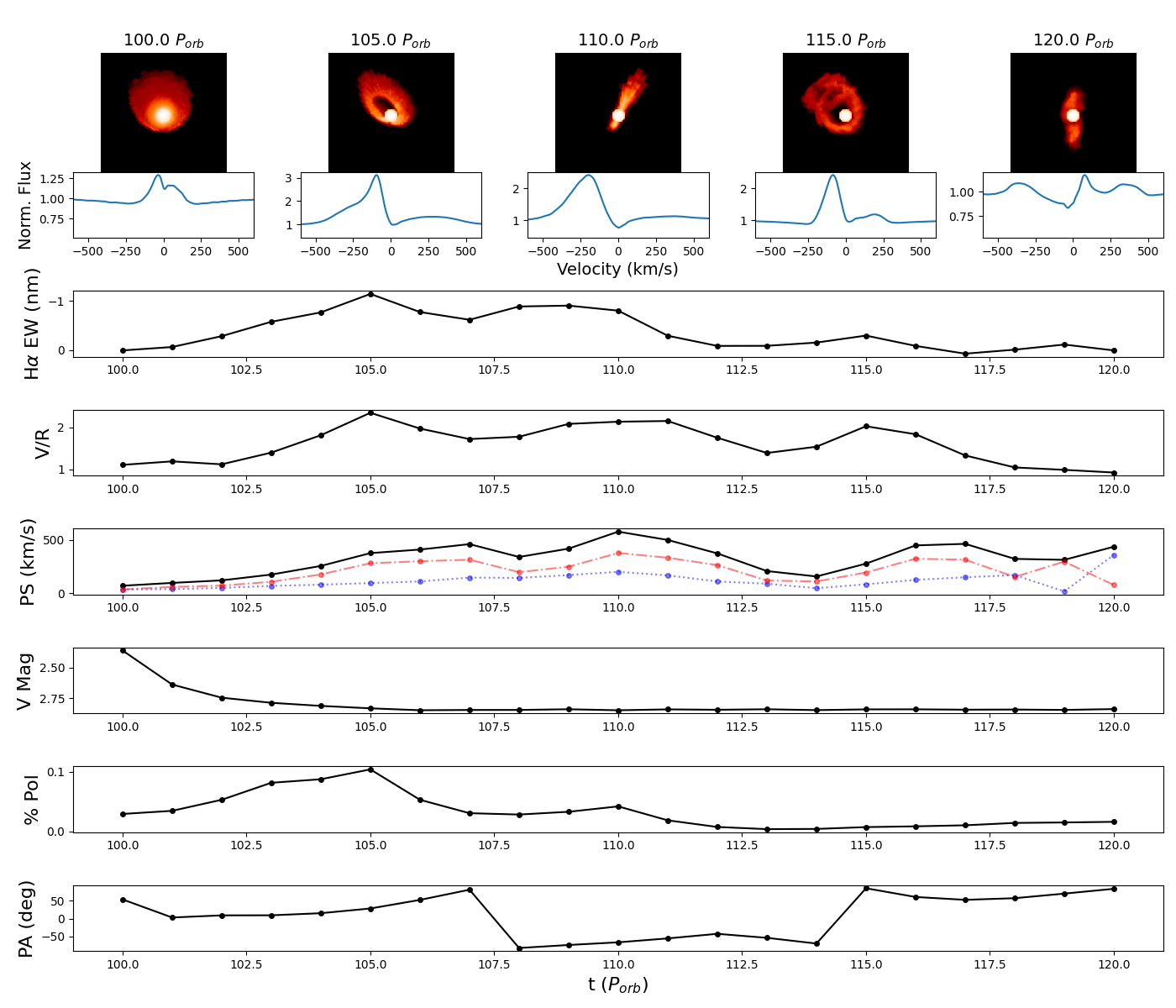}
    \caption{Top to bottom: images of the disc in H$\alpha$, created by \textsc{hdust}; the associated \Halpha line profiles; \Halpha EW; \Halpha V/R ratio; \Halpha peak separation, with the absolute values of the blue and red peak positions in blue and red respectively; $V$-band magnitude; polarization degree; and polarization position angle, from 100 to 120 $\rm P_{orb}$ for the base \textsc{sph} model ($q$ = 1, $\alpha_{\rm ss}$ = 0.1) with a $\ang{60}$ binary misalignment. The \Halpha images and associated line profiles are computed at the times indicated on top of the images. All measurements are computed from a pole-on ($\theta=\ang{0}$, $\phi=\ang{0}$) observing angle.}
    \label{fig:big_obs_0_0}
\end{figure*}

\begin{figure*}
    \centering  
    \includegraphics[scale = 0.5]{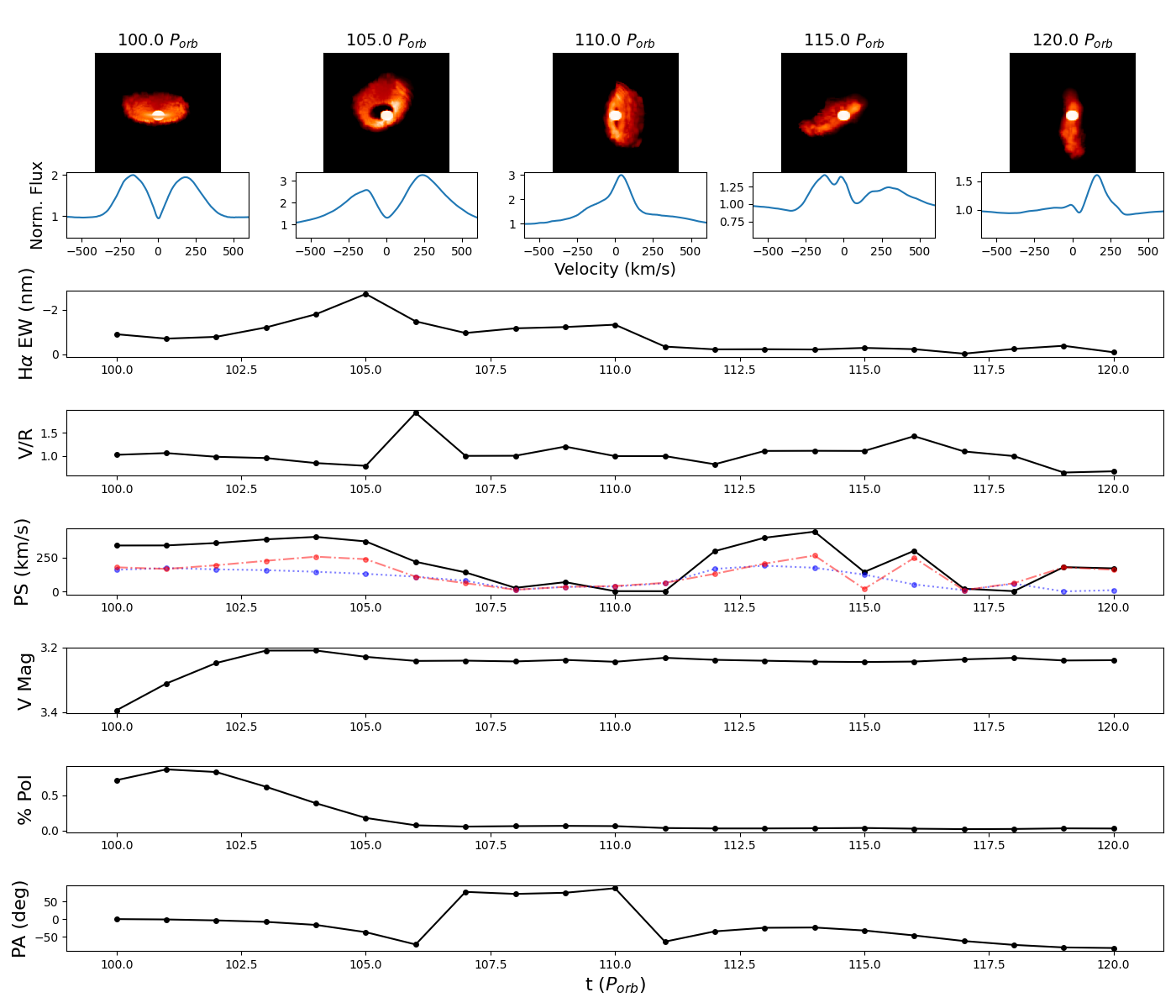}
    \caption{Same format as Figure \ref{fig:big_obs_0_0}, but for an edge-on ($\theta=\ang{90}$, $\phi=\ang{0}$) observing angle.}
    \label{fig:big_obs_90_0}
\end{figure*}

The observable trends of the KL oscillation from a pole-on and equator-on perspective are shown in Figures \ref{fig:big_obs_0_0} and \ref{fig:big_obs_90_0}, respectively. We chose to highlight the \Halpha line at five time steps by showing the \Halpha disc image, created by \textsc{hdust}, and the associated line profile, and then at the start of all 20 orbital periods we compute the \Halpha equivalent width (EW), violet-to-red (V/R) ratio, peak separation, $V$-band magnitude, polarization degree, and polarization position angle (PA).

For the pole-on case (Figure \ref{fig:big_obs_0_0}) the disc is close to face-on at 100 $\rm P_{orb}$, and then alternates between face-on and edge-on as the KL oscillation changes the disc inclination. The eccentricity changes in the disc due to the KL oscillation also creates an eccentric gap in the inner disc, seen in the image at 105 $\rm P_{orb}$. These changes are evidenced by the \Halpha emission line, where the EW, V/R ratio, and peak separation oscillate with the disc inclination. The fast increase in \Halpha EW before 105 \Porb is the result of the dissipation of the inner disc and the sharp dimming of the $V$ magnitude. The effect of changing disc eccentricity, and the eccentric gap in the disc, can be seen in the large V/R ratio that arises after a few orbital periods. The gap in the disc creates much more disc material on one side of the disc than the other, resulting in more emission, and hence a stronger peak, for one side of the line. The effect of the gap can also subtly be seen in the individual positions of the red and blue peaks of the \Halpha line. In the fifth row of Figure \ref{fig:big_obs_0_0}, we have plotted the absolute values of the red and blue peak positions along with the total separation. Here we can see that as the initial eccentric gap grows, the red peak moves to a much faster velocity than the blue peak - which would of course be reversed if the observer was looking from the other pole. This red/blue peak asymmetry also oscillates depending on the relative alignment of the disc to the observer.

We also see oscillations in the $V$-band photometry and polarization throughout the KL oscillation, in addition to the \Halpha line. The oscillations in the $V$-band are very slight ($\approx$0.01 mag), because the contribution of the disc relative to the star decreases due to the disc dissipating while the KL process occurs. The polarization, however does show significant oscillations up to 0.1\% and approaching $\ang{90}$ in the PA. Note that the abrupt jump in PA is due to a discontinuity in the calculation at +/-\ang{90}.

When the observer is equator-on (Figure \ref{fig:big_obs_90_0}) the disc initially is edge-on. The observables again oscillate as the disc inclination changes, however in contrast to the pole-on scenario, the edge-on case has much lower V/R variability, and approaches a single-peaked profile at around 110 $\rm P_{orb}$. At 115 \Porb we see a unique triple-peaked structure, with two low velocity peaks on the blue side of the line, and a third broad peak on the redward side. The peak separation still shows some asymmetry due to the eccentric hole, however the impact of the hole is reduced compared to the pole-on case. The $V$-band again oscillates with an amplitude of approximately 0.01 mag, while the polarization degree shows a steady decrease, with only very small (0.01\%) perturbations due to the tilting disc.

\subsection{Production of a Triple-Peaked Line Profile}
\label{sec:triple-peak}

As mentioned in the previous paragraph, our model at 115 \Porb when the observer is at ($\theta\,=\,\ang{90}$, $\phi\,=\,\ang{0}$), predicts a triple-peaked line profile. Triple peaks have been observed in Be star emission lines \citep[][for example]{Rivinius2006, Stefl2009, Moritani2013}, and it has been proposed that they are produced during the transition in a V/R cycle \citep{panoglou2018discs}, or from a warped disc due to a misaligned binary companion \citep{Moritani2013}. Despite this, there has never been a radiative transfer model that has produced a triple-peaked line profile.

\begin{figure*}
    \centering
    \includegraphics[width=0.5\linewidth]{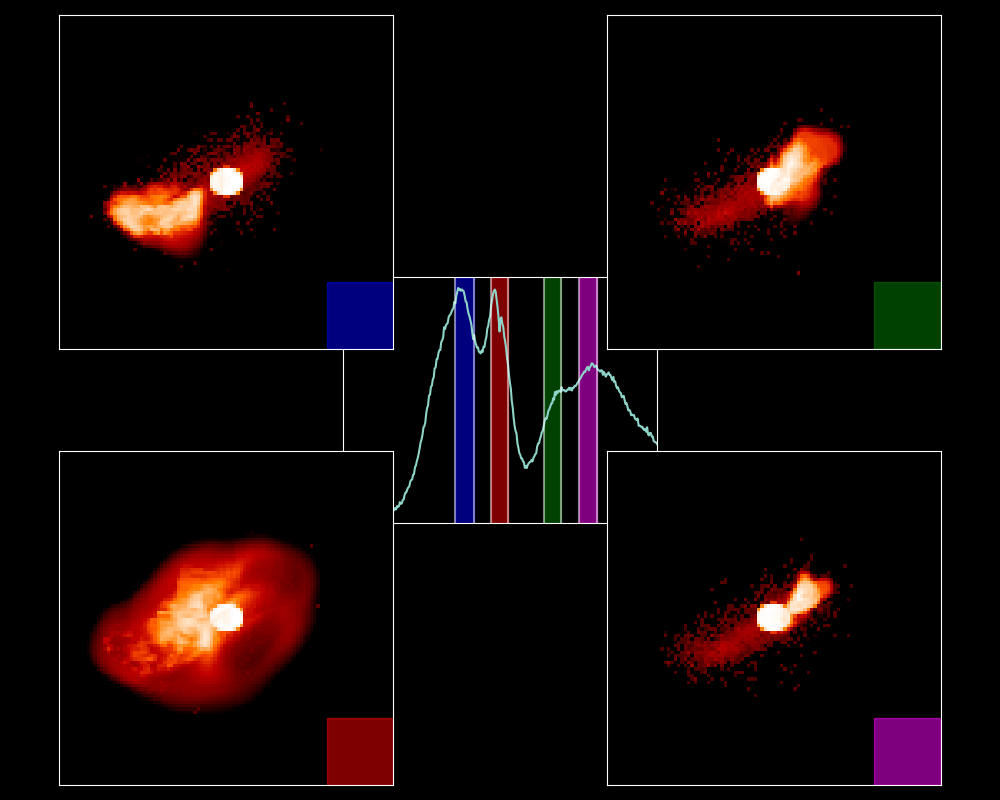}
    \caption{The various emitting regions of different portions of the triple-peaked \Halpha line produced during KL oscillations are shown. The center panel shows the normalized emission line, while the four outer panels show images in \Halpha of the disc and star, at the four coloured bands on the center line profile, as indicated by the coloured square in the bottom right corner of each outside panel.}
    \label{fig:triple_peak}
\end{figure*}

Therefore, to the best of our knowledge, our model is the first to predict a triple-peaked structure. Figure \ref{fig:triple_peak} shows the emission line along with four disc images produced in \textsc{hdust} showing where in the disc various parts of the line emission originate. From this observing angle, the disc is roughly edge-on with the observer, and is much more radially extended on one side than the other. The emission line consists of a slower double-peak on the blue side of the line, which is produced from the more extended side of the disc, and a third peak on the redward side, which is very broad and produced from the smaller side of the disc. This is consistent with emission line shape being largely due to the relative velocity of the disc material to the observer. The radially extended side of the disk has material moving at slower velocities, meaning its emission will trend closer towards line center than the emission from the small side of the disc, which would have mostly faster-moving material. There is also a shoulder feature on the redward side prior to this third peak. In our simulation this profile shape lasts for between one and two orbital periods in the midst of the ongoing KL oscillations.

While this line profile was produced in the midst of a KL oscillation, we want to emphasize that the key to a triple-peaked structure like the one produced here is disc asymmetry, as evidenced by the line formation loci in Figure \ref{fig:triple_peak}. Many possibilities exist as to how the disc may gain this asymmetry, such as our case with KL oscillations, a spiral density wave \citep{Stefl2009, panoglou2018discs}, a highly eccentric or highly misaligned binary companion, or a giant outburst \citep{Moritani2013}.

\section{Interferometric Predictions of Gaps in the Inner Disc}
\label{sec:interferometry}

In \cite{Suffak2024}, and the observables presented in Section \ref{sec:obsverables}, we have focused on predicting photometry, polarization, and the \Halpha line for a disc undergoing tearing and KL oscillations. This has neglected the powerful data that can be collected through interferometry, thus in this section we provide interferometric predictions for a torn disc, and when the disc is undergoing KL oscillations.

To compute the normalized complex visibility, $\bf V$, from \textsc{hdust} at a given wavelength, we calculate the 2D Fourier transform of an \textsc{hdust} image and use the formula \citep{Faes2013}
\begin{equation}
    \textbf{V}(u,v) = \frac{\int\int I(x,y,\lambda) \exp{(-2\pi i(ux+vy))}dxdy}{\int\int I(x,y,\lambda)dxdy},
\end{equation}
where $I$ is the image intensity with coordinates ($x$,$y$), $\lambda$ is the image wavelength, and ($u$, $v$) are the spatial frequency coordinates of the given baseline vector $\vec{B}$, defined as $u\,=\,\vec{B}\sin(PA_{\rm int})/\lambda$ and $v\,=\,\vec{B}\cos(PA_{\rm int})/\lambda$, with $PA_{\rm int}$ being the position angle of the baseline. We also extend this equation to integrate over the wavelength bandpasses of \Halpha (0.6550 $\mu$m to 0.658 $\mu$m) and Br$\gamma$ (2.161 $\mu$m to 2.171 $\mu$m).

\begin{figure*}
    \centering
    \begin{subfigure}[b]{0.4\textwidth}
        \centering
        \includegraphics[scale=0.25]{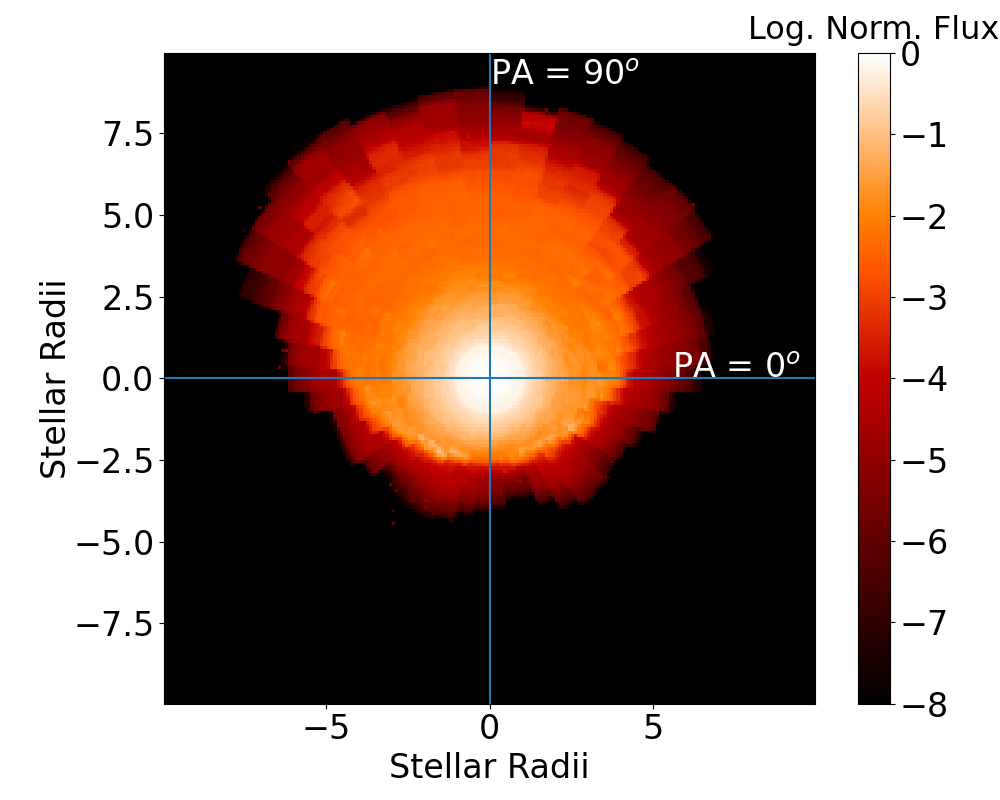}
        \caption{\Halpha image before KL.}
    \end{subfigure}
    \begin{subfigure}[b]{0.4\textwidth}
        \centering
        \includegraphics[scale=0.25]{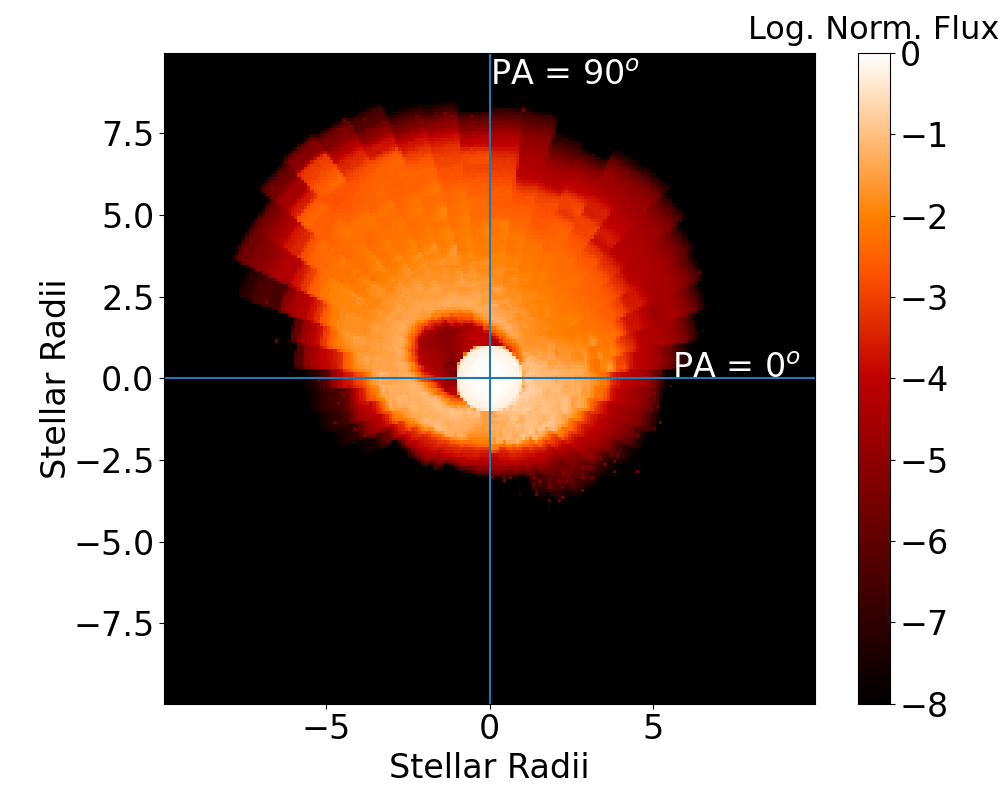}
        \caption{\Halpha image during KL.}
    \end{subfigure}\\
    \begin{subfigure}[b]{0.4\textwidth}
        \centering
        \includegraphics[scale=0.2]{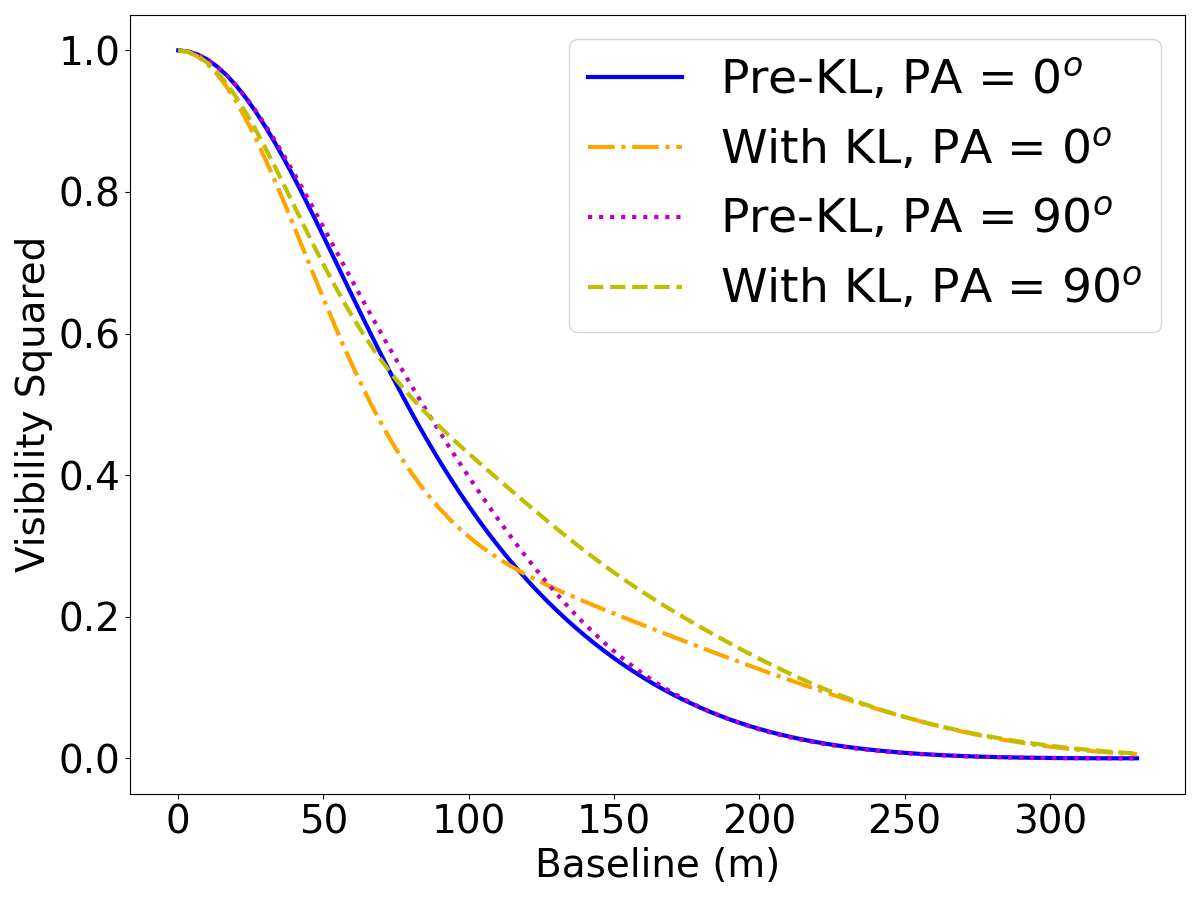}
        \caption{\Halpha visibility.}
    \end{subfigure}
    \begin{subfigure}[b]{0.4\textwidth}
        \centering
        \includegraphics[scale=0.2]{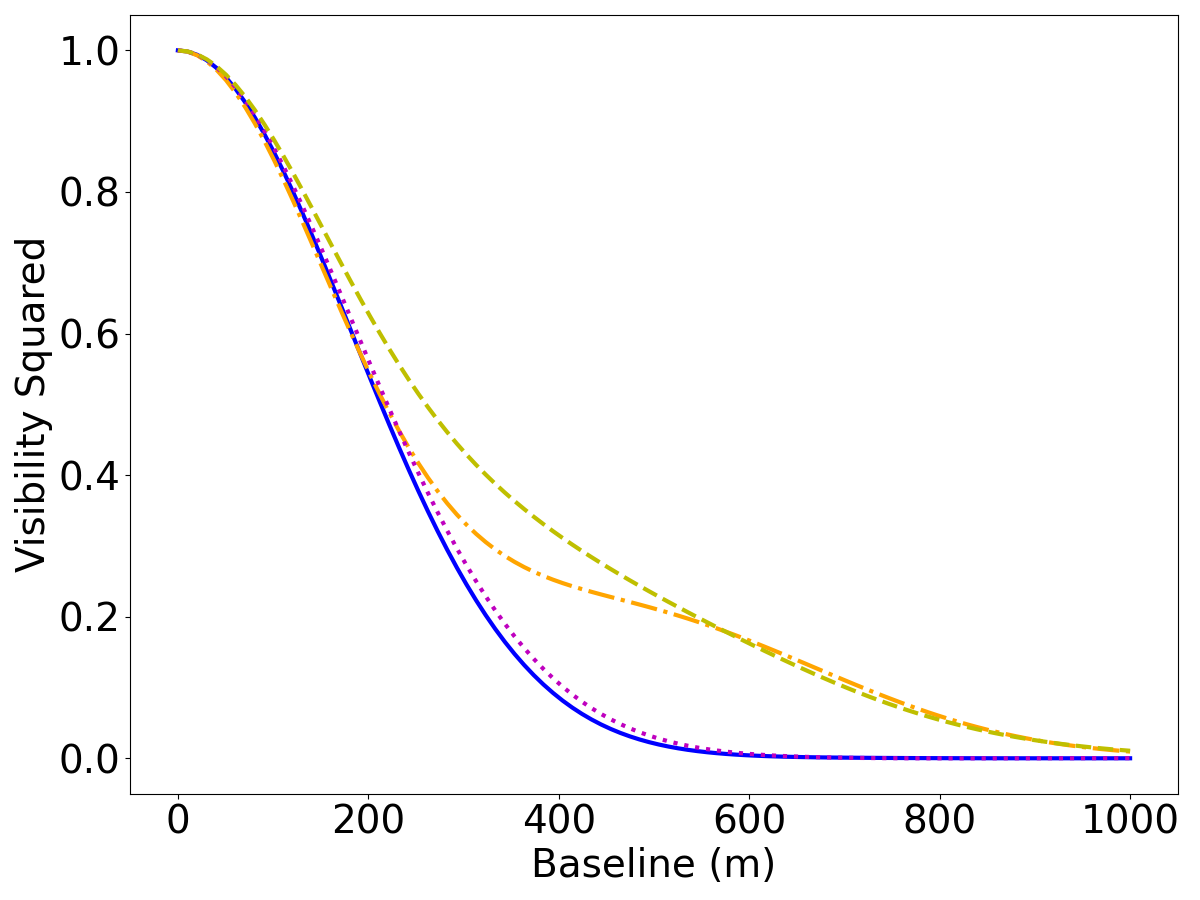}
        \caption{Br$\rm \gamma$ visibility.}
    \end{subfigure}
    \begin{subfigure}[b]{0.4\textwidth}
        \centering
        \includegraphics[scale=0.2]{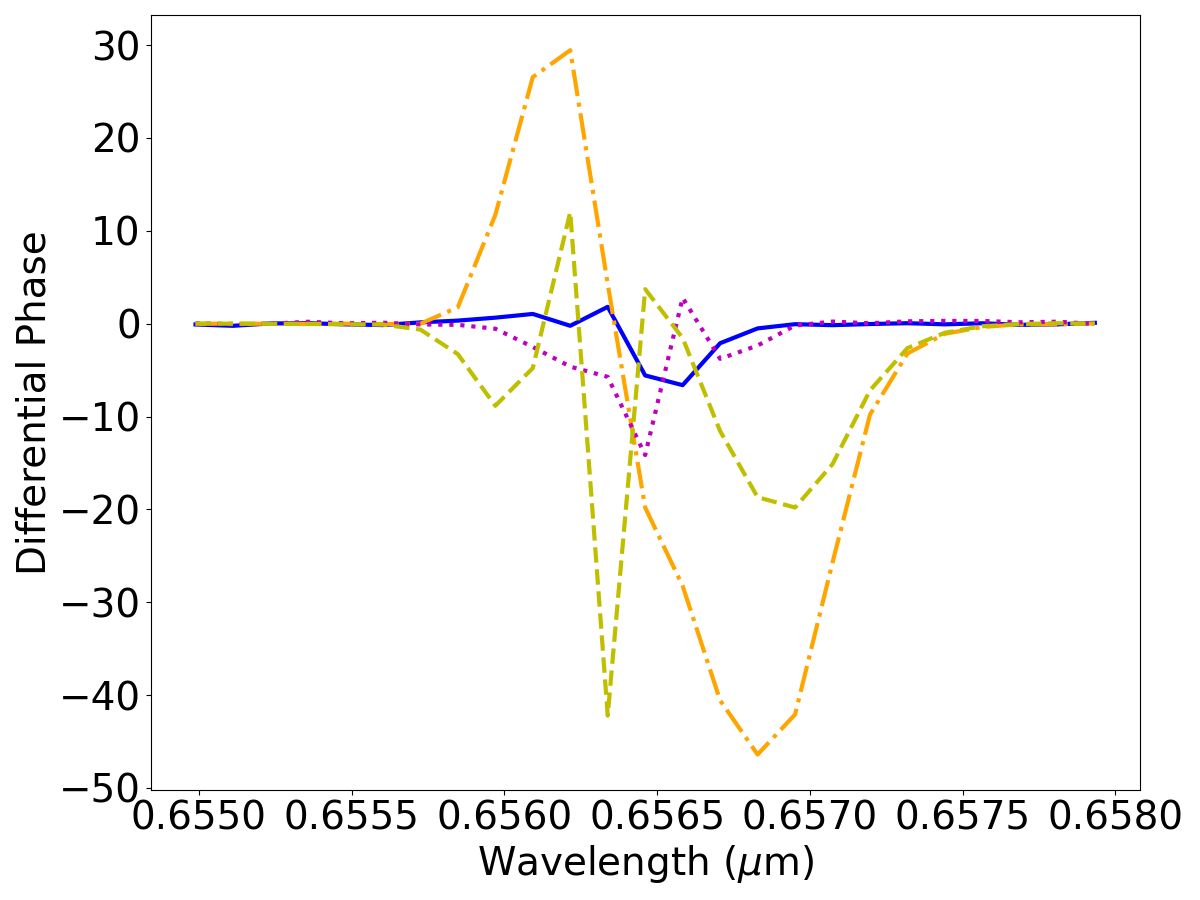}
        \caption{\Halpha differential phase.}
    \end{subfigure}
    \begin{subfigure}[b]{0.4\textwidth}
        \centering
        \includegraphics[scale=0.2]{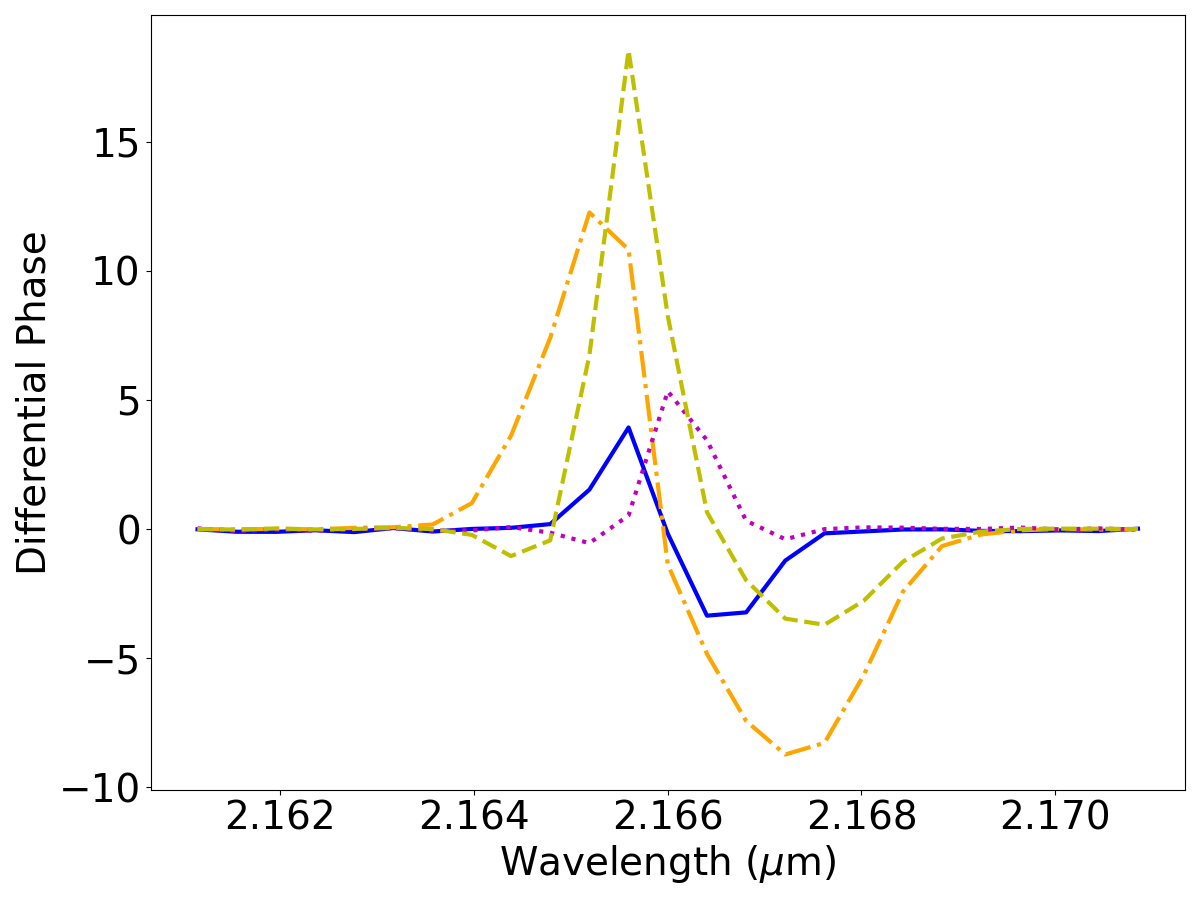}
        \caption{Br$\rm \gamma$ differential phase.}
    \end{subfigure}
    \caption{Top: \textsc{hdust} images in \Halpha for our $\ang{60}$ base \textsc{sph} model (a) before and (b) during KL oscillations. The baseline PAs of $\ang{0}$ and $\ang{90}$ are labelled on each image. Middle: Squared visibility versus baseline, and bottom: differential phase versus wavelength, before and during KL oscillations, at PAs of $\ang{0}$ and $\ang{90}$, for the \Halpha line (panels c,e), and the Br$\rm \gamma$ line (panels d,f). The legend in panel (c) is the same for panels (d-f). The baseline length for the phase plots is 100 m. Both visibility and phase calculations include light from the disc as well as the central star.}
    \label{fig:KL_vis}
\end{figure*}

\begin{figure*}
    \centering
    \begin{subfigure}[b]{0.4\textwidth}
        \centering
        \includegraphics[scale=0.25]{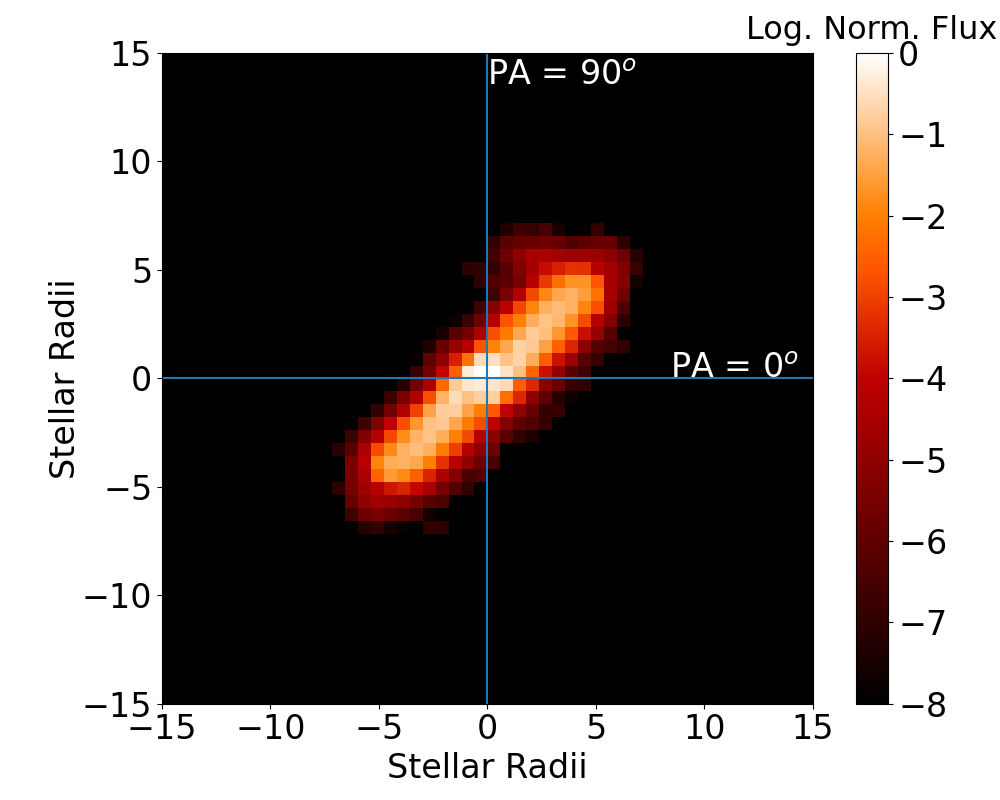}
        \caption{\Halpha image before disc tearing.}
    \end{subfigure}
    \begin{subfigure}[b]{0.4\textwidth}
        \centering
        \includegraphics[scale=0.25]{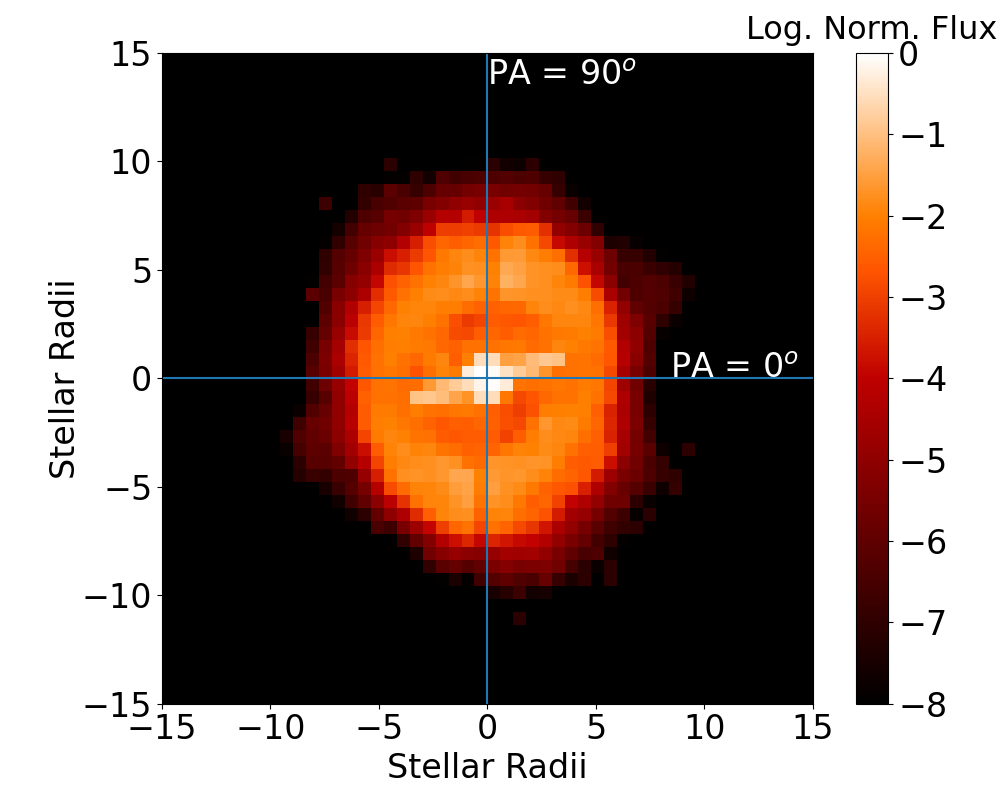}
        \caption{\Halpha image during disc tearing.}
    \end{subfigure}
    \begin{subfigure}[b]{0.4\textwidth}
        \centering
        \includegraphics[scale=0.2]{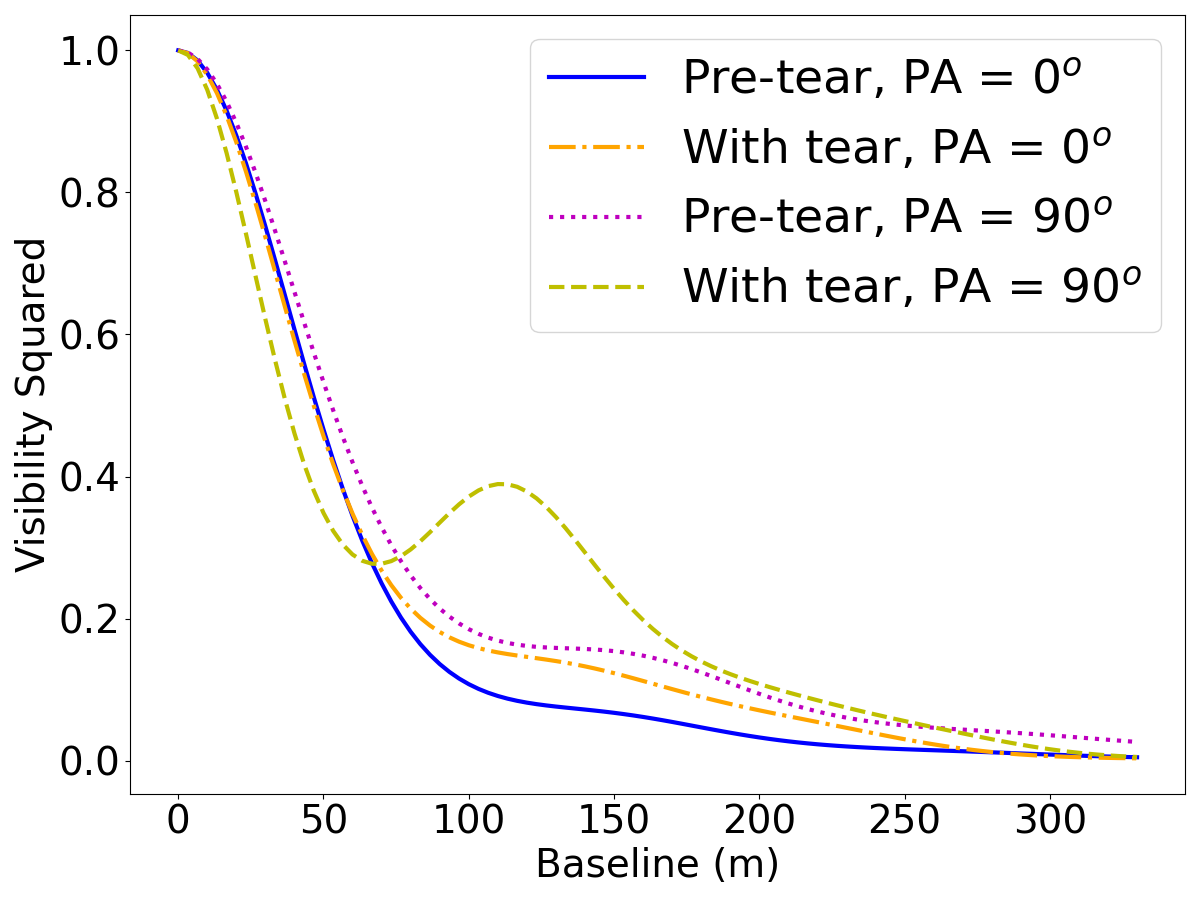}
        \caption{\Halpha visibility.}
    \end{subfigure}
    \begin{subfigure}[b]{0.4\textwidth}
        \centering
        \includegraphics[scale=0.2]{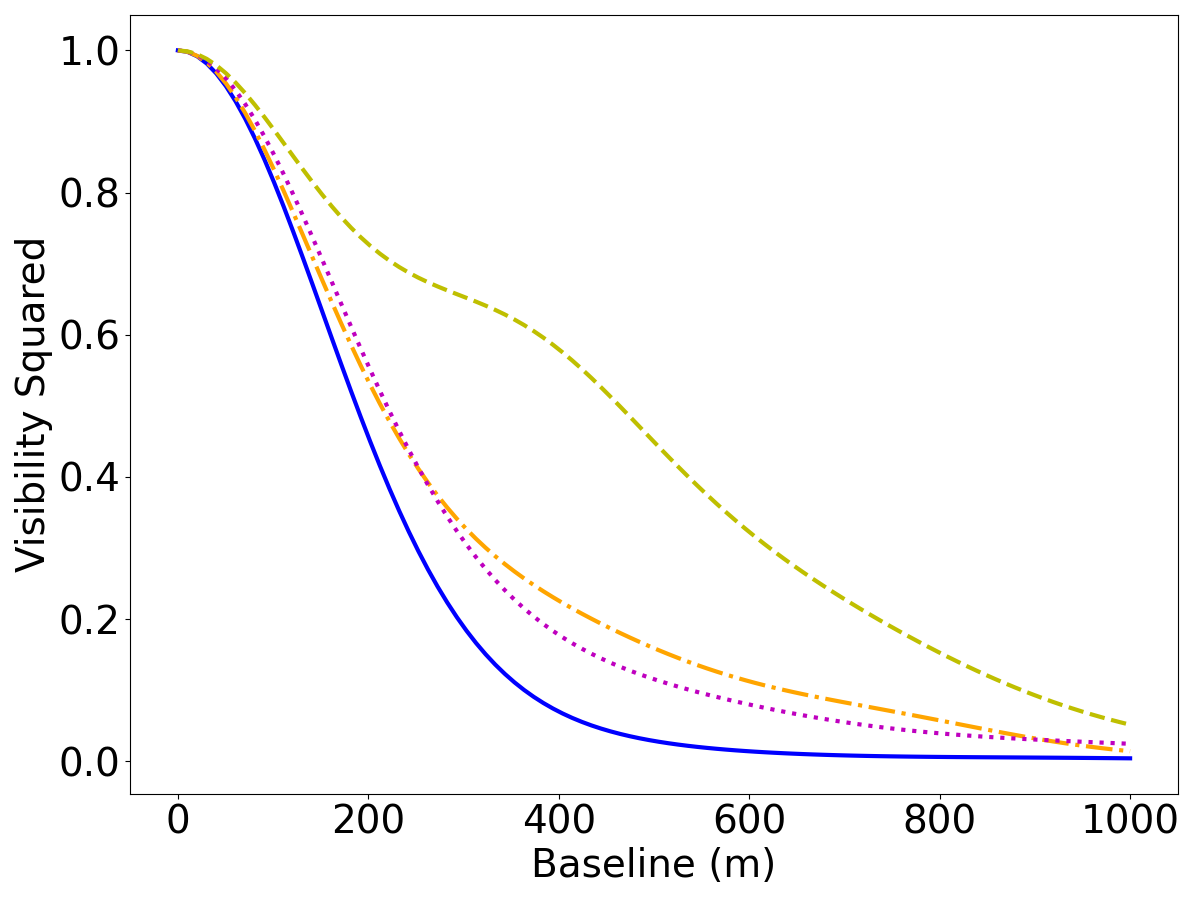}
        \caption{Br$\rm \gamma$ visibility.}
    \end{subfigure}
    \begin{subfigure}[b]{0.4\textwidth}
        \centering
        \includegraphics[scale=0.2]{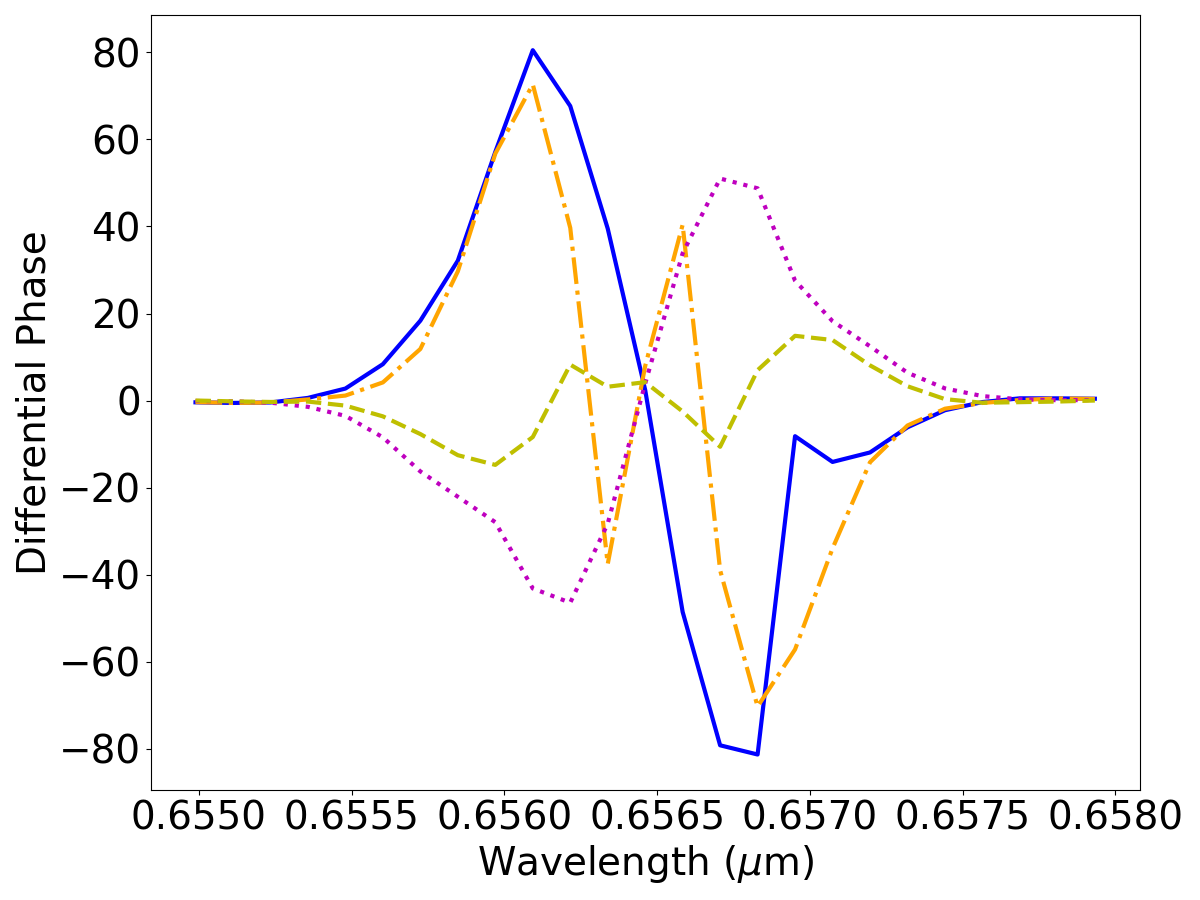}
        \caption{\Halpha differential phase.}
    \end{subfigure}
    \begin{subfigure}[b]{0.4\textwidth}
        \centering
        \includegraphics[scale=0.2]{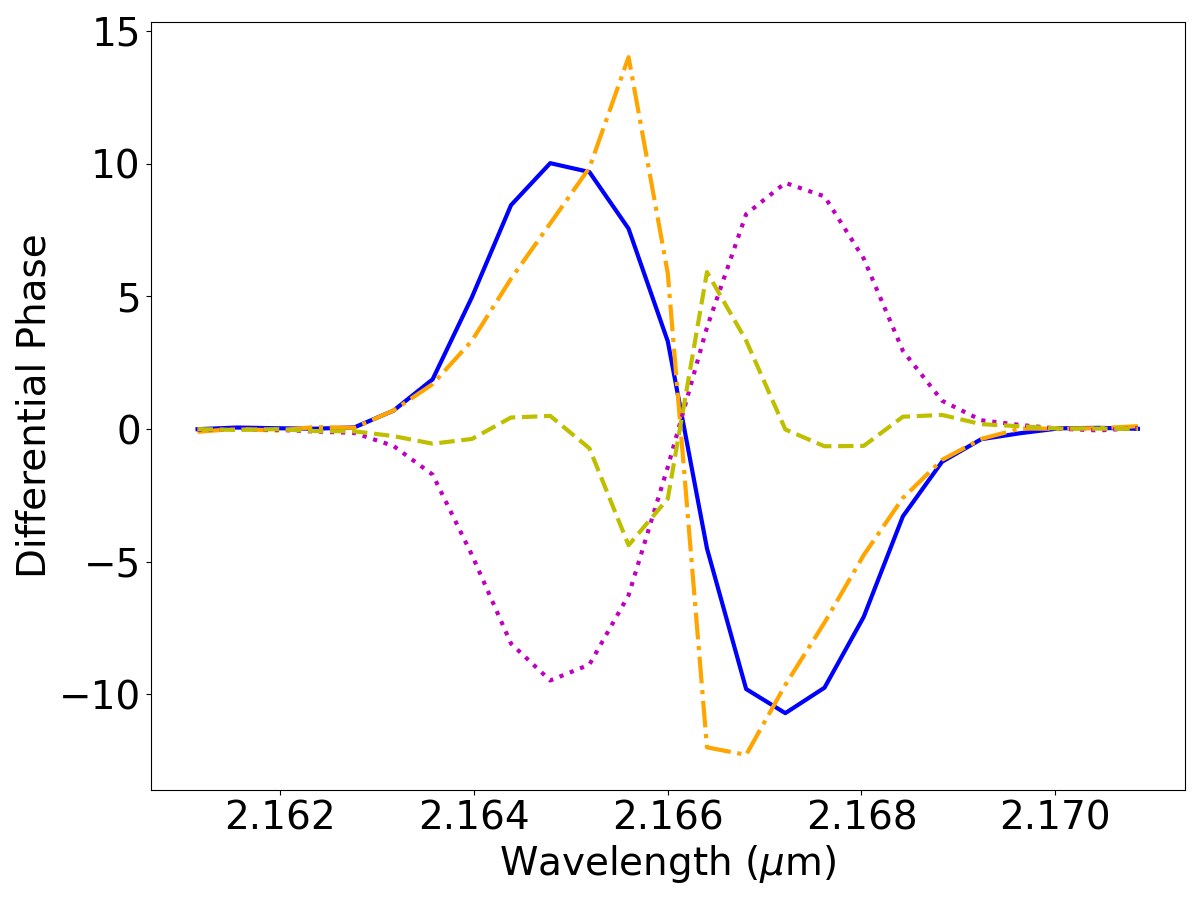}
        \caption{Br$\rm \gamma$ differential phase.}
    \end{subfigure}
    \caption{Same format as Figure \ref{fig:KL_vis}, but for before and during disc tearing, for our $\ang{40}$ base \textsc{sph} model.}
    \label{fig:tear_vis}
\end{figure*}

The results of this calculation are shown in Figures \ref{fig:KL_vis} and \ref{fig:tear_vis} for the disc undergoing KL oscillations, and a torn disc, respectively. The calculations include light from the central star as well as the disc, but not the signal of the companion. For both Figures, we have chosen a distance to the system of 100 pc, which sets the angular scale at $0.0465\,\rm mas/R_{\rm p}$. 

The top row of Figure \ref{fig:KL_vis} shows \Halpha \textsc{hdust} disc images (a) before and (b) during KL oscillations, where the disc has developed an eccentric gap near the primary star. The baselines for a $PA_{\rm int}$ of $\ang{0}$ and $\ang{90}$ are also shown on these images. The middle row presents the squared visibility versus baseline, and the bottom row shows the differential phases at a baseline of 100 m, before and during the KL oscillation. Both $PA_{\rm int}$ values are included, as indicated in the legend, for the \Halpha line (panels c,e), and the Br$\gamma$ line (panels d,f). We chose to end our \Halpha visibility curve at 330 m, as this is the maximum baseline of the CHARA array\footnote{\url{https://www.chara.gsu.edu/instrumentation/chara-array}}, while the Br$\gamma$ line required longer baselines to approach zero visibility, as expected due to its longer wavelength. 

At both wavelengths, both baselines show hump-like perturbations from the gap compared to the pre-KL visibility curves which show a smooth Gaussian profile. The \Halpha visibility is much larger at baselines from about 125 m to 250 m, while the major differences in Br$\gamma$ arise at baselines of 250 m to 900 m. Before the onset of the KL oscillations, the differential phases present a very minor S-shaped profile, due to the disc being nearly face-on with the observer and thus there are small relative velocity differences in the disc emission. However during KL oscillations, when the eccentric hole is present, these phases grow 2 to 3 times larger in the Br$\gamma$ line, and almost 10 times larger in H$\alpha$. 

Figure \ref{fig:tear_vis} is the same format as Figure \ref{fig:KL_vis}, but for before and during the disc-tearing process. At both wavelengths, we see the baseline at $PA_{\rm int}\,=\,\ang{90}$, which runs vertically through both gaps of the tear in the disc, presents a very large hump in the visibility curves, indicating a ring-like structure, with \Halpha showing a very significant increase around 125 m. The baseline at \ang{0} increases from the pre-tear curves, but does not show any differing perturbations since at this angle there is no intersection with the gap in the disc. The differential phases pre-tear show typical S-shaped curves. When the disc is torn, the phases reduce along $PA_{\rm int}\,=\,\ang{90}$ and show features of a central-quasi emission (CQE) phase shift \citep{Faes2013} due to being edge-on with the inner disc. The differential phases along $PA_{\rm int}\,=\,\ang{0}$ do not decrease in strength when the disc is torn, but do also show CQE in H$\alpha$. CQE appears in edge-on emission line profiles as a sharp central peak, and occurs due to a lack of absorbing disc material at zero relative velocity to the observer, and a large amount of material moving at slow radial velocities causing absorption on either side of this central emission \citep{Hanuschik1995}. This affects the differential phase measurements because the large amount of absorbing material at low velocities causes the photocenter position of the system to shift, which then corresponds to a changing signal in the differential phase \citep{Faes2013}.

\section{Discussion}
\label{sec:discussion}

\subsection{\textsc{sph} Simulations}

We have added eight new \textsc{sph} simulations onto our previous work \citep{Suffak2022}, taking our equal-mass binary system simulations that showed KL oscillations and disc-tearing, and using a lower mass ratio of 0.1 and 0.5, while also investigating a range of $\alpha_{\rm ss}$ of 0.1, 0.5 and 1.0, which covers the typical range of values found for Be star discs \citep{Rimulo2018}. These simulations represent a necessary step to understanding the behaviour of Be star discs with misaligned binary companions, as many Be stars have been found to have low-mass companions \citep[for some recent findings, see][]{Wang2021, Wang2023}, and the value of $\alpha_{\rm ss}$ is uncertain for individual systems, and varies between discs \citep{Rimulo2018}.

In varying the mass ratio of the $\ang{40}$ misalignment, we have found that the limit on disc-tearing is consistent with the findings of \cite{Dogan2015}. In the simulation with a mass ratio of 0.1, the low-mass companion does not provide sufficient precessional torque to overcome the viscous forces present in the disc and cause the disc to tear. However the discs with a $\ang{40}$ misalignment all precess after the mass-injection into the disc is turned off (see Figure \ref{fig:40deg_diffMR}). The precession period is longer with a lower mass ratio, consistent with equation 21 of \cite{Larwood1996}. The fact that all of the discs precess without mass-injection points to the mass-injection working as an anchor for the disc at the equator of the primary star - the strength of which would depend on the mass-injection rate into the disc.

Varying $\alpha_{\rm ss}$ instead of the mass-ratio for the $\ang{40}$ misalignment (Figure \ref{fig:40deg_diffalpha}) we find the higher viscous communication within the disc, that comes with a higher $\alpha_{\rm ss}$, dampens out oscillations while mass-injection is occurring, and a steady-state of the disc is reached within $\approx$25 orbital periods. Thus the window for disc-tearing to occur seems very sensitive to the value of $\alpha_{\rm ss}$. The simulations with higher $\alpha_{\rm ss}$ also precess once mass-injection ceases, but the discs dissipate at a much faster rate, and so any observable signal from such precession would also dampen quickly.

With respect to the $\ang{60}$ misalignment simulations, KL oscillations still occur during dissipation for the $q\,=\,0.5$, but for $q\,=\,0.1$ the disc simply precesses while dissipating. A higher $\alpha_{\rm ss}$ also permits KL oscillations to a degree, but for $\alpha_{\rm ss}\,=\,1.0$ no KL oscillations occur.

For both misalignment angles, with a larger $\alpha_{\rm ss}$ the disc does not grow as massive, with $\alpha_{\rm ss}$ of 0.5 and 1.0 only reaching about half the mass of the $\alpha_{\rm ss}\,=\,0.1$ simulations. This is a clear consequence of the accumulation effect \citep{Okazaki2002, panoglou2016discs}, where a less viscous disc is more susceptible to angular momentum being transferred by the resonant torque of the binary companion, and thus transports angular momentum slower. \cite{Cyr2017} showed this effect is also active in misaligned systems, however it is not as prominent as accumulation occurring in aligned systems \citep{panoglou2016discs}.

Conversion of the Shakura-Sunyaev viscosity parameter into \textsc{sph} artificial viscosity (\autoref{eq:alpha_conv}) has been implemented and used in many studies of Be star discs \citep{Okazaki2002, panoglou2016discs, Cyr2017, Brown2019,Cyr2020,Suffak2022}. However, the \textsc{sph} prescription from \cite{Monaghan1983} is widely used on its own as well \citep[][for example]{Martin2014,Franchini2021,Overton2024}, and does not include an assumption that the disc structure follows the theoretical scale height like the Shakura-Sunyaev viscosity does. In our comparison of the two prescriptions, we find we can recover the same qualitative disc evolution for a range of constant artificial viscosity as when $\alpha_{\rm ss}$ is held constant. Setting $\alpha_{\rm sph}$ to 10 times the value of $\alpha_{\rm ss}$ recovers most of the behaviour, however since $H/h\,<\,1$ in our simulations, as shown in \autoref{fig:alpha_v_r}, some oscillations, notably disc tearing, do not occur at this approximation and require lower values of $\alpha_{\rm sph}$ to reproduce the same phenomena.

We are not claiming in this work that one method of defining viscosity is preferred or more correct than another, however the differences between the two prescriptions for viscosity, and potential positives to alternate viscosity prescriptions for Be star discs, will be explored in future works.

\subsection{KL Oscillations}

KL oscillations are best known for occurring in hierarchical three-body systems where a small mass asteroid or planet experiences the changes in its orbital eccentricity and inclination \citep{Kozai1962, Lidov1962}. However in many recent publications, in addition to this work, KL oscillations have been seen in simulations and analytical investigations of astrophysical discs in binary star systems \citep{Martin2014, Fu2015a, Fu2015b, Lubow2017}, where the disc plays the role of the low-mass third body. While such dynamical simulations are essential to studying disc evolution, it is also necessary to know how KL oscillations will present themselves over time in observations. This is what we show in Section \ref{sec:obsverables}, using \textsc{hdust} to produce observables of our \textsc{sph} simulations over time. 

When a disc is dissipating (i.e., reaccreting onto the primary star), it is expected that the polarization drops first, as it is directly proportional to density and so is very sensitive to the inner disc (\citealp{Haubois2014}, see also figure 1 of \citealp{Carciofi2011}). The $V$ magnitude also changes quickly as it also originates in the inner disc, however whether it rises or falls depends on if the disc inclination is edge-on or pole-on to the observer \citep{Haubois2012}. The \Halpha EW may initially increase due to the drop in continuum, but then will also decrease with time, however at a much slower rate due to having a larger emitting area in the disc. See the star 66 Oph in \cite{Marr2021} for an example of this typical dissipation scenario. 

From the computed KL observables in Figures \ref{fig:big_obs_0_0} and \ref{fig:big_obs_90_0}, we see that the overall trends in \Halpha EW, $V$-band photometry and polarization degree show what is expected from a dissipating Be star disc. In the pole-on case (Figure \ref{fig:big_obs_0_0}), the photometric magnitude and polarization degree decrease due to the decreasing disc density, while the \Halpha EW initially increases due to the drop in continuum level, but then also decreases as the disc becomes more diffuse. When the observer is equator-on (Figure \ref{fig:big_obs_90_0}) the $V$ magnitude increases due to less of the star being blocked by the disc, and the polarization and \Halpha emission decrease with time. However, the change in disc inclination with time due to KL oscillations add an oscillation in addition to these standard trends in the observables. These oscillations are most noticeable in the \Halpha EW and in the polarization position angle. The abrupt flips seen in the position angle are simply due to a discontinuity in the calculation at $\pm \ang{90}$. Within the \Halpha line, there is also an oscillation in the V/R ratio and peak separation due to the asymmetry of the disc.

The effect of the eccentric gap created by the KL oscillation can also be seen in the differing positions of the violet and red peaks of the \Halpha line, particularly in the pole-on case. In a normal double-peaked \Halpha line created by an axisymmetric disc, the peaks would be an equal distance from line center due to the equivalent amount of material being on either side of the disc. However in the case of this eccentric disc created by the KL oscillation, the gap created means there is a lack of high velocity disc material on the side of the disc that is more radially extended and covers a larger area. This causes the stronger peak to be at a low velocity, while on the other side of the disc there is all high velocity material, and thus the other peak is at high velocity but is small in strength. This accounts for both the V/R ratio and the asymmetric peak separations in the \Halpha line.

Nearing 120 $\rm P_{orb}$ there is very little disc left, as the maximum density cell in the disc is roughly $1\times 10^{-14} \,\rm g/cm^3$, i.e., three orders of magnitude less than the density at 100 $\rm P_{orb}$, so the \Halpha line is nearly level with the continuum and peaks become difficult to identify. With a 30 day orbital period, the changes in observables here occur over 600 days, or roughly 1.6 years. So there would be a small window to detect these oscillations in this case, and such a system would also require frequent observation to see the oscillations.



\subsection{Interferometric Predictions}

Interferometric measurements are not only dependent on the combination of baseline length and position angle - of which there are infinite theoretical combinations - but also on the wavelength, disc density, and the distance and the orientation of the object on the sky, which defines the angular size of the object on the sky viewed from the theoretical interferometer. As such, we are not able to show every possible combination of parameters when predicting interferometry signatures of our disc models. However, based on our findings in Figures \ref{fig:KL_vis} and \ref{fig:tear_vis}, we can confidently say that a gap in the disc, created by KL oscillations or disc-tearing, will cause an increase in the squared visibility at large baselines if the baseline crosses the gap in the disc. We also see that the creation of a gap in the disc can potentially cause large changes to the measured differential phase. The appearance of a central-quasi emission in the phase shape is an indicator of a change in orientation of the disc to the observer, which both KL and disc-tearing cause, while the strengthening or weakening of the phase shift could be indicative of gaps in the disc, either broadening or narrowing the velocity profile of the line. A detection of such a gap would require multiple observations over time to see the increase in visibility measured from the same projected baseline, and likewise with a change in phase shift - and of course would require that the gap aligns with the direction of the observer. The modelling method displayed here could easily and effectively be applied to real-world data, where the wavelength, baseline, and position angle of the telescope, as well as distance to the object, are already known.

The best candidate to test this is on the Be star Pleione (28 Tau), which has been suggested to have a tearing disc \citep{Marr2022}, and has since been shown with modelling to be very likely to possess a tearing disc \citep{Martin2022, Suffak2024}. Pleione is at a distance of 133 pc \citep{Marr2022}, which is close to our chosen model distance of 100 pc, and thus the visibility curves we show may be representative of what Pleione would present, just at slightly longer baselines due to the radius of Pleione being about 70\% of our model stellar radius. The CHARA interferometer has a maximum baseline of 330 m and has been used in many Be star studies at $H$ and $K$ band wavelengths \cite[for example,][]{millan2010spectro, Klement2024}. From our visibility curve in Figure \ref{fig:tear_vis} in the Br$\gamma$ wavelength, which sits in the middle of the $K$ band, CHARA could detect a change due to the tearing disc at its longest baselines as the squared visibility increases by 0.1 to 0.2 when the disc is torn at our $PA_{int}$ of $\ang{90}$. For high resolution differential phase measurements, the GRAVITY instrument at ESO VLTI \citep{Gravity2017} could detect these changes and has also been used for other Be star studies \citep[see][for example]{Klement2024}.

\section{Conclusions}
\label{sec:conclusions}

This paper has been dedicated to expanding our previous works on misaligned Be star binary systems, while also investigating the observational effect of gaps in the discs in our Be star simulation that arise due to the phenomena of KL oscillations and disc tearing. In this paper, we have shown that lowering the binary mass ratio, or raising the disc viscosity, can hinder the dynamical phenomena of disc-tearing and KL oscillations in Be star discs. We have also predicted observables of a KL oscillating disc, presenting the first triple-peaked emission line profile produced from radiative transfer modelling of Be stars, in addition to showing that the observables will oscillate in tandem with the changing disc geometry, and present interferometric predictions due to gaps that form in the disc during these phenomena. The Be star, Pleione, could be a prime candidate to detect an increase in the interferometric visibility due to a torn disc.

The fully 3D models presented in this work reveal a new level of detail to the phenomena of KL oscillations and disc-tearing. Such techniques will be invaluable when modelling specific star-disc systems, and will only become more powerful as model advancements progress and more detail is added into the codes used here. Future work on this subject can include expanding the suite of predictions via the combination of \textsc{sph} and radiative transfer, but also should focus on modelling specific systems in order to rigorously test these techniques, the VDD model, different viscosity prescriptions, and work towards constantly improving these models.

\section*{Acknowledgements}

The authors thank the anonymous referee for their comments which have improved this paper. The authors thank Chris Tycner for valuable feedback on our interferometric modelling results. M.W.S. acknowledges support via the Ontario Graduate Scholarship program. C.E.J. acknowledges support through the National Science and Engineering Research Council of Canada. A. C. C. acknowledges support from CNPq (grant 314545/2023-1) and FAPESP (grants 2018/04055-8 and 2019/13354-1). This work was made possible through the use of the Shared Hierarchical Academic Research Computing Network (SHARCNET).

\section*{Data Availability}

No new data was generated, however the models computed in this work can be made available upon request.

\bibliographystyle{mnras}
\bibliography{aastexBeStarbib}

\begin{appendices}

\section{KL Oscillations from other Observing Angles}
\label{sec:appendix}

Here we present figures similar to Figures \ref{fig:big_obs_0_0} and \ref{fig:big_obs_90_0}, showing how observables can change during a KL oscillation for an observer inclined by $\ang{30}$, $\ang{60}$, or $\ang{90}$ to the stellar pole.

\begin{figure*}
    \centering
    \begin{subfigure}[b]{0.7\textwidth}
        \centering
        \includegraphics[width=\textwidth]{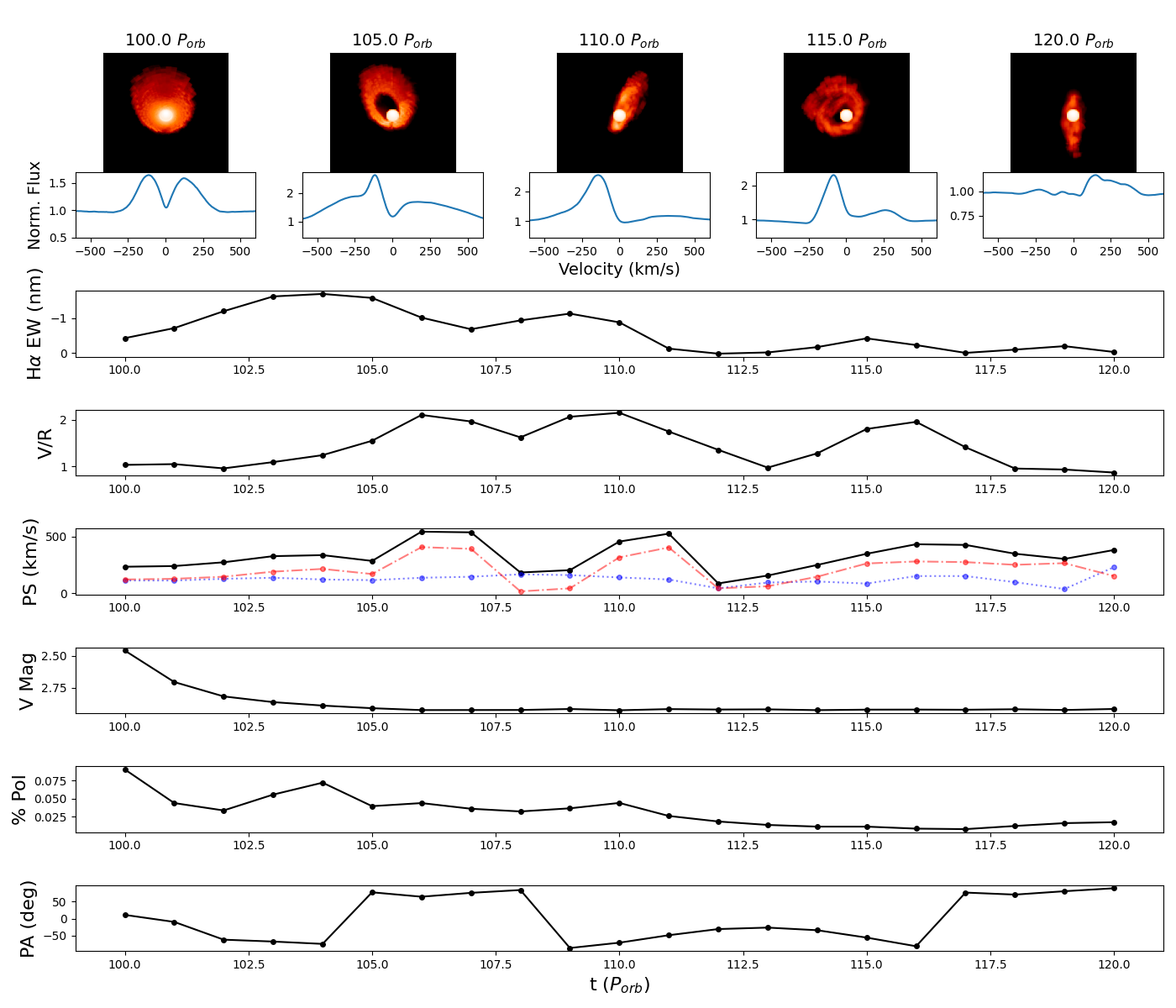}
        \subcaption{$\theta\,=\,\ang{30}$, $\phi\,=\,\ang{0}$}
        \vspace{3pt}
    \end{subfigure} 
    \hspace{5pt}
    \begin{subfigure}[b]{0.7\textwidth}
        \centering
        \includegraphics[width=\textwidth]{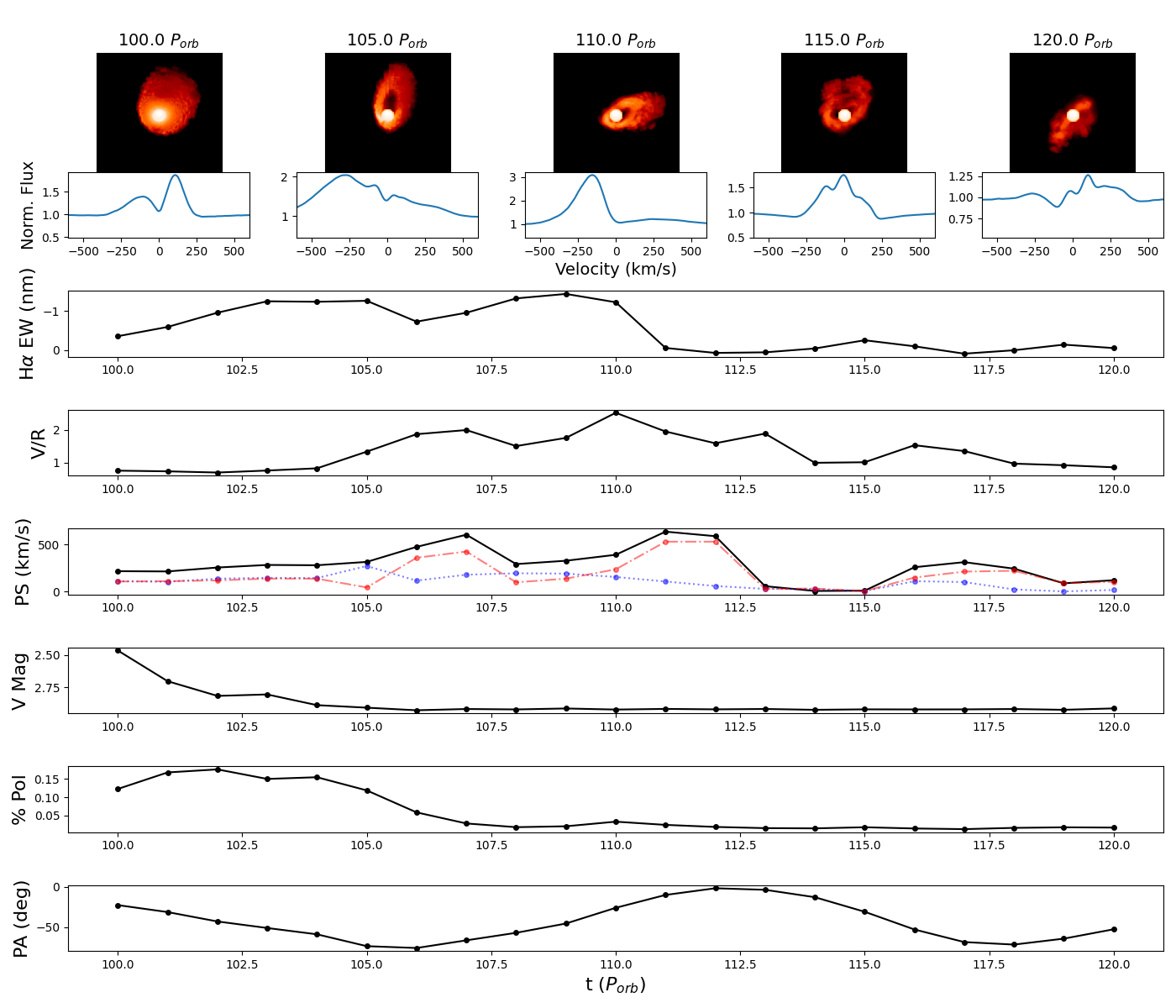}
        \subcaption{$\theta\,=\,\ang{30}$, $\phi\,=\,\ang{45}$}
        \vspace{3pt}
    \end{subfigure} 
    \caption{Same format as Figure \ref{fig:big_obs_0_0}, for polar observing angles of $\ang{30}$ and varying azimuthal angles as indicated in the various captions.}
\end{figure*}

\begin{figure*}\ContinuedFloat
    \centering
    \begin{subfigure}[b]{0.7\textwidth}
        \centering
        \includegraphics[width=\textwidth]{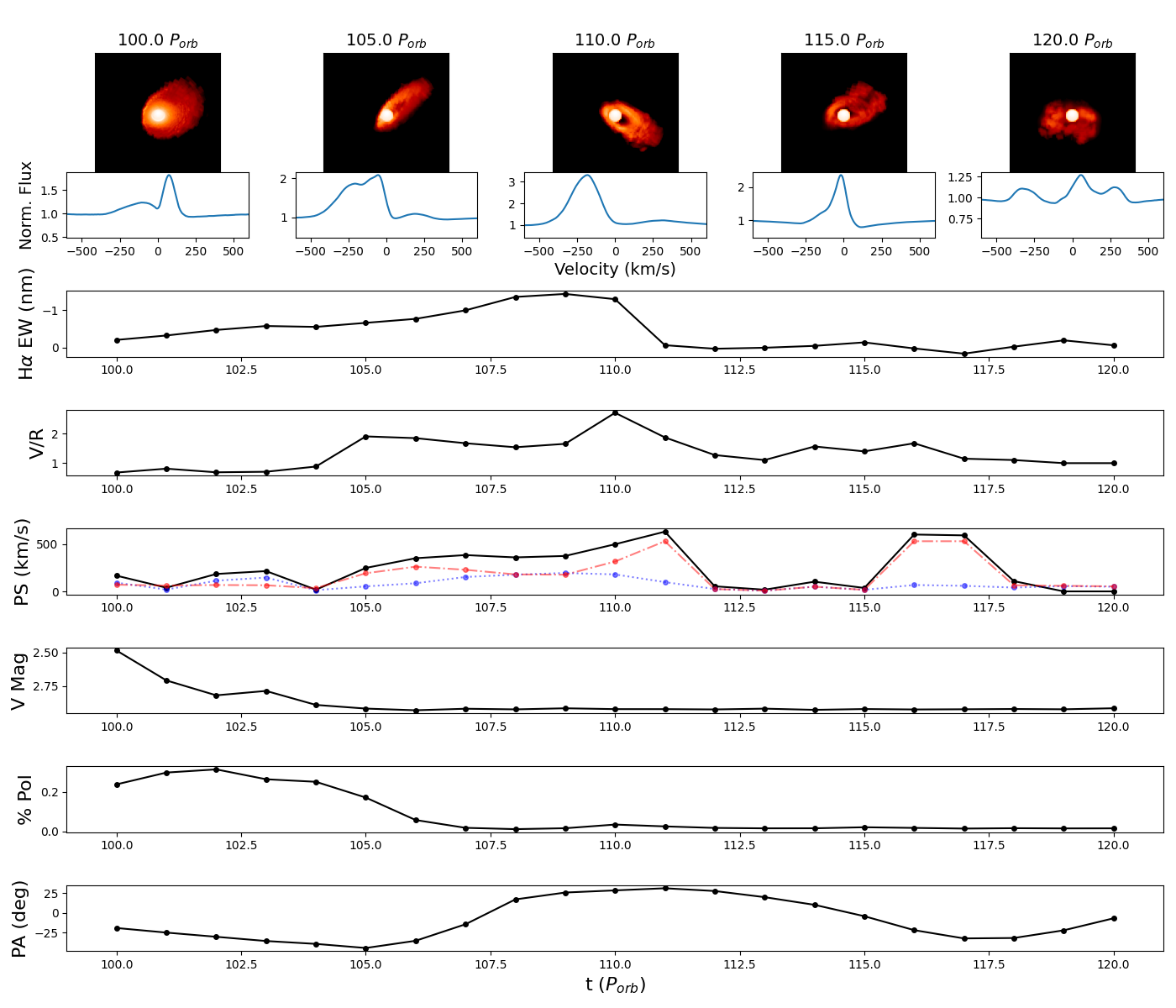}
        \subcaption{$\theta\,=\,\ang{30}$, $\phi\,=\,\ang{90}$}
        \vspace{3pt}
    \end{subfigure} 
    \hspace{5pt}
    \begin{subfigure}[b]{0.7\textwidth}
        \centering
        \includegraphics[width=\textwidth]{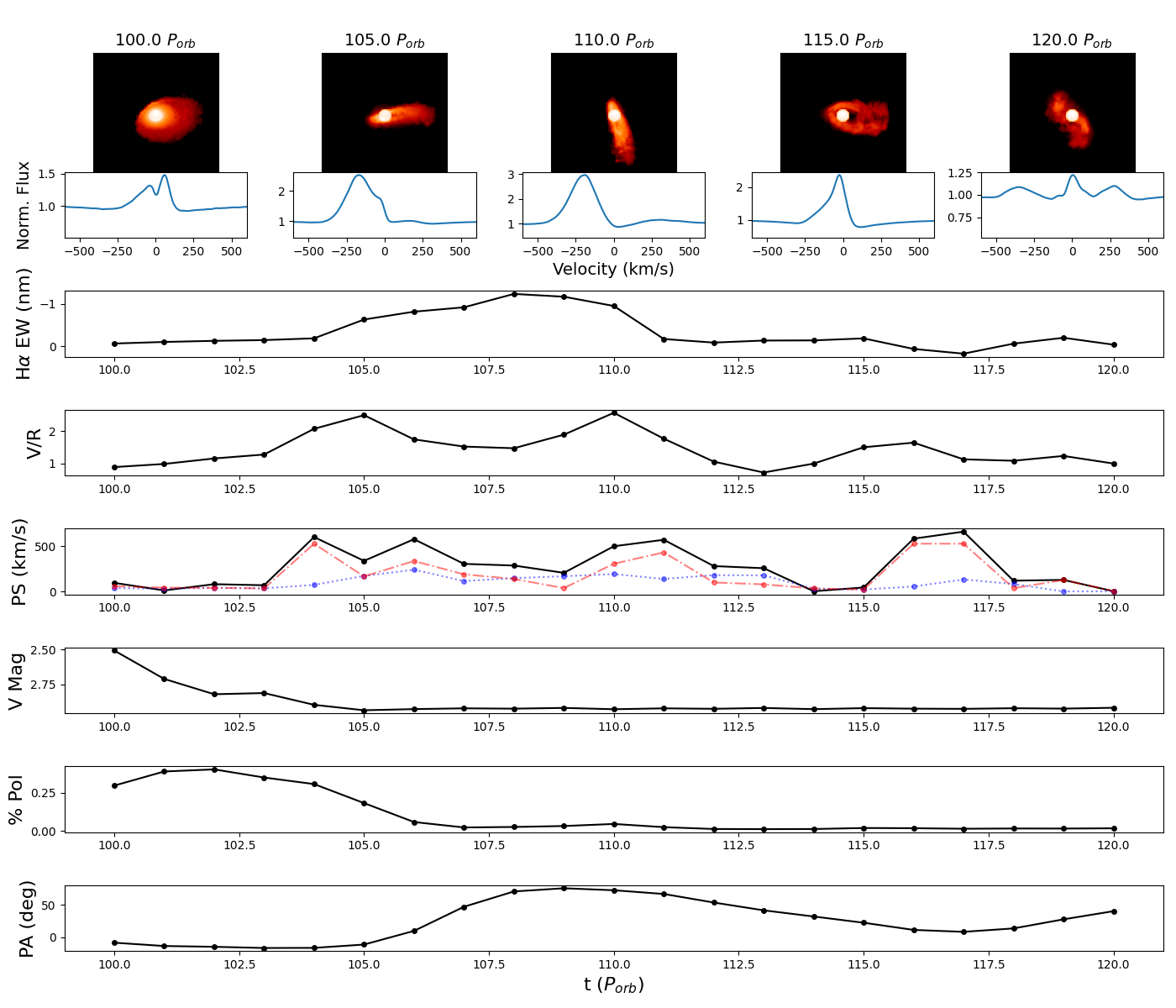}
        \subcaption{$\theta\,=\,\ang{30}$, $\phi\,=\,\ang{135}$}
        \vspace{3pt}
    \end{subfigure} 
    \caption{Continued}
\end{figure*}

\begin{figure*}\ContinuedFloat
    \centering
    \begin{subfigure}[b]{0.7\textwidth}
        \centering
        \includegraphics[width=\textwidth]{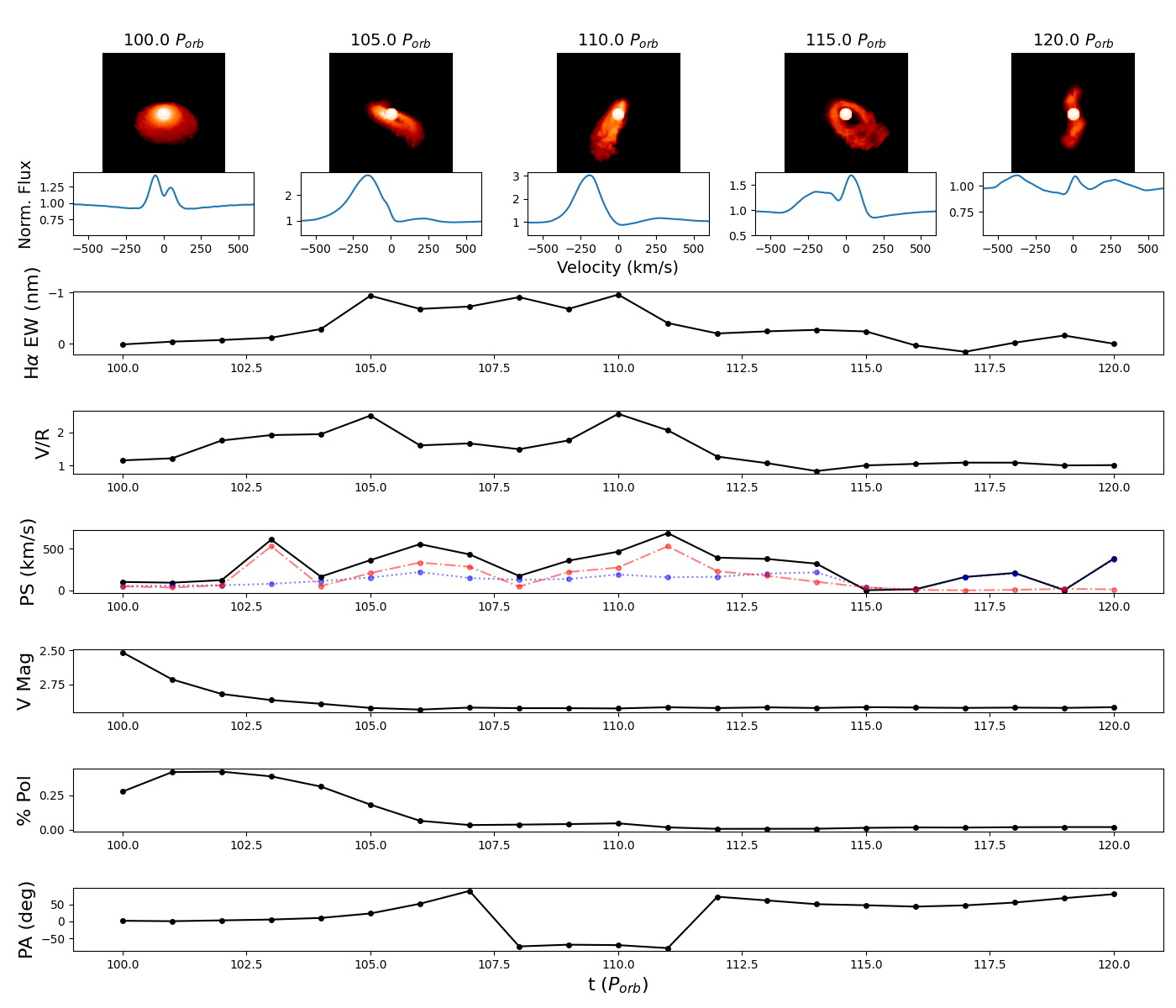}
        \subcaption{$\theta\,=\,\ang{30}$, $\phi\,=\,\ang{180}$}
        \vspace{3pt}
    \end{subfigure} 
    \hspace{5pt}
    \begin{subfigure}[b]{0.7\textwidth}
        \centering
        \includegraphics[width=\textwidth]{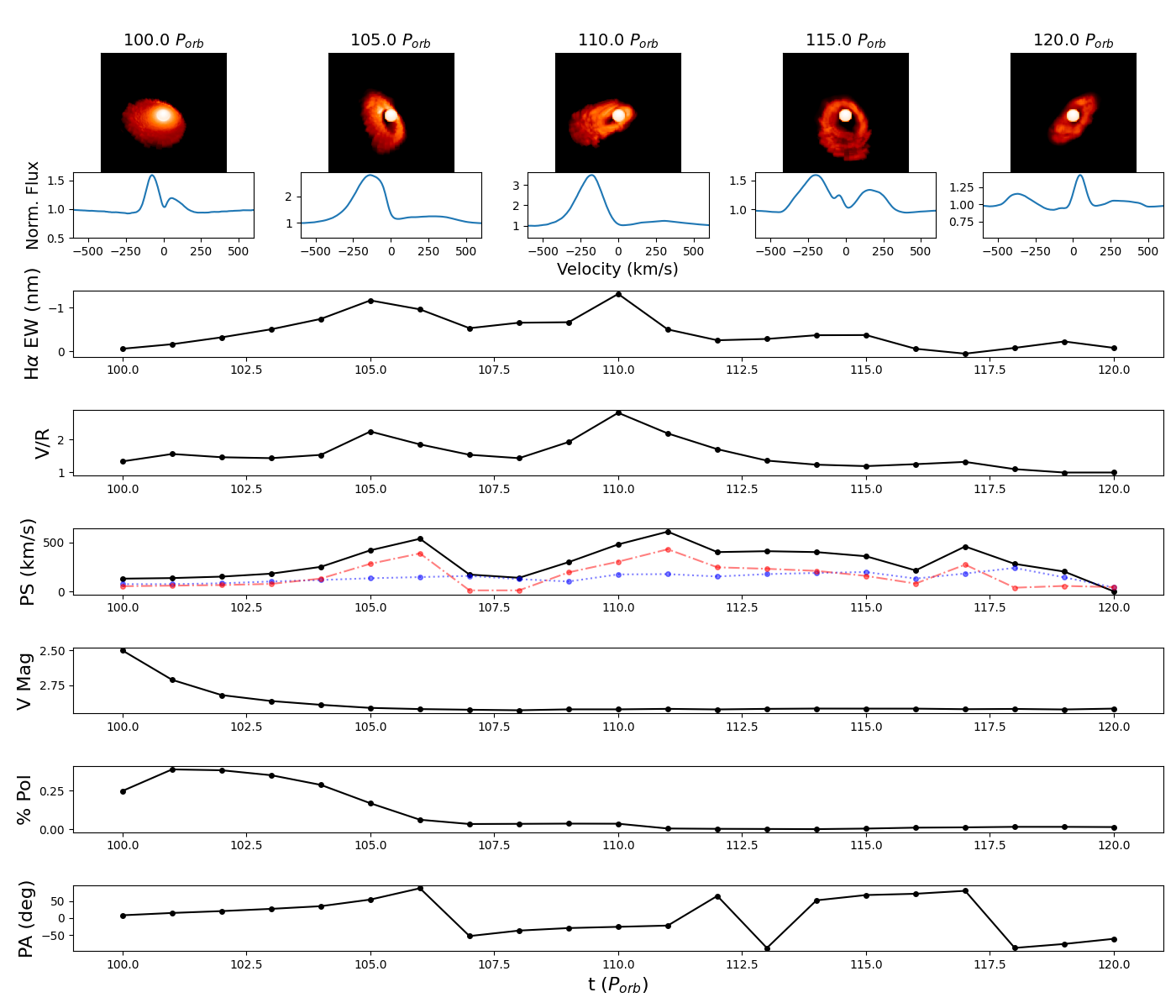}
        \subcaption{$\theta\,=\,\ang{30}$, $\phi\,=\,\ang{225}$}
        \vspace{3pt}
    \end{subfigure} 
    \caption{Continued}
\end{figure*}

\begin{figure*}\ContinuedFloat
    \centering
    \begin{subfigure}[b]{0.7\textwidth}
        \centering
        \includegraphics[width=\textwidth]{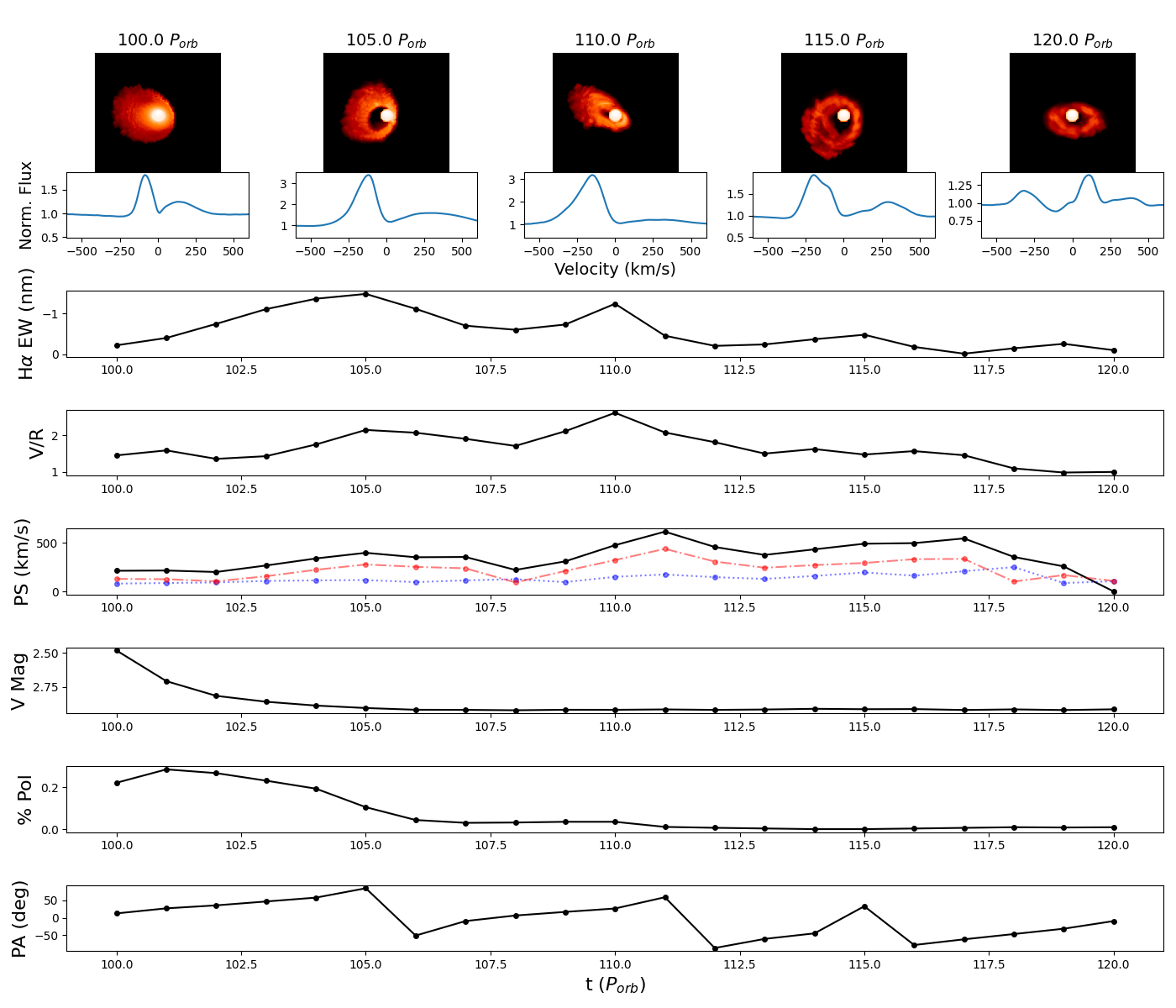}
        \subcaption{$\theta\,=\,\ang{30}$, $\phi\,=\,\ang{270}$}
        \vspace{3pt}
    \end{subfigure} 
    \hspace{5pt}
    \begin{subfigure}[b]{0.7\textwidth}
        \centering
        \includegraphics[width=\textwidth]{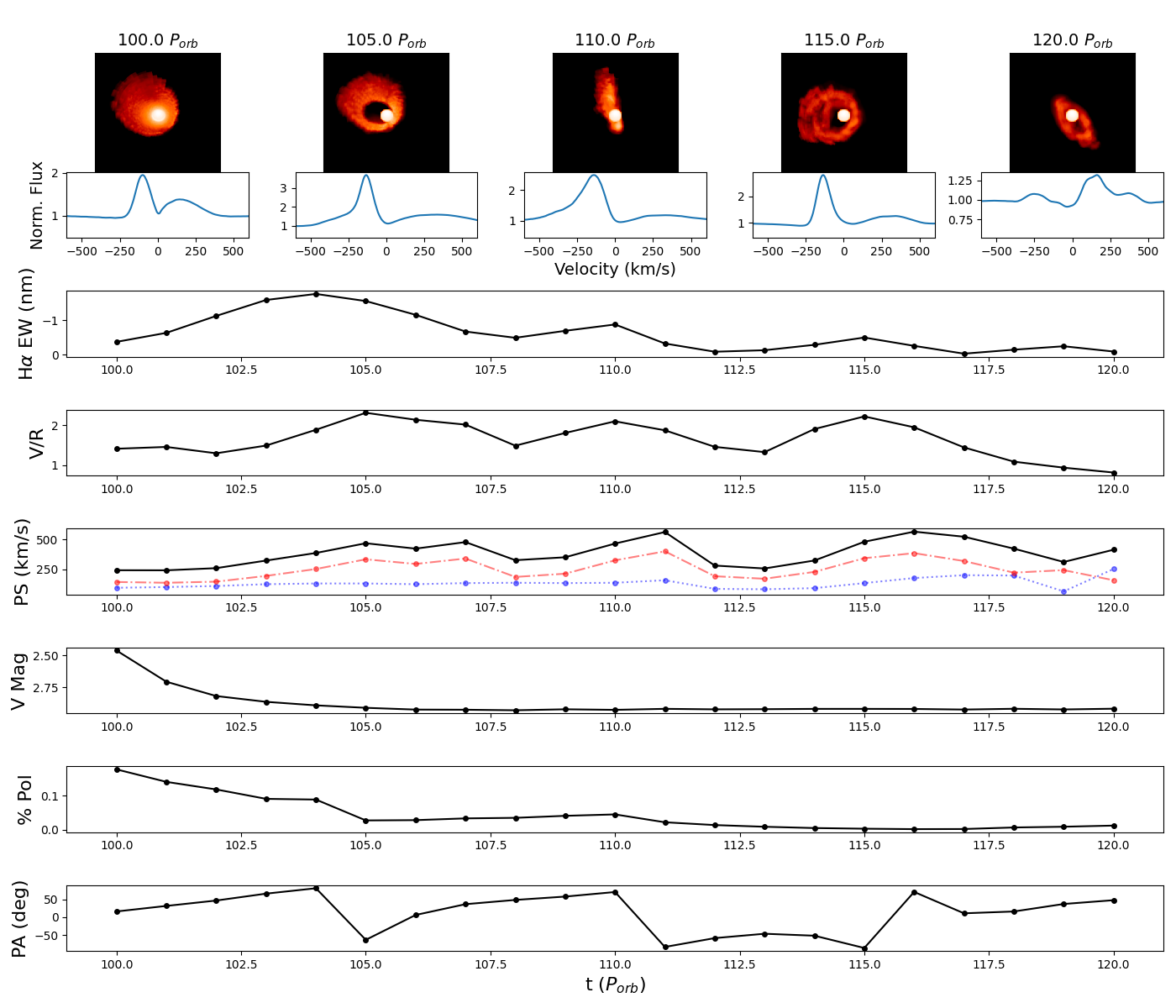}
        \subcaption{$\theta\,=\,\ang{30}$, $\phi\,=\,\ang{315}$}
        \vspace{3pt}
    \end{subfigure} 
    \caption{Continued}
\end{figure*}

\begin{figure*}
    \centering
    \begin{subfigure}[b]{0.7\textwidth}
        \centering
        \includegraphics[width=\textwidth]{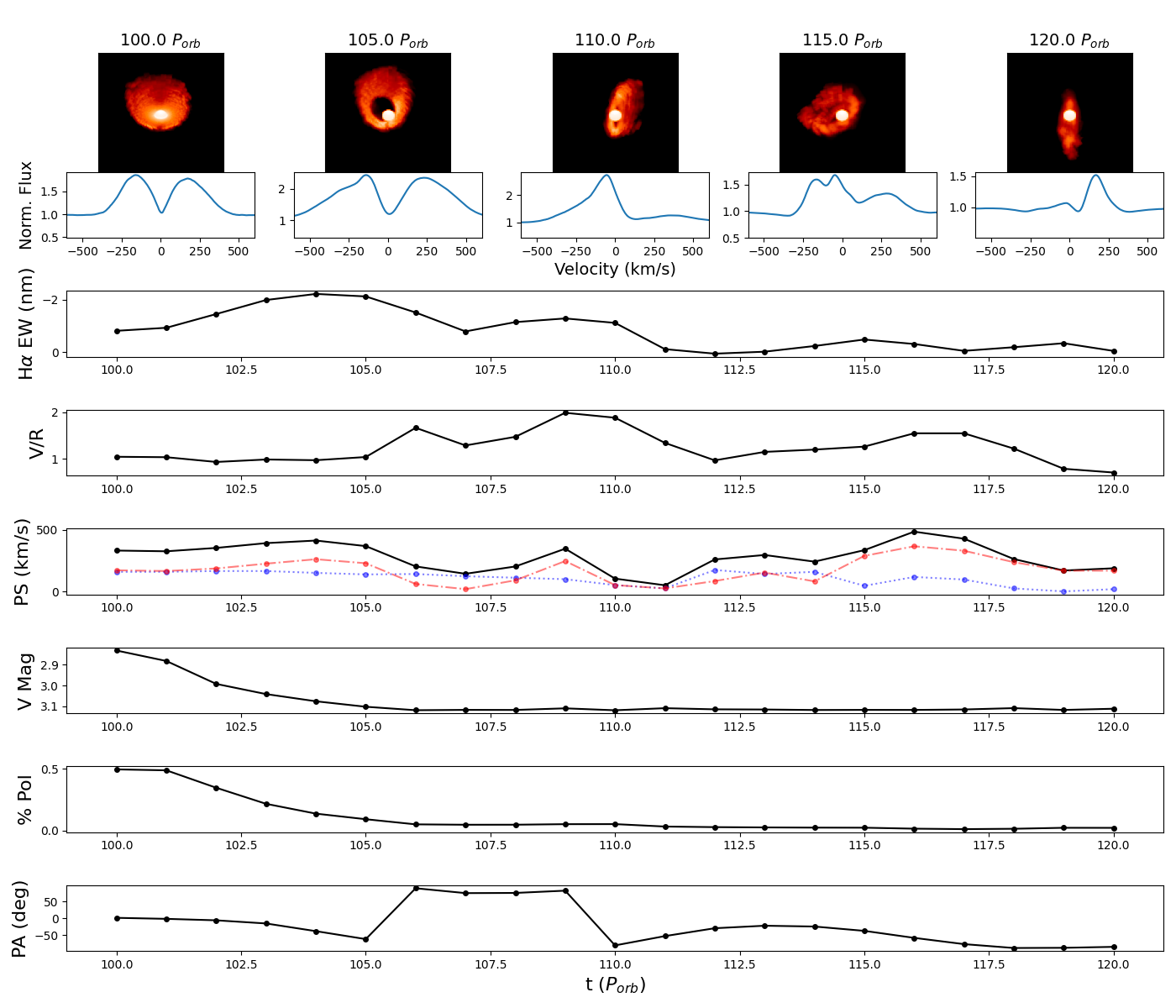}
        \subcaption{$\theta\,=\,\ang{60}$, $\phi\,=\,\ang{0}$}
        \vspace{3pt}
    \end{subfigure} 
    \hspace{5pt}
    \begin{subfigure}[b]{0.7\textwidth}
        \centering
        \includegraphics[width=\textwidth]{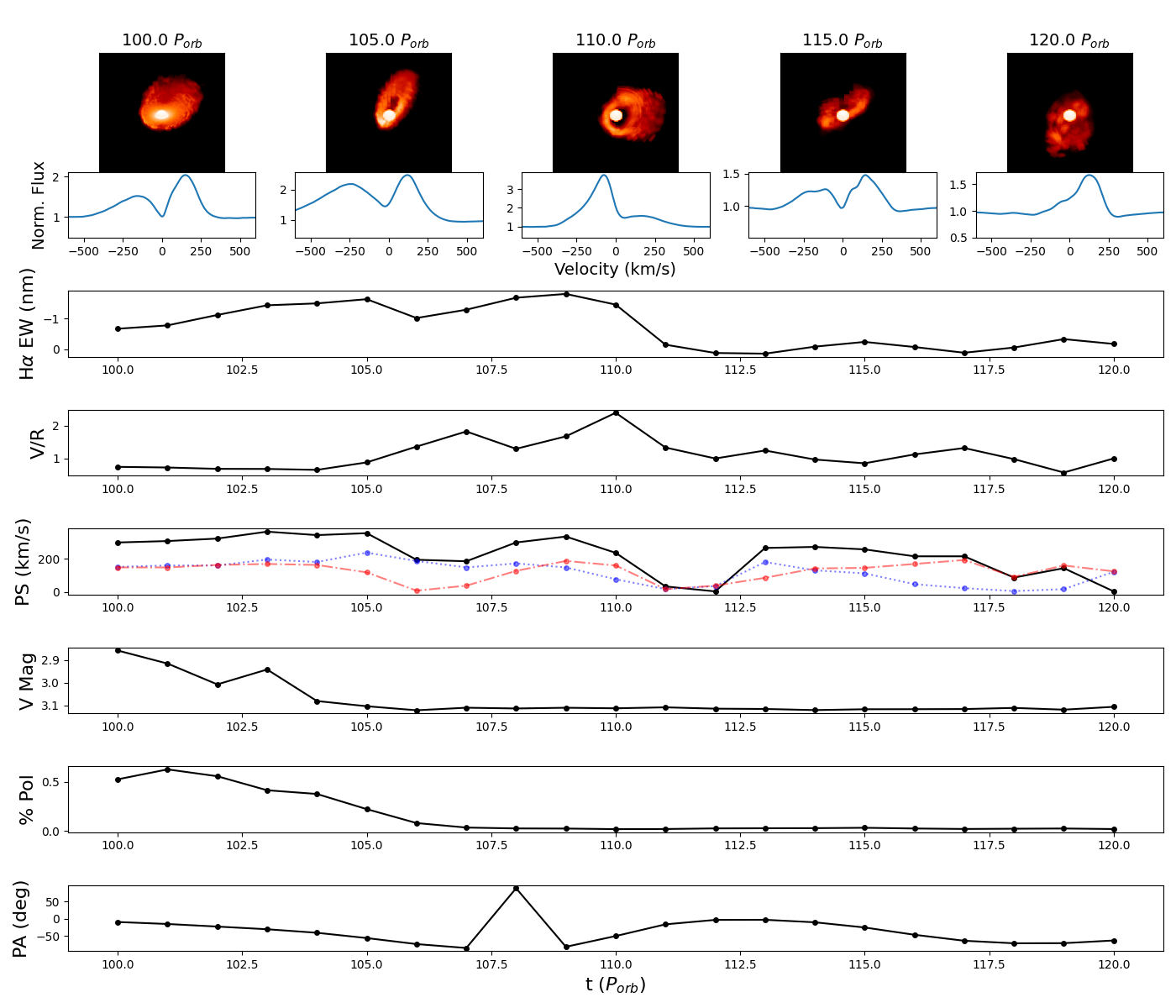}
        \subcaption{$\theta\,=\,\ang{60}$, $\phi\,=\,\ang{45}$}
        \vspace{3pt}
    \end{subfigure} 
    \caption{Same format as Figure \ref{fig:big_obs_0_0}, for polar observing angles of $\ang{60}$ and varying azimuthal angles as indicated in the various captions.}
\end{figure*}

\begin{figure*}\ContinuedFloat
    \centering
    \begin{subfigure}[b]{0.7\textwidth}
        \centering
        \includegraphics[width=\textwidth]{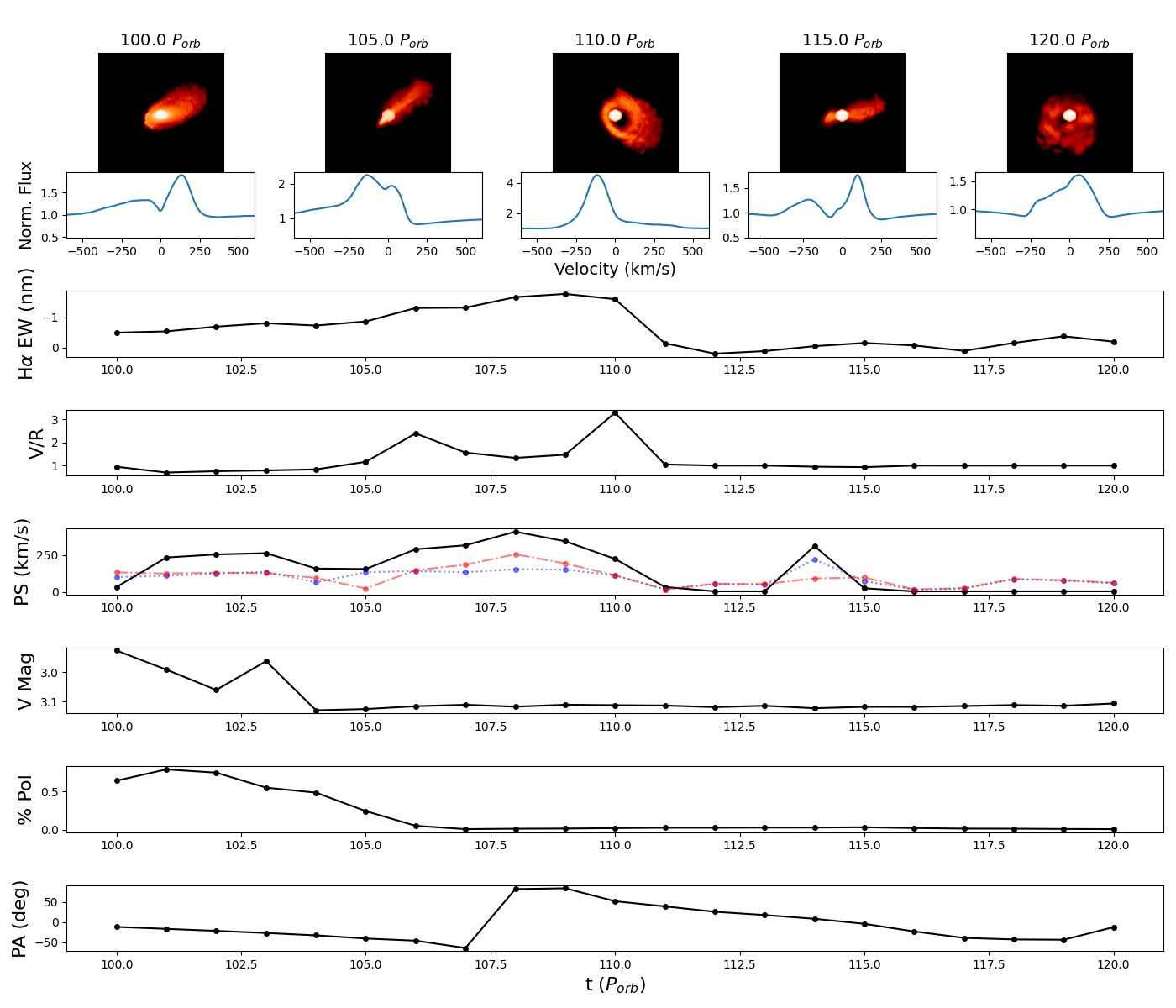}
        \subcaption{$\theta\,=\,\ang{60}$, $\phi\,=\,\ang{90}$}
        \vspace{3pt}
    \end{subfigure} 
    \hspace{5pt}
    \begin{subfigure}[b]{0.7\textwidth}
        \centering
        \includegraphics[width=\textwidth]{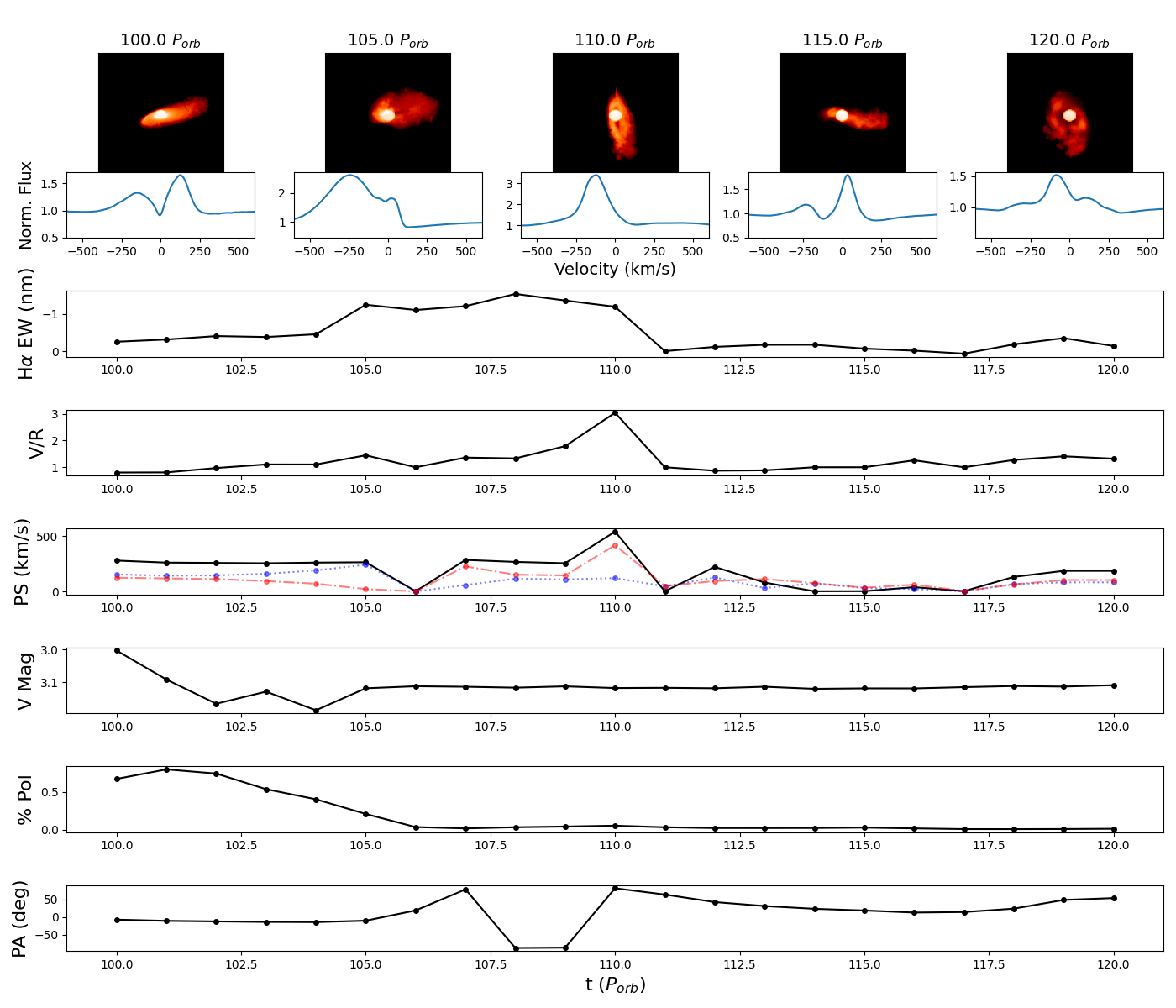}
        \subcaption{$\theta\,=\,\ang{60}$, $\phi\,=\,\ang{135}$}
        \vspace{3pt}
    \end{subfigure} 
    \caption{Continued}
\end{figure*}

\begin{figure*}\ContinuedFloat
    \centering
    \begin{subfigure}[b]{0.7\textwidth}
        \centering
        \includegraphics[width=\textwidth]{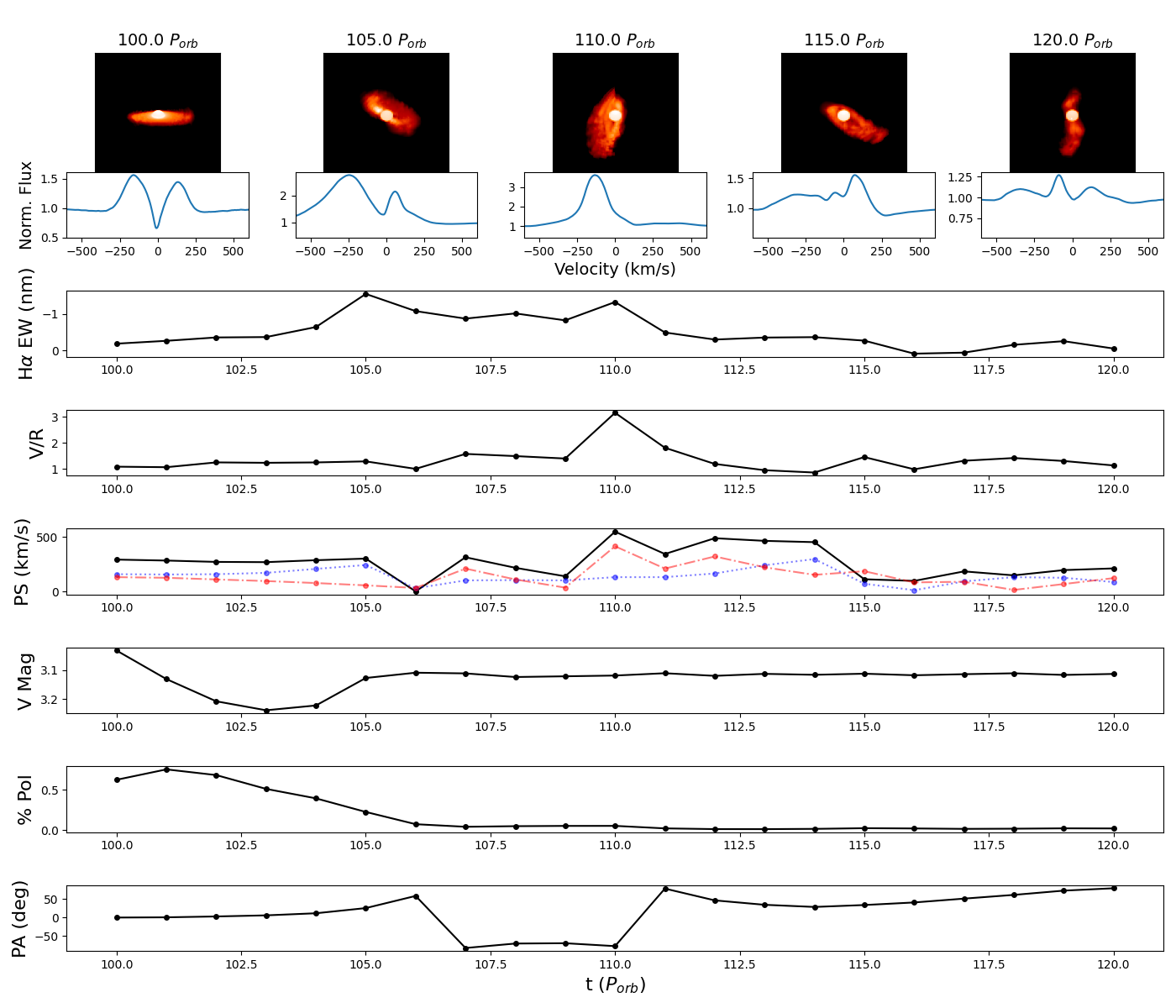}
        \subcaption{$\theta\,=\,\ang{60}$, $\phi\,=\,\ang{180}$}
        \vspace{3pt}
    \end{subfigure} 
    \hspace{5pt}
    \begin{subfigure}[b]{0.7\textwidth}
        \centering
        \includegraphics[width=\textwidth]{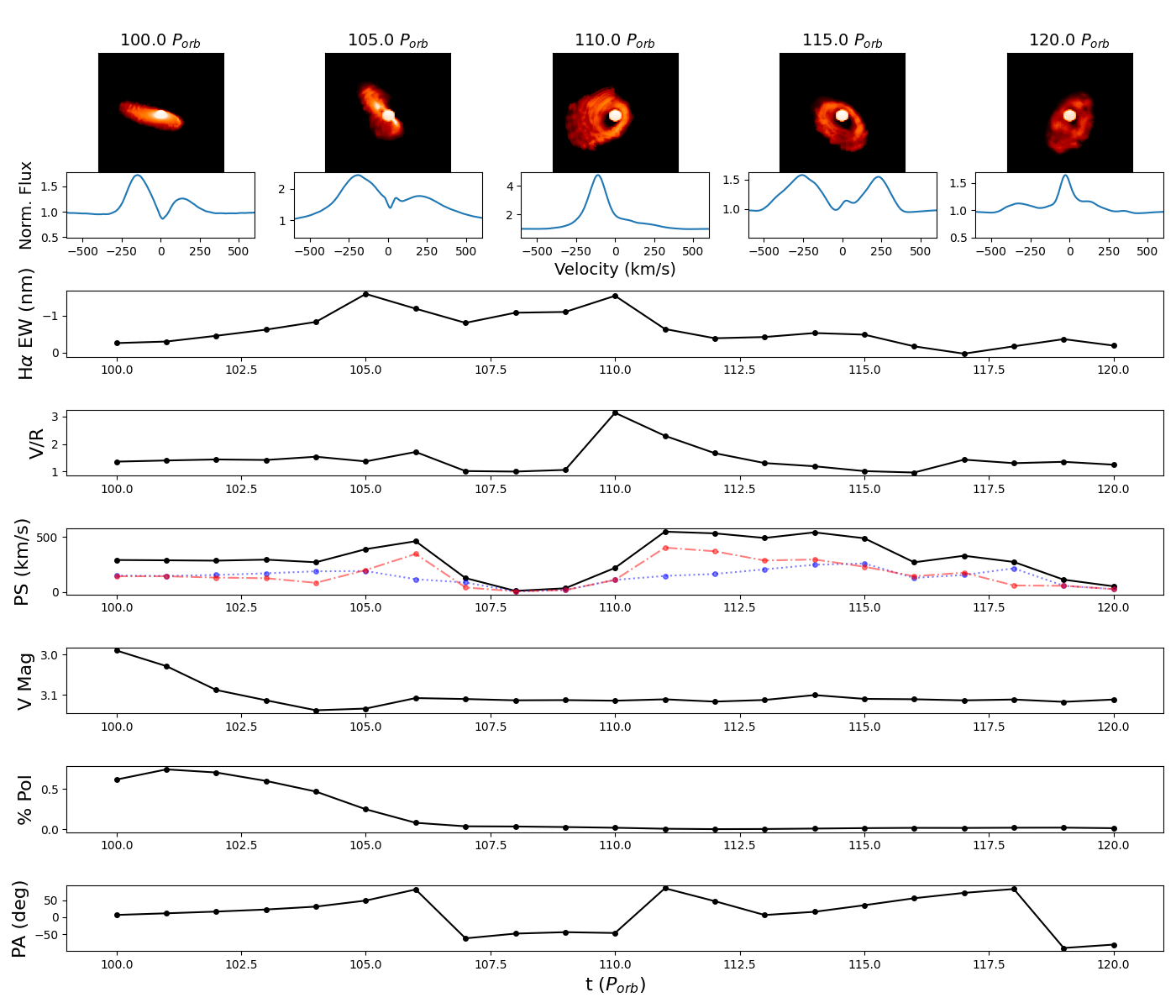}
        \subcaption{$\theta\,=\,\ang{60}$, $\phi\,=\,\ang{225}$}
        \vspace{3pt}
    \end{subfigure} 
    \caption{Continued}
\end{figure*}

\begin{figure*}\ContinuedFloat
    \centering
    \begin{subfigure}[b]{0.7\textwidth}
        \centering
        \includegraphics[width=\textwidth]{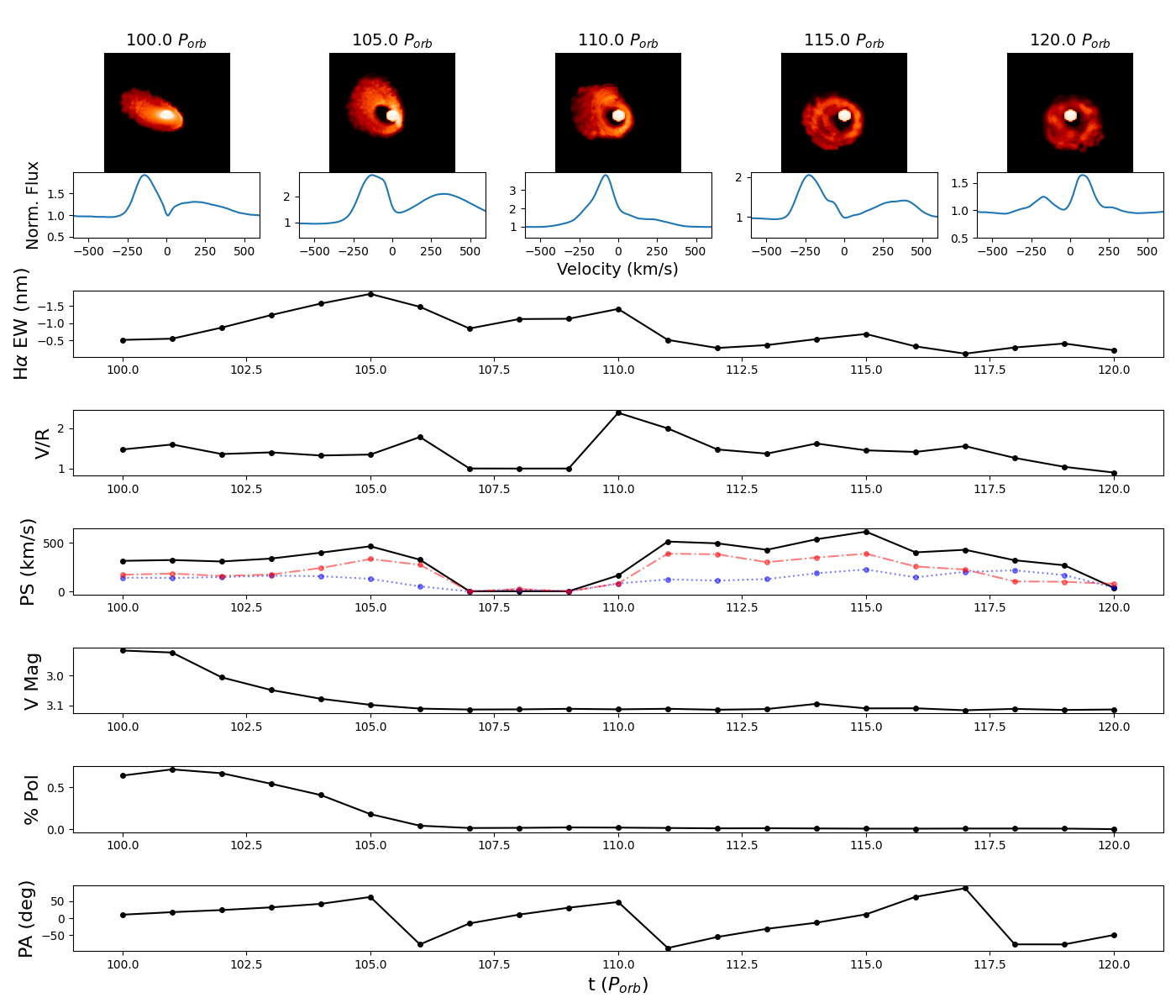}
        \subcaption{$\theta\,=\,\ang{60}$, $\phi\,=\,\ang{270}$}
        \vspace{3pt}
    \end{subfigure} 
    \hspace{5pt}
    \begin{subfigure}[b]{0.7\textwidth}
        \centering
        \includegraphics[width=\textwidth]{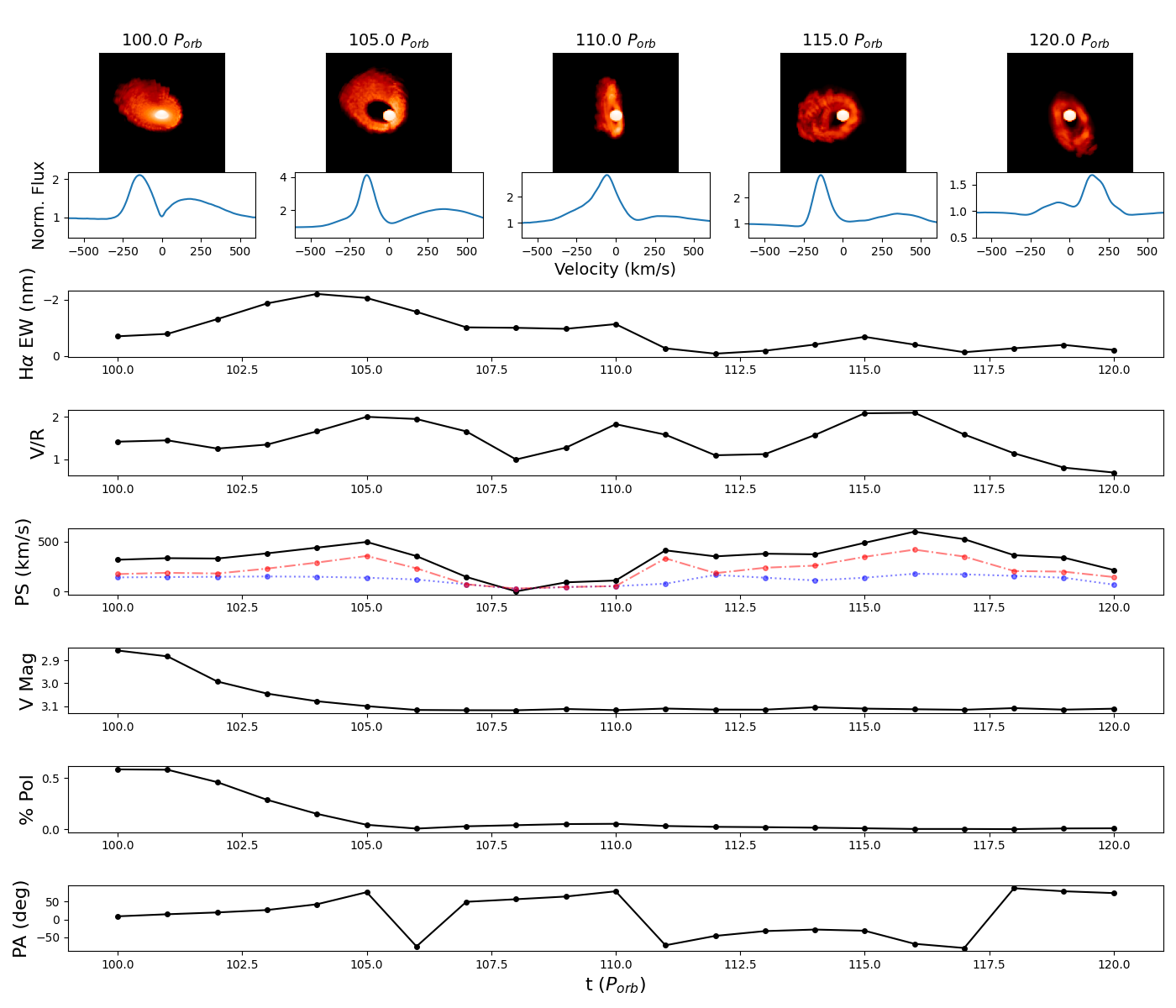}
        \subcaption{$\theta\,=\,\ang{60}$, $\phi\,=\,\ang{315}$}
        \vspace{3pt}
    \end{subfigure} 
    \caption{Continued}
\end{figure*}

\begin{figure*}
    \centering
    \begin{subfigure}[b]{0.7\textwidth}
        \centering
        \includegraphics[width=\textwidth]{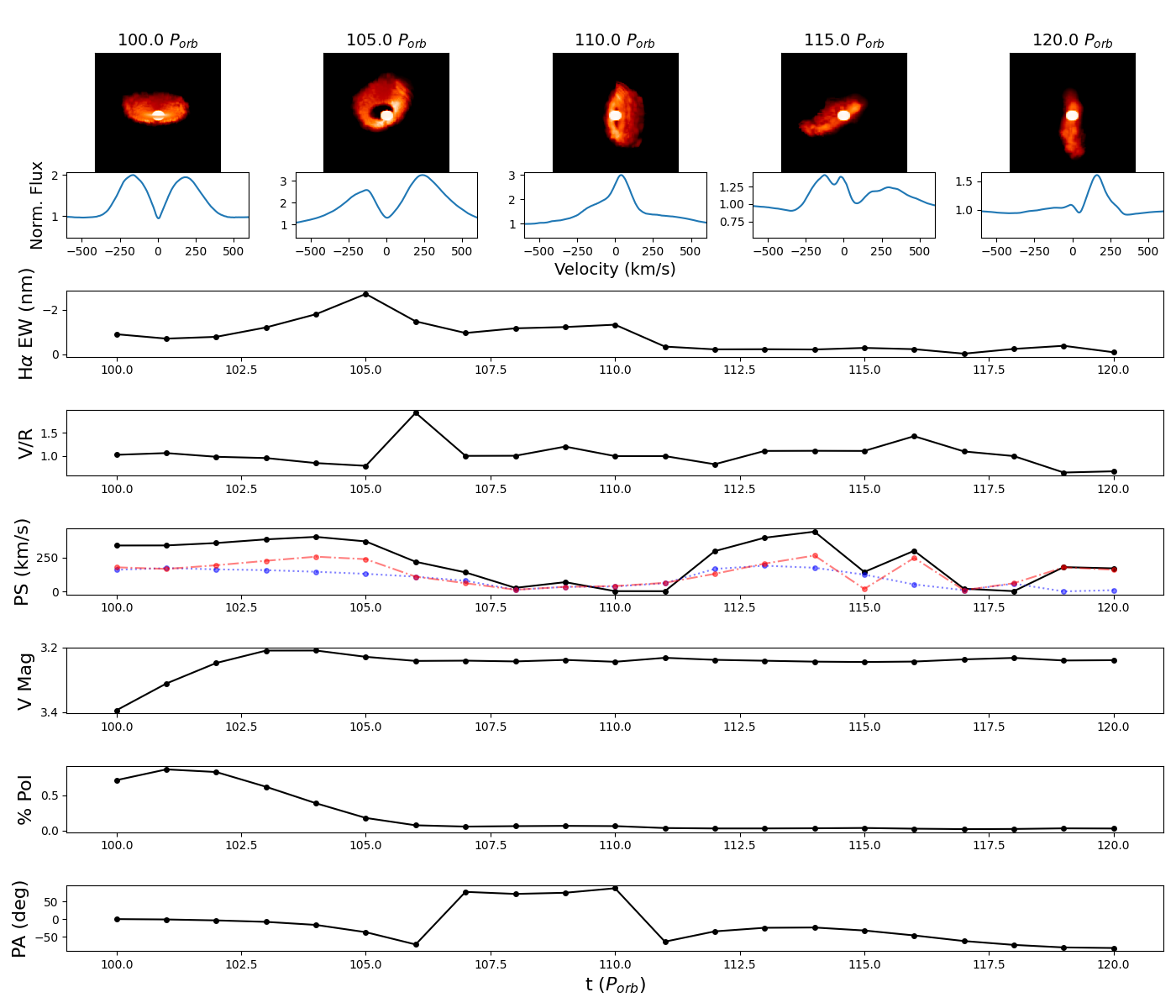}
        \subcaption{$\theta\,=\,\ang{90}$, $\phi\,=\,\ang{0}$}
        \vspace{3pt}
    \end{subfigure} 
    \hspace{5pt}
    \begin{subfigure}[b]{0.7\textwidth}
        \centering
        \includegraphics[width=\textwidth]{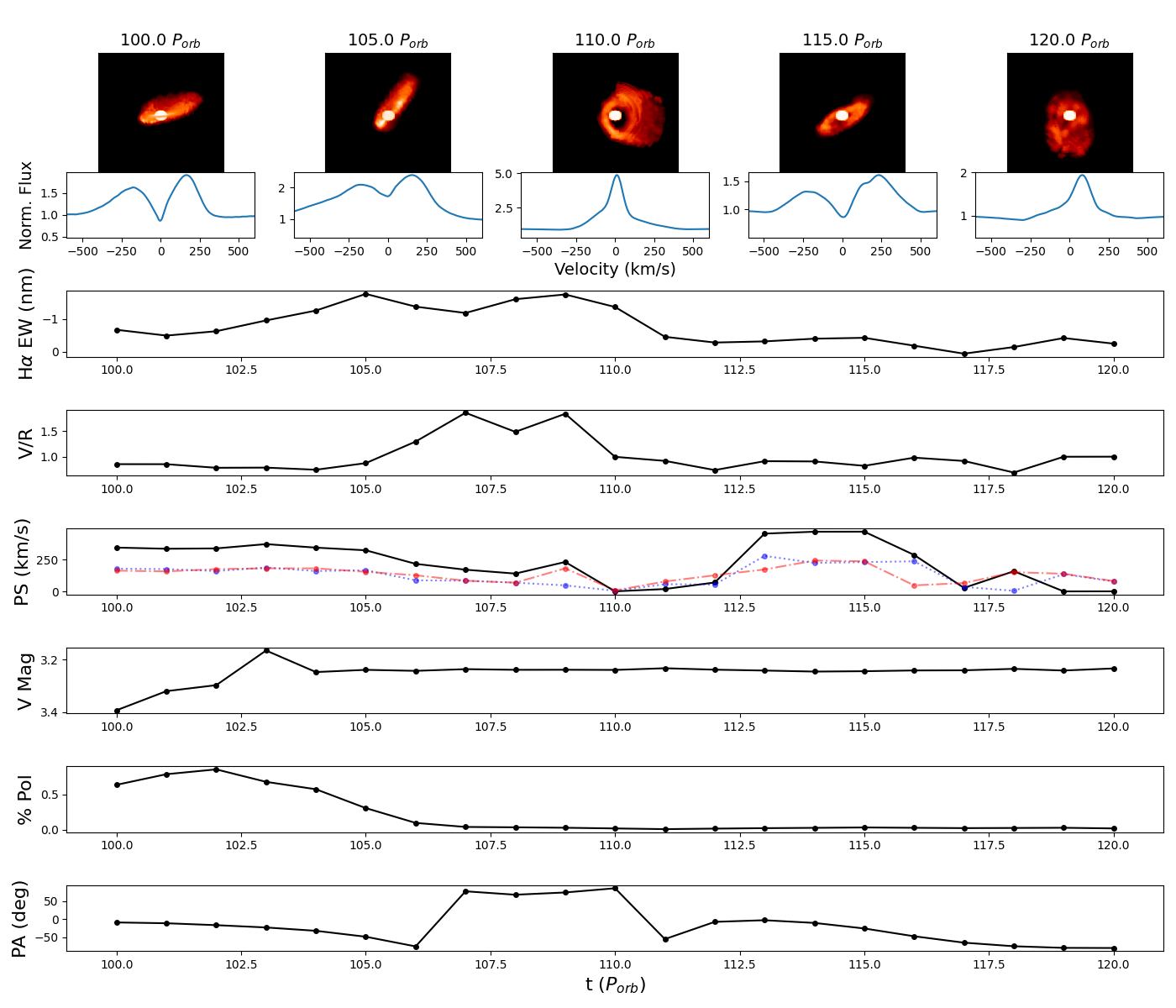}
        \subcaption{$\theta\,=\,\ang{90}$, $\phi\,=\,\ang{45}$}
        \vspace{3pt}
    \end{subfigure} 
    \caption{Same format as Figure \ref{fig:big_obs_0_0}, for polar observing angles of $\ang{90}$ and varying azimuthal angles as indicated in the various captions.}
\end{figure*}

\begin{figure*}\ContinuedFloat
    \centering
    \begin{subfigure}[b]{0.7\textwidth}
        \centering
        \includegraphics[width=\textwidth]{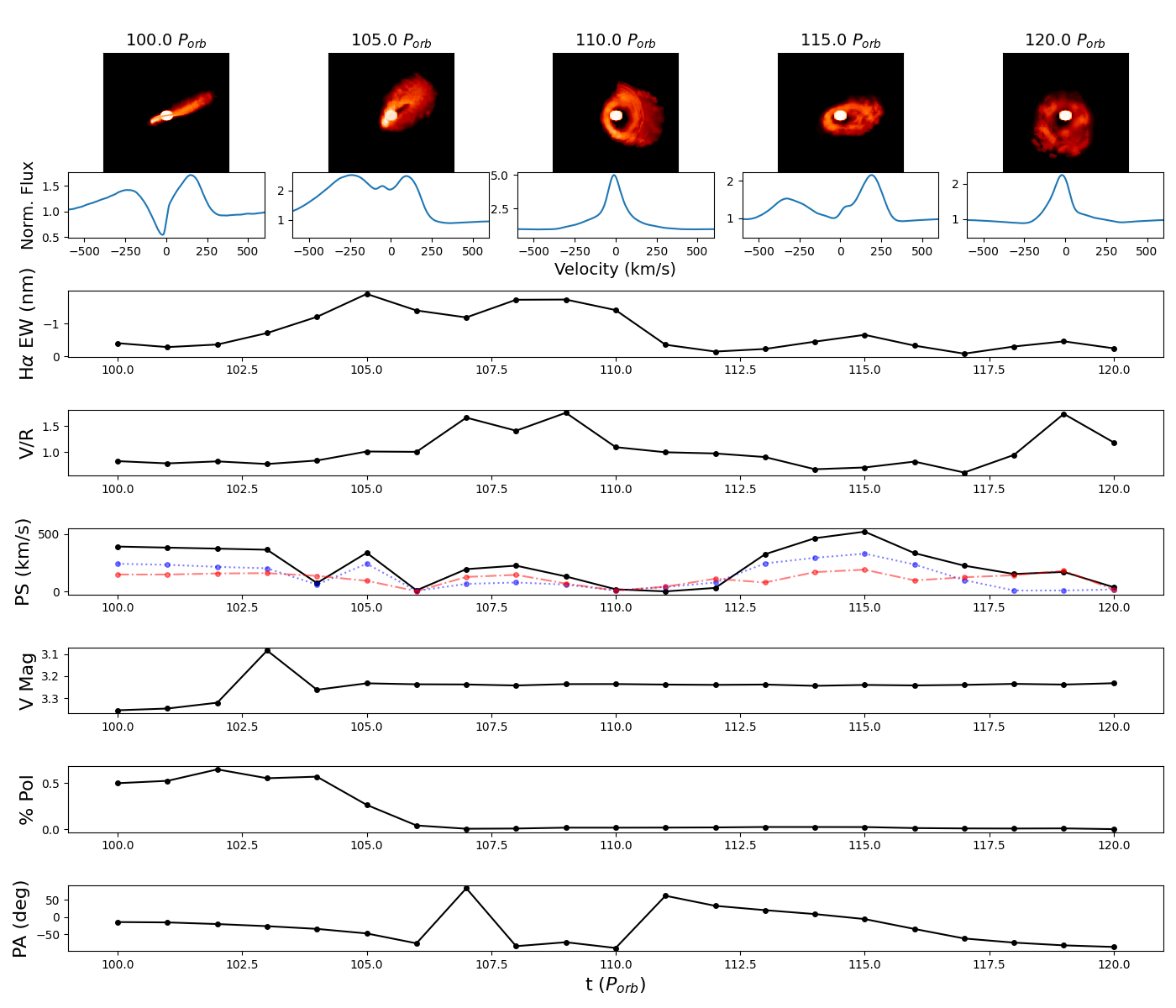}
        \subcaption{$\theta\,=\,\ang{90}$, $\phi\,=\,\ang{90}$}
        \vspace{3pt}
    \end{subfigure} 
    \hspace{5pt}
    \begin{subfigure}[b]{0.7\textwidth}
        \centering
        \includegraphics[width=\textwidth]{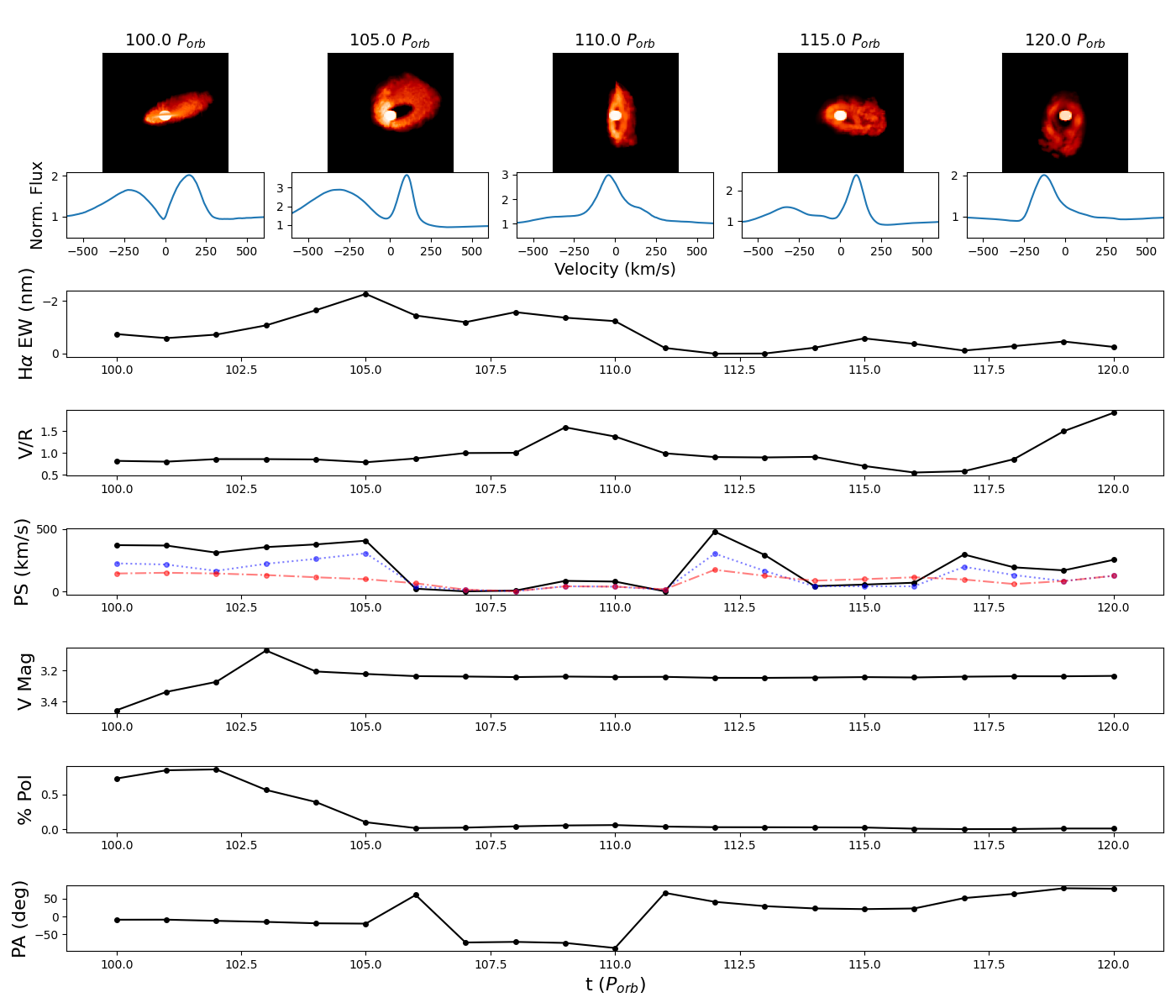}
        \subcaption{$\theta\,=\,\ang{90}$, $\phi\,=\,\ang{135}$}
        \vspace{3pt}
    \end{subfigure} 
    \caption{Continued}
\end{figure*}

\begin{figure*}\ContinuedFloat
    \centering
    \begin{subfigure}[b]{0.7\textwidth}
        \centering
        \includegraphics[width=\textwidth]{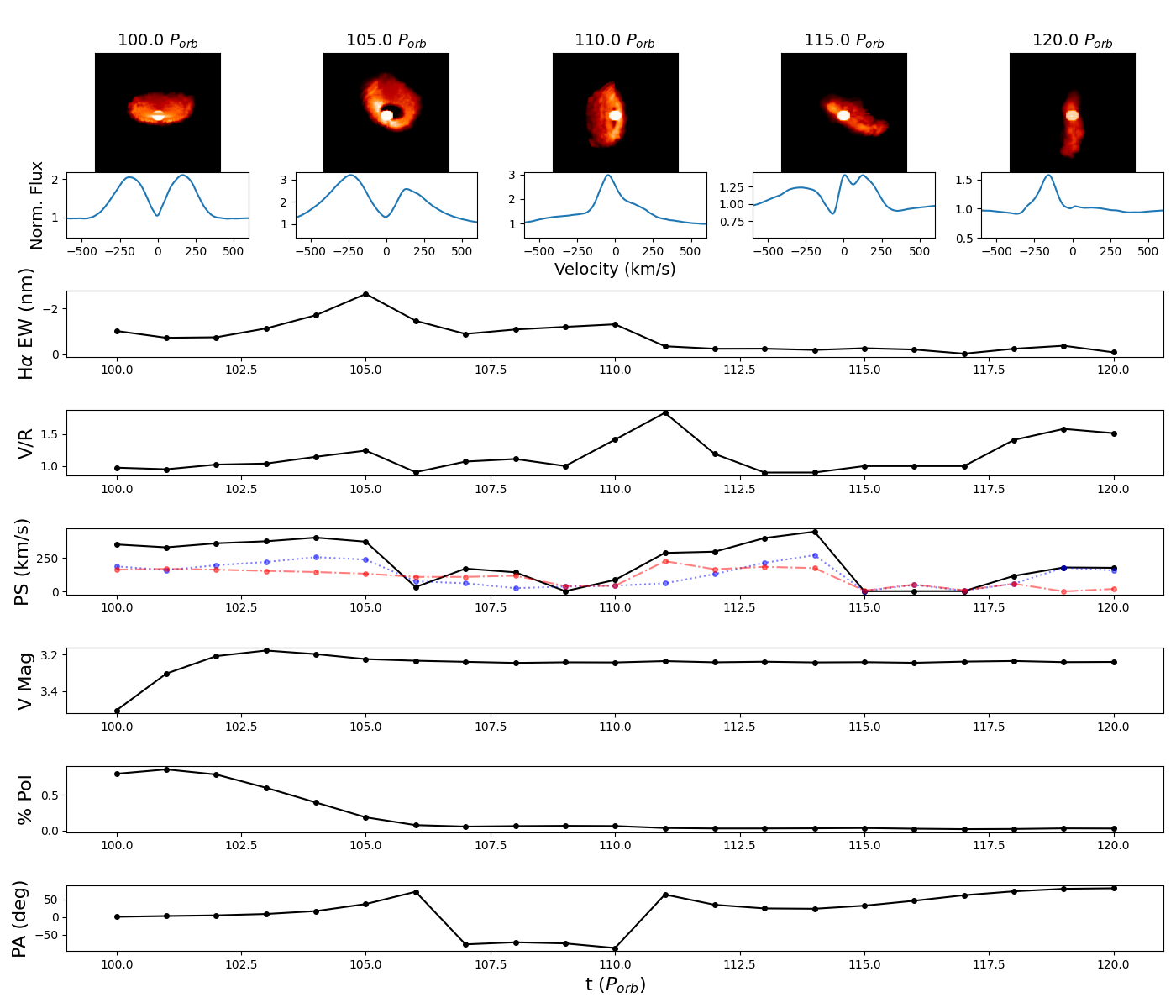}
        \subcaption{$\theta\,=\,\ang{90}$, $\phi\,=\,\ang{180}$}
        \vspace{3pt}
    \end{subfigure} 
    \hspace{5pt}
    \begin{subfigure}[b]{0.7\textwidth}
        \centering
        \includegraphics[width=\textwidth]{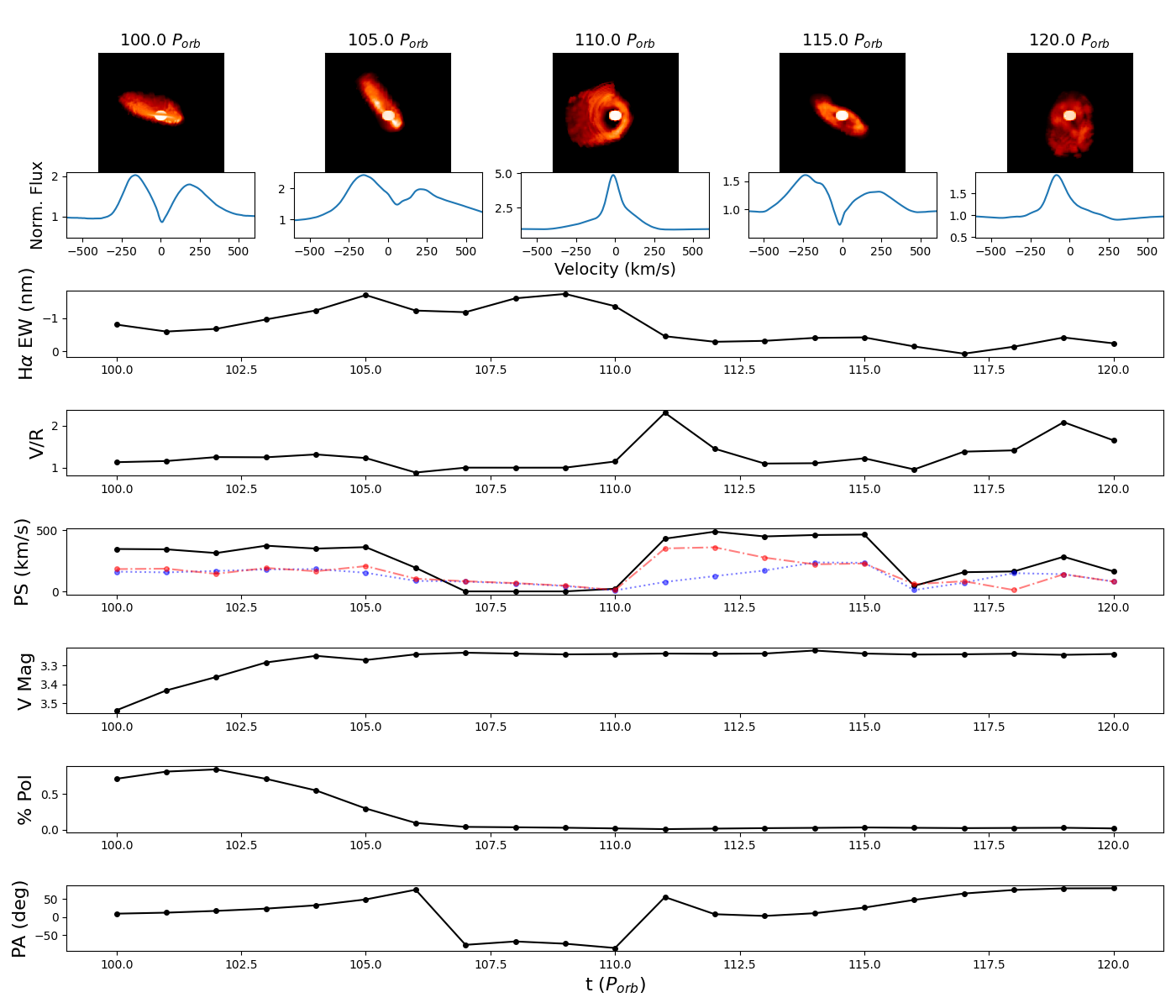}
        \subcaption{$\theta\,=\,\ang{90}$, $\phi\,=\,\ang{225}$}
        \vspace{3pt}
    \end{subfigure} 
    \caption{Continued}
\end{figure*}

\begin{figure*}\ContinuedFloat
    \centering
    \begin{subfigure}[b]{0.7\textwidth}
        \centering
        \includegraphics[width=\textwidth]{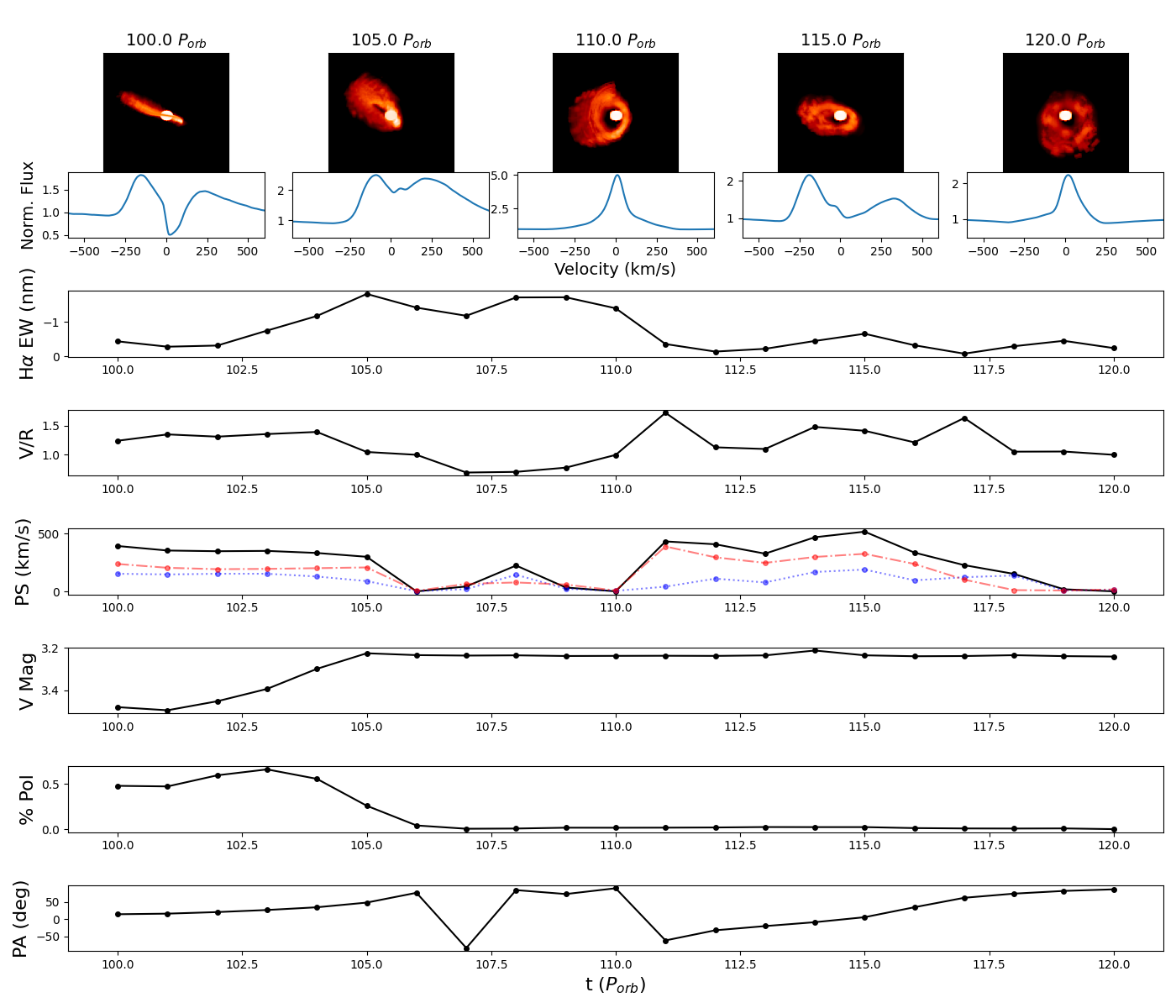}
        \subcaption{$\theta\,=\,\ang{90}$, $\phi\,=\,\ang{270}$}
        \vspace{3pt}
    \end{subfigure} 
    \hspace{5pt}
    \begin{subfigure}[b]{0.7\textwidth}
        \centering
        \includegraphics[width=\textwidth]{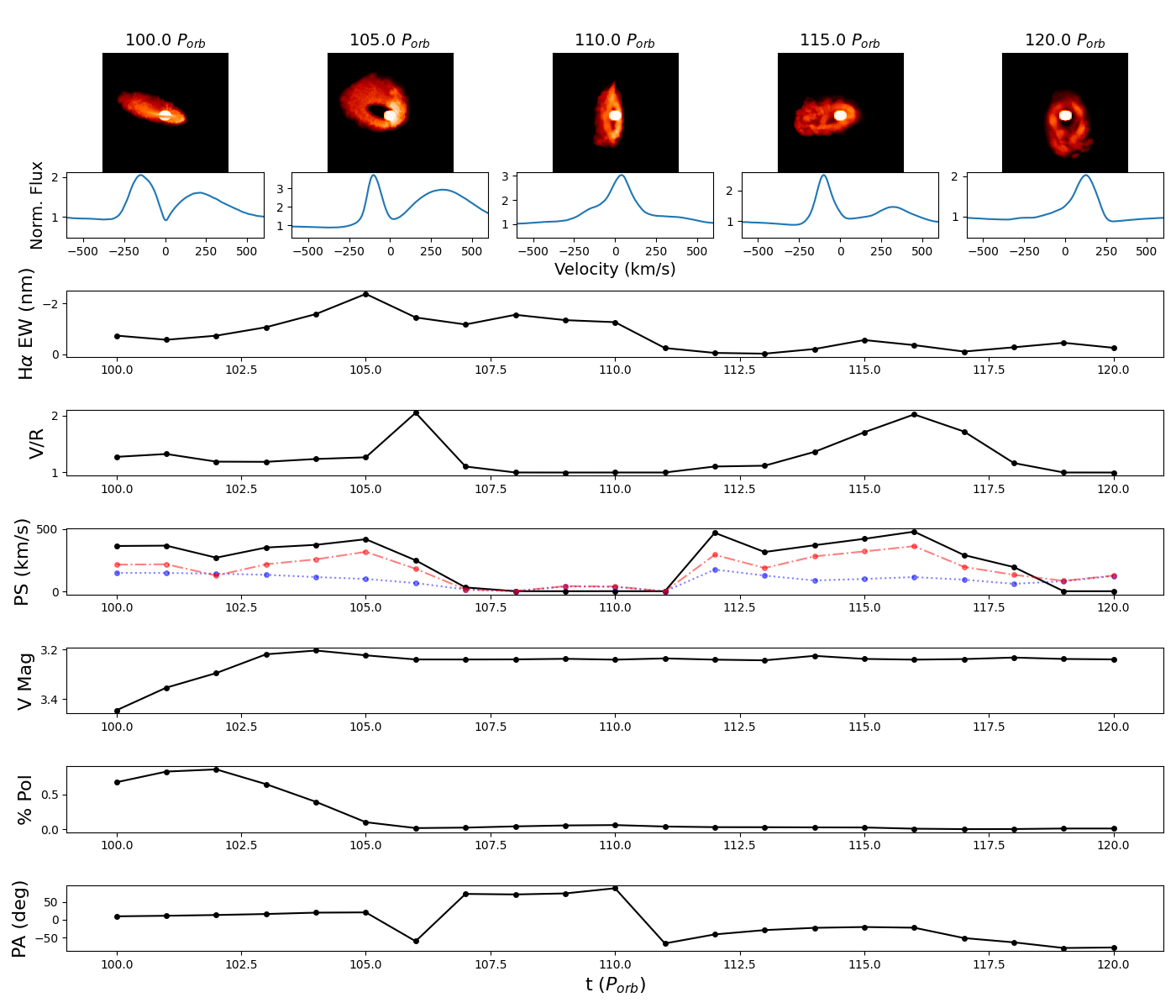}
        \subcaption{$\theta\,=\,\ang{90}$, $\phi\,=\,\ang{315}$}
        \vspace{3pt}
    \end{subfigure} 
    \caption{Continued}
\end{figure*}

\end{appendices}

\bsp	
\label{lastpage}
\end{document}